%% file: zjets.tex
\documentclass[showpacs,preprintnumbers,amsmath,amssymb,nofootinbib]{revtex4}

\usepackage{setspace}
\usepackage{graphicx}  
\usepackage{dcolumn}   
\usepackage{bm}        
\usepackage{rotating}
\usepackage{hyperref}

\newcommand{\ptjet}{p_{ T}}

\newcommand{\etajet}{\eta}
\newcommand{\phijet}{\phi}
\newcommand{\rapjet}{y}
\newcommand{\pthat}{\hat{p}_{T}}
\newcommand{\akt}{\hbox{anti-${k_t}$} }
\newcommand{\njet}{N_{\rm jet}}

\newcommand{\mjj}{{m}^{jj}}
\newcommand{\phijj}{\Delta \phi^{jj}}
\newcommand{\rjj}{\Delta R^{jj}}
\newcommand{\rapjj}{\Delta y^{jj}}
\newcommand{\zee}{Z/\gamma^*(\rightarrow e^+ e^-)}
\newcommand{\zll}{Z/\gamma^*(\rightarrow \ell^+ \ell^-)}
\newcommand{\zmm}{Z/\gamma^*(\rightarrow \mu^+ \mu^-)}
\newcommand{\ztt}{Z/\gamma^*(\rightarrow \tau^+ \tau^-)}
\newcommand{\zees}{Z/\gamma^*\rightarrow e^+ e^-}
\newcommand{\zlls}{Z/\gamma^*\rightarrow \ell^+ \ell^-}
\newcommand{\zmms}{Z/\gamma^*\rightarrow \mu^+ \mu^-}
\newcommand{\ztts}{Z/\gamma^*\rightarrow \tau^+ \tau^-}

\newcommand{\wens}{W\rightarrow e \nu}
\newcommand{\wmns}{W\rightarrow \mu \nu}
\newcommand{\ete}{E_{\rm T}^{e}}
\newcommand{\etae}{\eta^{e}}
\newcommand{\ptm}{p_{\rm T}^{\mu}}
\newcommand{\etam}{\eta^{\mu}}
\newcommand{\mee}{m_{e^+ e^-}}
\newcommand{\mmm}{m_{\mu^+ \mu^-}}
\newcommand{\mll}{m_{\ell^+ \ell^-}}

\begin{document}

\preprint{CERN-PH-EP-2011-162}
\preprint{Submitted to Physical Review D}


\title{Measurement of the production cross section for $Z/\gamma^*$ in association with jets in $pp$ collisions at $\sqrt{s}=7~$TeV with the ATLAS detector}
\author{(The ATLAS Collaboration)}
\date{\today}


\begin{abstract}

Results are presented on the production of jets of particles in association with a 
$Z/\gamma^*$ boson, in proton-proton collisions at $\sqrt{s} = 7$~TeV
with the ATLAS detector. The analysis includes the full 2010 data set, collected with 
a low rate of multiple proton-proton collisions in the accelerator, corresponding to 
an integrated luminosity of  $36 \ \rm pb^{-1}$.    
Inclusive jet cross sections in $Z/\gamma^*$ events, 
with $Z/\gamma^*$ decaying into electron or muon  pairs, 
are measured for jets with 
transverse momentum $\ptjet >$~30~GeV and jet rapidity 
$|\rapjet | < 4.4$.  The measurements are compared to
next-to-leading-order perturbative QCD calculations, and to predictions from 
different Monte Carlo generators implementing leading-order matrix elements 
supplemented by parton showers.

\end{abstract}

\pacs{12.38.Aw, 12.38Qk, 13.87.Ce, 14.70.Hp}  

\maketitle


\section{Introduction}
\label{sec:intro}

The study of the production of jets of particles 
in association with a $Z/\gamma^*$ boson in 
proton-proton collisions provides a stringent 
test of perturbative quantum chromodynamics (pQCD).
In addition, the proper understanding of these processes 
in the Standard Model (SM) is a fundamental element of the LHC physics program, 
since they 
constitute backgrounds in searches for new physics. 
These SM 
background contributions are estimated using next-to-leading order (NLO) pQCD calculations, and 
Monte Carlo (MC) predictions that include 
leading-order (LO) matrix elements supplemented by parton showers. The latter  
are affected by large scale uncertainties and need to be tuned and validated using data. 
Measurements of $Z/\gamma^*$+jets production have been previously reported in 
proton-antiproton collisions at $\sqrt{s} = 1.96$~TeV~\cite{tevatron} and in 
proton-proton collisions at $\sqrt{s} = 7$~TeV~\cite{cms}.  

This article presents measurements of jet production in events with 
a $Z/\gamma^*$ boson in the final state, 
using   $36 \pm 1 \ \rm pb^{-1}$ of data collected by the ATLAS experiment in 2010 at $\sqrt{s} = 7$~TeV.
In this period, the accelerator operated with a moderate instantaneous luminosity of up to $2.1 \times 10^{32} \ \rm cm^{-2} s^{-1}$,  
and a long spacing of 150~ns between proton bunches, leading to relatively low collision rates and    
low rates of multiple proton-proton interactions per bunch crossing (pileup) and out-of-time pileup, which makes 
this data sample especially suitable for cross section measurements at low jet transverse 
momentum $\ptjet$~\cite{coord}. 

Events are selected with 
a $Z/\gamma^*$ decaying into a pair of  electrons $(e^+ e^-)$ or muons $(\mu^+ \mu^-)$, and the measurements are 
corrected for detector effects.      
Inclusive jet differential cross sections are measured as functions of jet transverse momentum, $\ptjet$,  and rapidity,
$|\rapjet|$, and total cross sections as functions of jet multiplicity, $\njet$, in well-defined kinematic
regions for the leptons and jets in the final state. 
Differential cross sections are also measured as functions of $\ptjet$ and $|\rapjet|$ of the 
leading jet (highest $\ptjet$) and second leading jet in $Z/\gamma^*$ events with at least one and two jets in the final state, respectively.
For the latter, the cross section is measured as a function of the invariant mass and the angular separation of the 
two leading jets.   
The data are  compared to NLO pQCD predictions~\cite{mcfm,black}, including non-perturbative contributions, and to predictions from 
several MC programs.


The paper is organized as follows. The detector is described in the next Section. Section~\ref{sec:evt} 
discusses the event selection, while
Section~\ref{sec:sim} provides details
of the simulations used in the measurements and
Sections~\ref{sec:jetrec} and~\ref{sec:leprec} 
describe the reconstruction of jets and leptons, respectively.
The estimation of background contributions is described in Section~\ref{sec:backg}. 
Selected uncorrected distributions 
are presented in Section~\ref{sec:uncorr}, and 
the procedure used to correct the measurements for detector effects is explained in Section~\ref{sec:unfold}.
The study of systematic uncertainties is discussed in Section~\ref{sec:sys}. The NLO pQCD predictions are 
described in Section~\ref{sec:nlo}. The  measured cross sections are presented separately for the electron and muon channels
in Section~\ref{sec:results},   
where the combination of the electron and muon results is also discussed.
Finally, Section~\ref{sec:sum} provides a summary.


\section{Experimental setup}
\label{sec:atlas}


The ATLAS detector~\cite{atlas} covers almost the whole solid angle around
the collision point with layers of tracking detectors, calorimeters and muon
chambers. The ATLAS inner detector (ID) has full coverage  in $\phi$
and covers the pseudorapidity range $|\eta|<2.5$. It consists of a silicon pixel detector, a silicon microstrip detector (SCT), and a straw tube tracker (TRT) which also measures transition radiation for particle identification, all immersed in a 2 tesla axial  magnetic field produced by a solenoid.

High-granularity liquid-argon (LAr) electromagnetic sampling calorimeters, with very good
energy and position resolution~\cite{lar}, cover the pseudorapidity
range $|\eta|<$~3.2. The hadronic calorimetry in the range $|\eta|<$~1.7 is provided by a scintillator-tile calorimeter, consisting of a large barrel and two smaller extended barrel cylinders, one on either side of
the central barrel. In the end caps ($|\eta|>$~1.5), LAr hadronic
calorimeters match the outer $|\eta|$ limits of the end cap electromagnetic calorimeters. The LAr
forward calorimeters provide both electromagnetic and hadronic energy measurements, and extend
the coverage to $|\eta| < 4.9$.

 The muon spectrometer measures the deflection of muon tracks in the large superconducting air-core toroid magnets
in the pseudorapidity range  $|\eta|<2.7$, instrumented with separate trigger and high-precision tracking chambers. 
Over most of the $\eta$-range, a precision measurement of the track coordinates in the principal bending direction of the magnetic field is provided by monitored drift tubes. At large pseudorapidities, cathode strip chambers  with higher granularity are used in the innermost plane over $2.0 < |\eta| < 2.7$.
The muon trigger system,  which covers the pseudorapidity range $|\eta| < 2.4$, consists of resistive plate chambers in the barrel ($|\eta|<1.05$) and thin gap chambers  in the end cap regions ($1.05 < |\eta| < 2.4$), with a small overlap in the $|\eta| = $1.05 region.


\section{$\zlls$ selection}
\label{sec:evt}

The data samples considered in this paper were collected with
tracking detectors, calorimeters, muon chambers, and magnets 
fully operational, and correspond 
to a total integrated luminosity of $36 \ \rm pb^{-1}$. 

In the case of the $\zees$ analysis, events are selected online using a
trigger that requires the presence of at least one identified electron candidate 
in the calorimeter with transverse energy above 15 GeV in the region $|\eta|<2.5$. 
The events are then selected to 
have two oppositely charged reconstructed electrons ({\it{medium}} quality electrons, as described in Ref.~\cite{elec}) with transverse energy $\ete > 20$~GeV,  
pseudorapidity in the range $|\etae|<2.47$ (where the transition 
region between calorimeter sections $1.37 < |\etae| < 1.52$ is excluded), and a dilepton 
invariant mass in the range $66$~GeV$ < \mee < 116$~GeV, which optimizes the signal sensitivity.

The $\zmms$ sample is collected online using a trigger that requires the 
presence of at least one muon candidate reconstructed in the muon spectrometer, consistent with having 
originated from the interaction region with $p_{\rm T} > 10$~GeV or $p_{\rm T} > 13$~GeV, 
depending on the data period, and with the majority of the data taken with the higher threshold,  
and $|\eta|<2.4$. The muon 
candidates are associated with track segments reconstructed 
in the inner detectors which, combined with the muon spectrometer information,  
define the final muon track. Combined muon tracks with $\ptm > 20$~GeV and $|\etam| < 2.4$ are 
selected. A number of quality requirements are applied to the muon candidates~\cite{muon}: the associated 
inner detector track segment is required to have a minimum number of hits in the pixel, SCT and TRT detectors; and 
the muon transverse and longitudinal impact parameters, $d_{\rm{0}}$ and $z_{\rm{0}}$, with respect to the reconstructed primary vertex  
are required to be $d_{\rm{0}}/\sigma(d_{\rm{0}}) < 3$ and $z_{\rm{0}}<10$~mm  in the $r - \phi$ and $r - z$ planes, respectively, where $\sigma(d_{\rm{0}})$ denotes the $d_{\rm{0}}$ resolution. 
The muons are required to be isolated:  
 the scalar sum of the transverse momenta of the tracks in an $\eta - \phi$ cone
of radius 0.2 around the muon candidate is required to be less than 10$\%$ of the muon $p_{\rm T}$. 
Events are selected with two oppositely charged muons and an invariant mass $66$~GeV$ < \mmm < 116$~GeV. 

In both analyses, events are required to 
have a reconstructed primary vertex of the interaction with at least 3 tracks associated to it, which 
suppresses beam-related background contributions and cosmic rays. The selected dilepton samples
contain a total of 9705 and 12582 events for the electron and muon channels, respectively.


\section{Monte Carlo simulation}
\label{sec:sim}

Monte Carlo event samples are used to 
compute detector acceptance and reconstruction efficiencies,
determine background contributions, correct the measurements for detector effects, 
and estimate systematic uncertainties on the final results. 

Samples of simulated $\zee$+jets and $\zmm$+jets events with a dilepton invariant mass above 40~GeV are generated using 
ALPGEN v2.13~\cite{alpgen} (including  LO matrix elements for up to $2 \to 5$ parton scatters) interfaced to HERWIG v6.510~\cite{herwig} for parton shower
and fragmentation into particles, and to JIMMY v4.31~\cite{jimmy} to model underlying event (UE) contributions. 
Similar samples are generated using Sherpa 1.2.3~\cite{sherpa} with an UE modeling according to Ref.~\cite{sherpaUE}. 
For the ALPGEN samples CTEQ6L1~\cite{cteq} parton density functions (PDFs) are employed, while for 
Sherpa  CTEQ6.6~\cite{cteq66} is used.  
The ALPGEN and Sherpa samples  are normalized to the next-to-next-to-leading order (NNLO) pQCD  
inclusive Drell-Yan prediction  of $1.07 \pm 0.05$~nb,   
 as determined by the FEWZ~\cite{nnlo} program
 using the MSTW2008 PDFs.
In addition, $Z/\gamma^*$+jets samples ($q\bar{q} \to Z/\gamma^* g$ and $qg \to Z/\gamma^* q$ processes  
 with $\pthat > 10$~GeV, where $\pthat$
is the transverse momentum defined in the rest frame of the hard interaction)   
are produced using PYTHIA v6.423~\cite{pythia} and HERWIG plus JIMMY with 
MRST2007LO${}^*$~\cite{mrst2007lo} PDFs. 
For the ALPGEN and HERWIG plus JIMMY MC samples the  AUET1~\cite{auet1} tuned set of parameters is used 
to model the UE activity in the final state. 
In the case of the PYTHIA samples, the AMBT1~\cite{ambt1} tune is employed.

Background samples from $W$+jets and $\ztt$+jets final states, and diboson 
($WW$, $WZ$, $ZZ$) processes are generated using 
ALPGEN with CTEQ6L1 PDFs normalized to NNLO~\cite{nnlo}  and NLO~\cite{mcfm}  
pQCD predictions, respectively. TAUOLA v1.0.2~\cite{taus} is used for tau decays.
Simulated top-quark production samples are generated  
using MC@NLO~\cite{mcatnlo} and CTEQ6.6 PDFs. 

The MC samples are generated with minimum bias interactions from PYTHIA overlaid on top of the hard-scattering event 
in order to account for the presence of the pileup 
experienced in the data. The number of minimum bias (MB) interactions follows a Poisson distribution with 
a mean of two, which is appropriate for the 2010 data. The MC generated samples are then passed through a full simulation~\cite{atlas_sim} 
of the ATLAS detector and trigger system, based on GEANT4~\cite{geant}. The simulated events are
reconstructed and analyzed with the same analysis chain as for the data, using the same trigger and event 
selection criteria, and  re-weighted such 
that the distribution of the number of primary vertices matches that of the data. 

The multi-jets background contributions in the electron and muon channels are determined using data, as discussed 
in Section~\ref{sec:backg}.


\section{Jet reconstruction}
\label{sec:jetrec}

Jets are defined using the $\akt$ jet algorithm~\cite{akt} with the distance parameter 
set to $R=0.4$. 
Energy depositions reconstructed as calorimeter clusters are the inputs to the jet algorithm 
in data and MC simulated events. The same jet 
algorithm is applied to final state particles in the MC generated events to define jets at particle 
level~\cite{hadron}. The jet kinematics in data and MC simulated events are corrected to account for the following
effects: the presence of
additional proton-proton interactions per bunch crossing, leading to an additional energy offset of  
$(500 \pm 160)$~MeV within the jet cone for each extra interaction~\cite{offset};  
the position of the primary vertex of the interaction; and the measurement biases induced by calorimeter non-compensation, additional dead material, and out-of-cone 
effects.  
The measured 
jet $\ptjet$ is corrected for detector effects back to the true jet energy~\cite{jet_pro_atlas} using an average correction, computed as a function of the jet transverse momentum and pseudorapidity, and extracted from inclusive jet MC samples.
The measured jet $\ptjet$ is reconstructed with a resolution of about 10$\%$ at low $\ptjet$ which improves to  
6$\%$ for  $\ptjet$ about 200~GeV.
The measured jet angular variables $\rapjet$ and $\phijet$ are reconstructed with no significant shift and a resolution better than 0.05, which improves as the jet transverse momentum  increases. 

In this analysis, jets are selected with corrected  $\ptjet > 30$~GeV and $|\rapjet| < 4.4$ to ensure
full containment in the instrumented region.
Events are required to have at least one jet well separated  
from the final state leptons from the $Z/\gamma^*$ decay. 
Jets within a cone of radius 0.5 around any selected lepton are not considered.
Additional quality criteria are  applied to ensure that jets are not produced by 
noisy calorimeter cells, and to avoid problematic detector regions.

The final sample for $\zee$+jets contains 1514, 333, 62, and 15  
events with at least one, two, three, and four jets in the final state, respectively.      
Similarly, the $\zmm$+jets sample contains 1885, 422, 93, and 20  
events with at least one, two, three, and four jets in the final state, respectively.

\section{Lepton reconstruction}
\label{sec:leprec}

Samples of $\zees$ and $\zmms$  events in  data and MC simulation, together with
the world average values for the $Z$ boson mass and width, are
used to determine
the absolute scale and 
resolution of the energy/momentum of the leptons, to validate calibration- and alignment-related constants in data, 
and to check the MC description~\cite{lepton_scales}.
In addition,  the trigger and offline lepton reconstruction efficiencies are studied 
using control samples in data, and the results are compared to the simulation. The differences 
observed between data and MC predictions define scale factors which are 
applied in the analysis to the simulated samples before they are used to correct 
the measurements for detector effects.

For the electron channel, the trigger and offline electron reconstruction and identification 
efficiencies for single electrons 
are  estimated using $\wens$ and $\zees$ events in data and compared to MC predictions.
In the kinematic range for the electrons considered in the analysis (see Section~\ref{sec:evt}), 
the trigger and offline efficiencies per 
electron are above 99$\%$ and 93$\%$, respectively.
The study indicates a good agreement between data and simulated trigger efficiencies with a MC-to-data scale
factor of $0.995 \pm 0.005$. The simulation tends
to  overestimate the offline efficiencies.  Scale factors in the range between $0.901 \pm 0.045$ and  $0.999 \pm 0.016$, depending on $\etae$ and $\ete$, for $\ete > 20$~GeV, are applied per lepton to the MC samples to 
account for this effect.  

In the muon analysis, the trigger and offline muon reconstruction efficiencies are also 
estimated using the data and are compared to simulation. 
The measured average single muon trigger efficiency is about $85\%$, independent of $\ptm$, and   
varies from 80$\%$ for $|\etam|< 0.63$ and 73$\%$ for $0.63 < |\etam|< 1.05$  to 94$\%$ 
for $1.05 < |\etam|< 2.4$, limited mainly by the trigger chamber geometric acceptance. 
The measured average offline muon reconstruction efficiency is about 92$\%$
and approximately independent of $\ptm$. 
The MC simulation predicts efficiencies very similar to those in the data, but tends to 
overestimate the average offline reconstruction 
efficiency by about 1$\%$. This originates from the transition region  between the barrel part and 
the endcap wheels at $|\eta| \sim 1$, where the simulation 
overestimates the offline reconstruction efficiency by about $6\%$. The latter is attributed to
the limited accuracy of the magnetic field map used in this region which leads to a small 
mismeasurement of the stand-alone muon momentum and an  overestimation in the simulated 
efficiency. Scale factors are applied in the analysis that take this effect into account. 

\section{Background estimation}
\label{sec:backg}

The background contribution to the electron and muon  analyses from SM processes 
is estimated using MC simulated samples, as discussed in Section~\ref{sec:sim}, with the exception of 
the multi-jets background that is estimated using data. 

The multi-jets background contribution in the $\zee$+jets analysis is estimated using a 
control data sample with two electron candidates 
which pass a loose selection   
but fail to pass the  {\it{medium}} identification requirements. This sample is dominated by jets faking electrons in the final state and 
is employed to determine
the shape of the multi-jets background  under each of the measured distributions. 
The normalization  of the multi-jets background events in the signal region is extracted
from a fit to the measured inclusive dilepton invariant mass spectrum with nominal lepton requirements, 
using as input the observed shape of the multi-jets contribution in data and the   
MC predictions for the shape of the signal and the rest of the SM background processes. 
The multi-jets background contribution to the measured inclusive jet multiplicity varies between
$3.2 \pm 0.5 (\rm stat.) {}^{+ 0.3}_{-0.2} (\rm{syst.}) \%$ for $\njet \geq 1$ and  
$4.5 \pm 1.9 (\rm stat.) {}^{+ 0.4}_{-0.2} (\rm{syst.}) \%$ for $\njet \geq 4$. 
The quoted total systematic 
uncertainty includes: uncertainties related to the details of the parameterization and the mass range  
used to fit the measured dilepton invariant mass spectrum; uncertainties on the shape of the dilepton 
invariant mass distribution, as determined in the control sample; and uncertainties 
on the shape of the simulated dilepton invariant mass distribution for the other SM processes. 

In the $\zmm$+jets case, the multi-jets background  mainly originates from heavy-flavour jet production 
processes, with muons from bottom and charm quark decays, as well as from the decay-in-flight of pions and kaons, which
are highly suppressed by the isolation requirement applied to the  muon candidates.
The  isolation criterion of the muon pair, 
defined as the isolation of the least-isolated muon candidate, is used together with the 
dimuon invariant mass to estimate the remaining multi-jets 
background contribution. The MC simulation indicates that, for multi-jet processes, the muon isolation is not
correlated with the dimuon invariant mass, and so the ratio of isolated to non-isolated muon pairs (as defined 
with an inverted isolation criterion) does not depend on the dimuon mass. 
The multi-jets background with isolated muons with  $66$~GeV$ < \mmm  < 116$~GeV is therefore extracted from data
as the ratio between the number of isolated and non-isolated dimuon candidates in the region $40$~GeV$ < \mmm < 60 $~GeV multiplied by the number of non-isolated dimuon candidates in the range $66$~GeV$ < \mmm  < 116$~GeV. A small contribution 
from top pair production processes is subtracted from the data according to MC predictions. 
The multi-jets background contribution to the $\zmm$+jets analysis is of the order of one per mille 
and therefore negligible. 

In the electron channel, the total background increases from 5$\%$ to 17$\%$  as 
the jet multiplicity increases and is dominated by multi-jet processes, followed
by contributions from $t\bar{t}$ and diboson production at large jet multiplicities. 
In the muon channel, the SM background contribution increases  
from $2\%$ to $10\%$ as the jet multiplicity increases, dominated by $t\bar{t}$ and diboson processes.
%
%
Table~\ref{tab:events} shows, for the electron and muon analyses separately, the observed number of events 
for the different jet multiplicities in the final state compared to predictions for signal and background 
processes.

\begin{table}[htbp]
\begin{center}
\begin{footnotesize}
\begin{tabular}{|c c c c c c|} \hline\hline
%
%
\multicolumn{6}{|c|}{$\zees$ channel} \\ \hline\hline
               &               & $\geq 1$~jet  &  $\geq 2$~jets  & $\geq 3$~jets  & $\geq 4$~jets       \\
\hline
$\zees$             & ALPGEN           &   1357   &  307   & 64.4    &  12.7   \\
$\wens$             & ALPGEN           &    4.3   &  1.0   & 0.31    &  0.11   \\
$\ztts$             & ALPGEN            &   0.9  &  0.25   & 0.03    & 0.005   \\
$WW,WZ,ZZ$         & ALPGEN           &     9.6   &  4.8   & 1.7    &  0.45   \\
$t\bar{t}$         & MC@NLO           &    11.7   &  9.2   &  4.3    &  1.3   \\
multi-jets           & from data      &      49   & 12.6   &  2.2    &   0.7   \\
\hline
SM prediction      &           &   1432   &   334   &   72.9    &   15.2   \\
data (36 pb${}^{-1}$)  &        &   1514             &   333               &   62               &   15 \\ \hline\hline
%
%
\multicolumn{6}{|c|}{$\zmms$ channel} \\ \hline\hline
              &      & $\geq 1$~jet  &  $\geq 2$~jets  & $\geq 3$~jets  & $\geq 4$~jets       \\ 
\hline
$\zmms$        & ALPGEN       & 1869 &     421 &  87.2   & 17.7   \\
$\wmns$        & ALPGEN       &  0.3 &    0.06 &  0.04   & 0.04           \\
$\ztts$        & ALPGEN       & 0.68 &    0.11 &  0.03   & $<$0.01     \\
$WW,WZ,ZZ$      & ALPGEN     &  12.8 &     6.8  & 2.3   & 0.57    \\
$t\bar{t}$      & MC@NLO     &  13.6 &    10.7  &  4.6  &  1.4    \\
multi-jets     &  from data   &    1 &     0.3 &   0.1   & 0.01             \\
\hline
SM prediction   &   &   1898     &  439   &  94.2   &  19.8    \\
data (36 pb${}^{-1}$) &       & 1885  &  422  &  93  &   20 \\
\hline\hline
\end{tabular}
\end{footnotesize}
\caption{\small 
Number of events for the $\zees$ and $\zmms$ analyses as a function of inclusive jet multiplicity. The data are compared to 
the predictions for the signal (as determined by ALPGEN) and background processes (see Sections~\ref{sec:sim} and ~\ref{sec:backg}). 
No uncertainties are indicated. The statistical uncertainty on the total 
prediction is negligible, and the corresponding systematic uncertainty  
varies between $10\%$ and $23\%$ with increasing $\njet$.
}
\label{tab:events}
\end{center}
\end{table}


\section{Uncorrected distributions}
\label{sec:uncorr}

\begin{figure}[h]
\begin{center}
\mbox{
\includegraphics[width=0.495\textwidth]{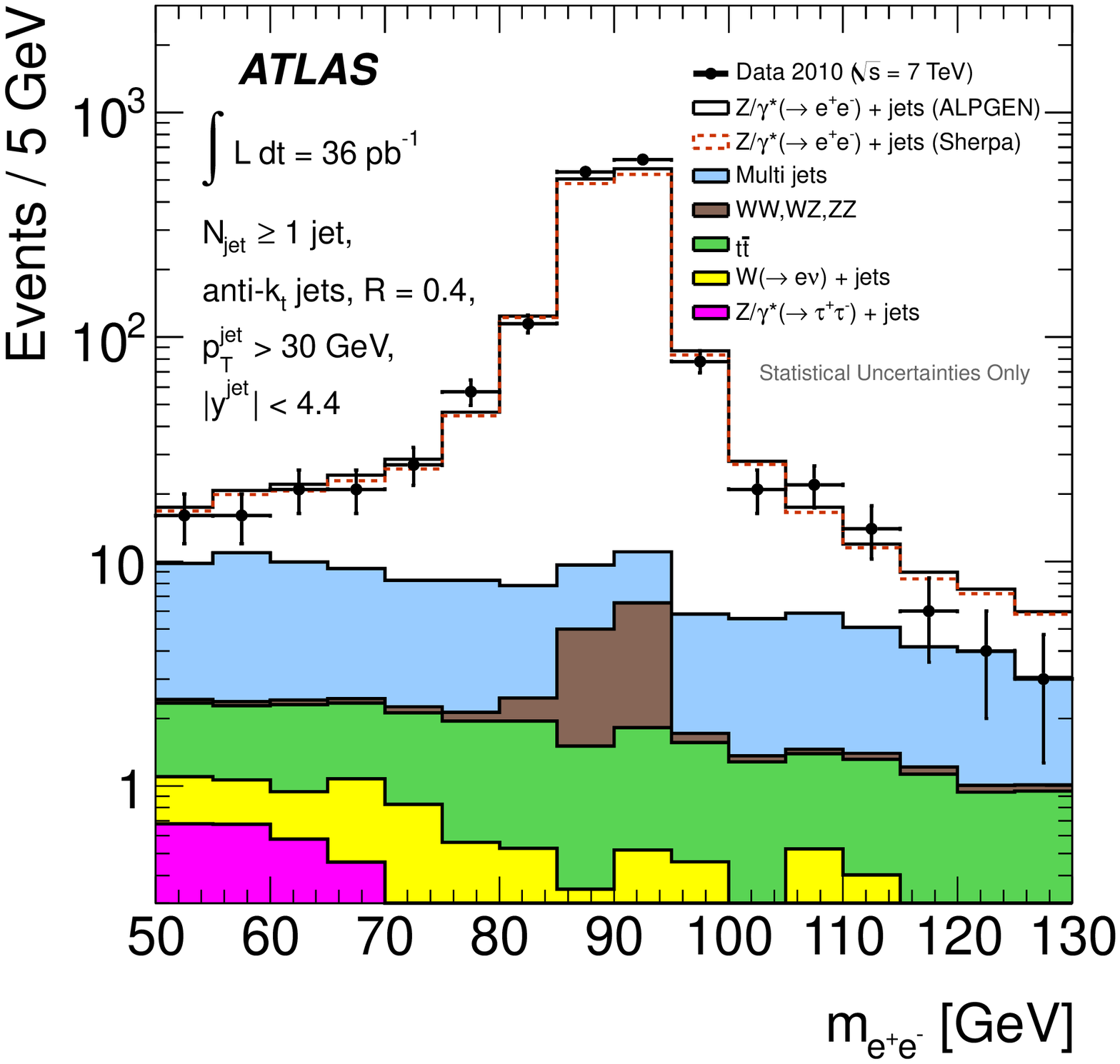}
\includegraphics[width=0.495\textwidth]{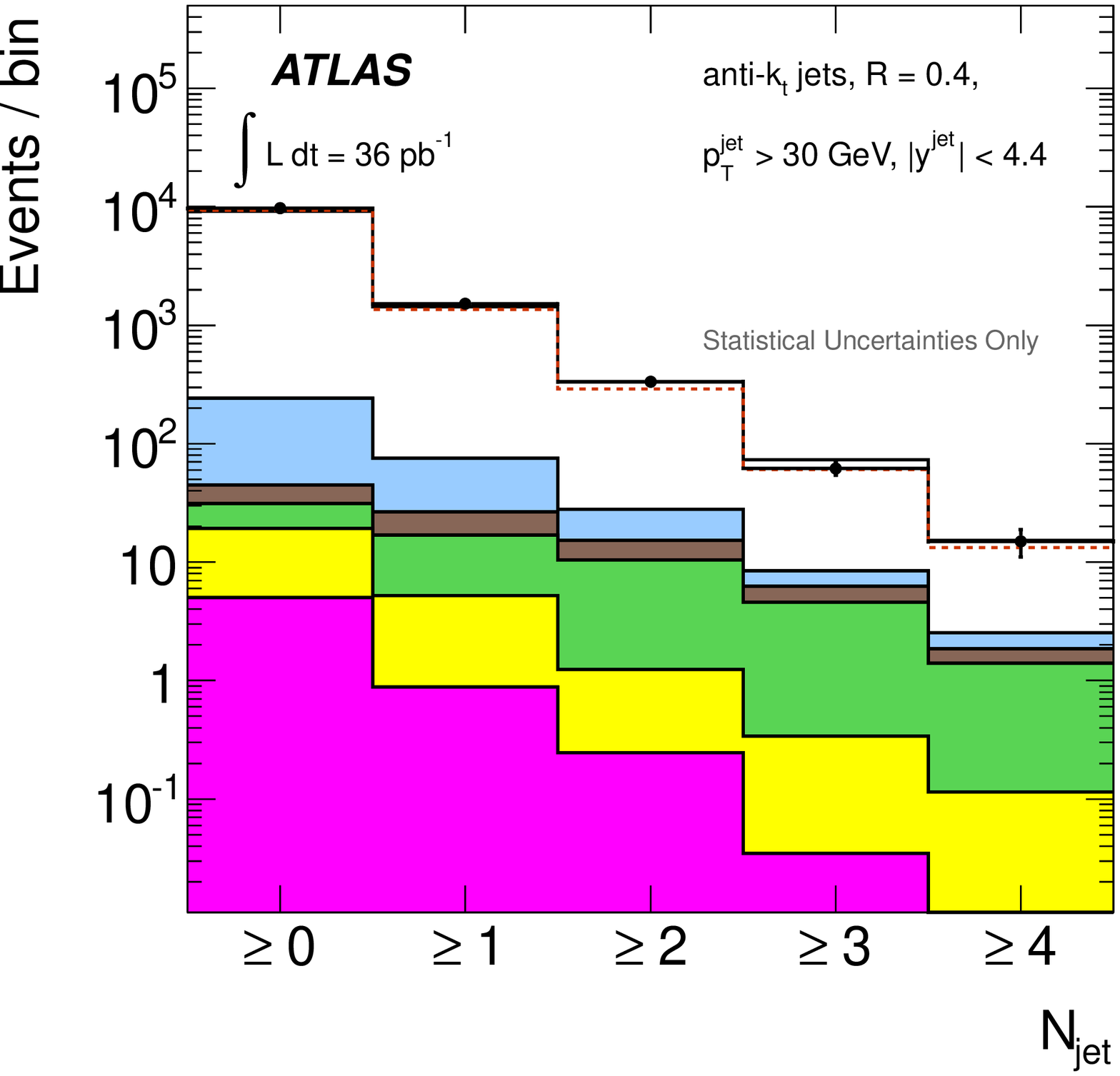}
}
\mbox{
\includegraphics[width=0.495\textwidth]{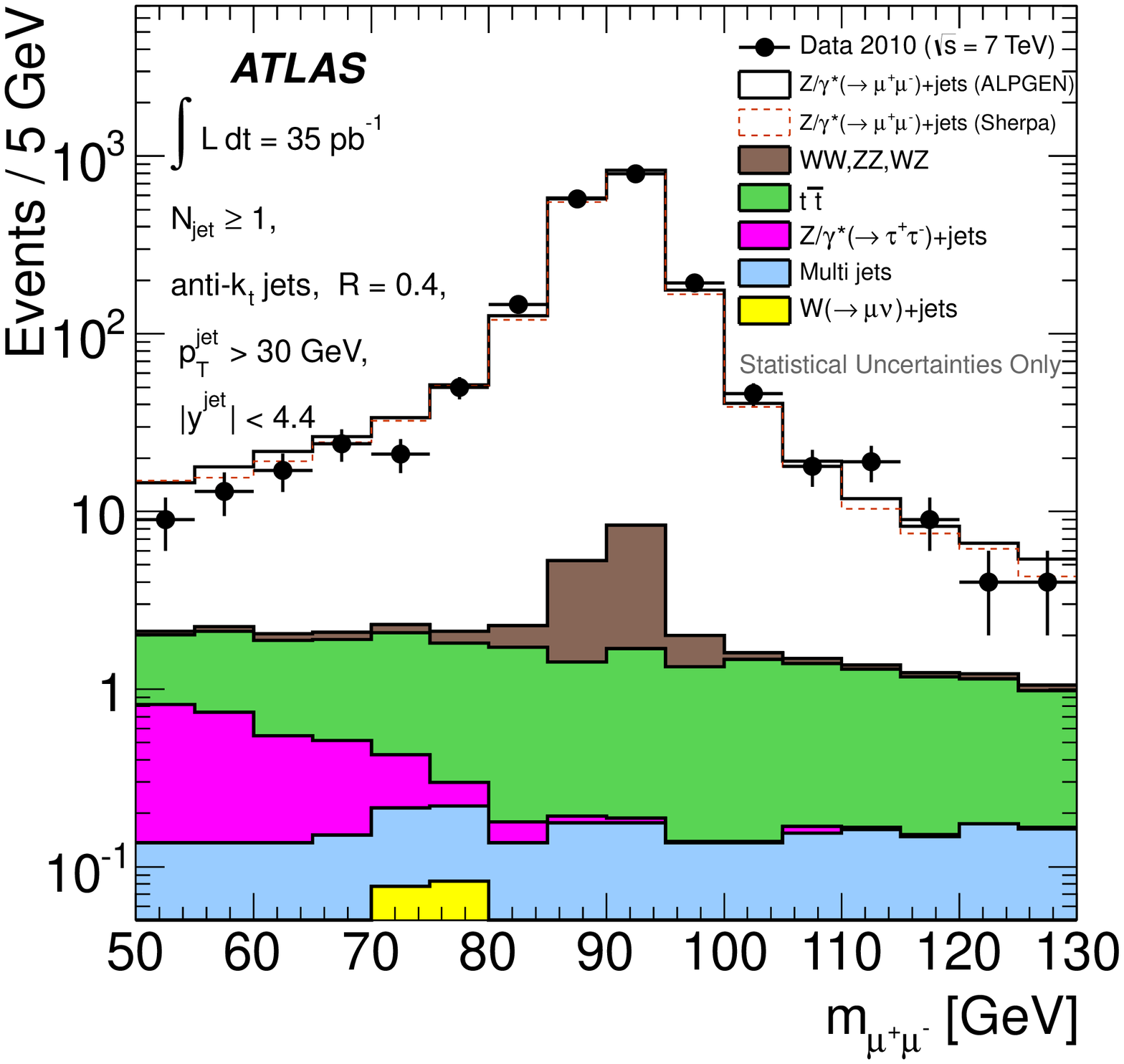}
\includegraphics[width=0.495\textwidth]{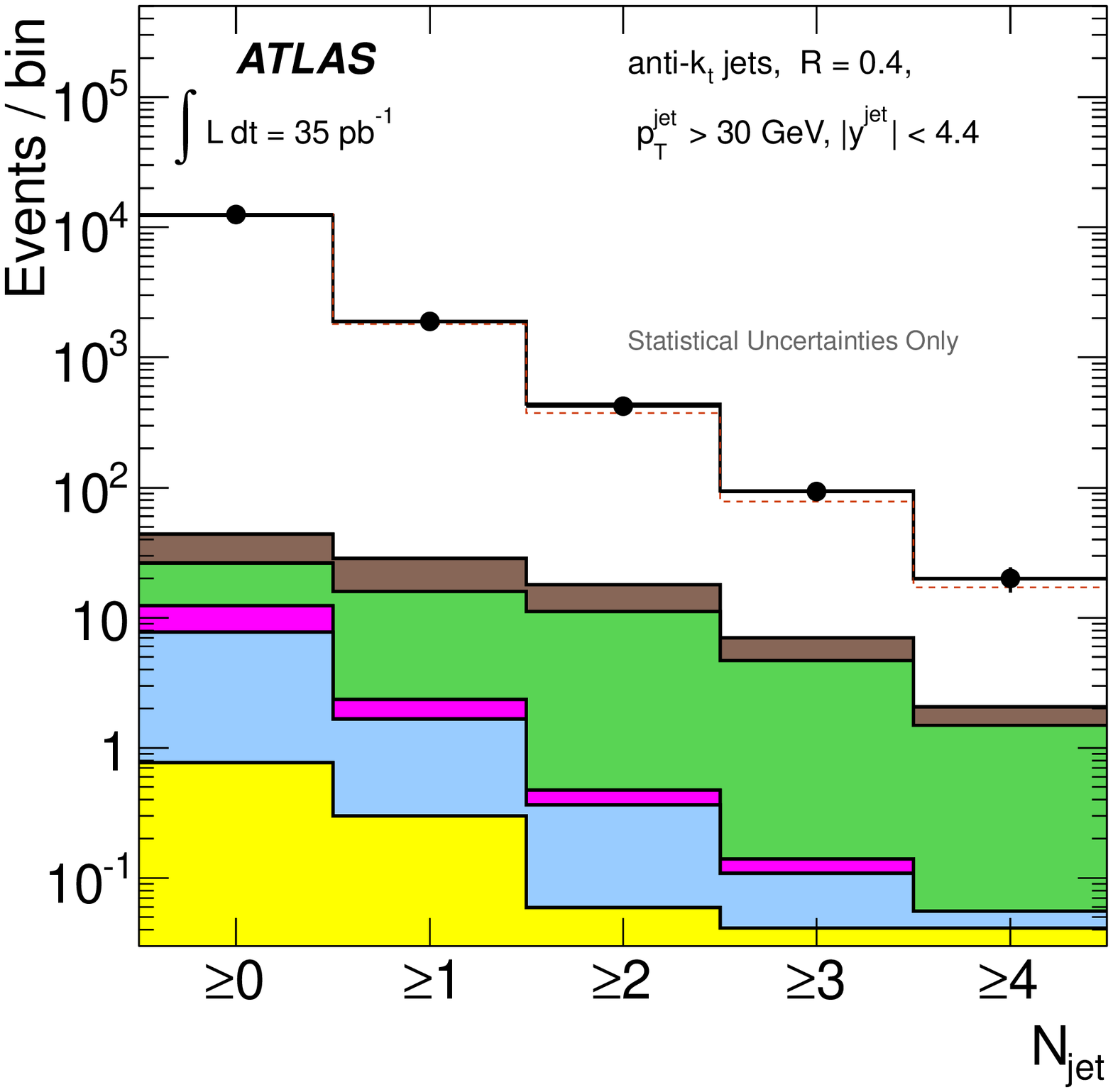}
}
\end{center}
\caption{\small
Uncorrected dilepton invariant mass in (top) $\zees$ and  (bottom) $\zmms$ events
with at least one jet in the final state, shown in 
a wider dilepton mass region than the one selected
(left), and  
uncorrected  inclusive jet multiplicity (right), 
for jets with $\ptjet > 30$~GeV and $|\rapjet|<4.4$ (black dots), and in the mass range $66$~GeV$< \mll < 116$~GeV ($\ell = e, \mu$).  
Only statistical uncertainties are shown.
The data are compared to 
predictions for signal (ALPGEN and Sherpa, both normalized to the FEWZ value for the total cross section) and background processes (filled histograms). 
}
\label{fig:uncorr_e_m_1}
\end{figure}

The uncorrected $\zee$+jets  and $\zmm$+jets 
data are compared to the predictions for signal and background contributions.
For the signal, both ALPGEN and Sherpa predictions 
are considered.  
As an example, Fig.~\ref{fig:uncorr_e_m_1} shows, separately for the electron and muon channels, 
the measured dilepton invariant mass in events with at least one 
jet in the final state, as well as the measured uncorrected inclusive jet multiplicity. 
Other observables considered include:  
the uncorrected inclusive jet $\ptjet$, $\rapjet$, and $\phijet$
distributions; the corresponding $\ptjet$, $\rapjet$, and $\phijet$ 
distributions of the leading, second-leading and third-leading 
jet in events with at least one, two and three jets in the final state, respectively;  
  the 
invariant mass of the two leading jets, $\mjj$, and their rapidity difference, $\rapjj$, their azimuthal separation, $\phijj$,  
and the angular separation in $y -\phi$ space, $\rjj = \sqrt{(\rapjj)^2 + (\phijj)^2}$, in 
 events with at least two jets in the final state.  
In all cases, the data yields are described, within statistical uncertainties, by the MC predictions for the signal 
plus the estimated SM background contributions.

\section{Correction for detector effects}
\label{sec:unfold}

The jet measurements are 
corrected for detector effects back to the particle level 
using a bin-by-bin correction procedure, based on MC simulated samples, 
that corrects for jet selection efficiency and resolution effects and also 
accounts for the efficiency of the $Z/\gamma^*$ selection. 

The corrected measurements refer to particle level jets identified using the 
$\akt$ algorithm with $R=0.4$, for jets with $\ptjet >30$~GeV and $|\rapjet|<4.4$.
At particle level, the 
lepton kinematics in the MC generated samples     
include the contributions from the photons radiated within a cone of radius 0.1  
around the lepton direction. 
The measured cross sections are defined in a limited kinematic range for the $Z/\gamma^*$ decay products.
\begin{itemize}
\item
In the electron channel, the measured cross sections refer to the region: $66$~GeV$ < \mee < 116$~GeV, 
$\ete >20$~GeV, $|\etae| < 1.37$ or $1.52 < |\etae| < 2.47$, and $\Delta R(\rm jet - electron) > 0.5$. 
\item Similarly, in the muon case the
measurements are presented in the region: $66$~GeV$ < \mmm < 116$~GeV, $\ptm > 20$~GeV, $|\etam|<2.4$, and 
$\Delta R(\rm jet - muon) > 0.5$.
\end{itemize}

\noindent
The ALPGEN samples for $Z/\gamma^*$+jets processes provide a satisfactory 
description of both lepton and jet distributions in data and are employed to compute the  
correction factors. 
For each observable $\xi$ the bin-by-bin correction factors $U(\xi)$ are defined as the ratio between the simulated 
distribution, after all selection criteria are applied, and the corresponding distribution at 
the particle level defined in a limited fiducial kinematic region  
for the generated leptons and jets, as detailed above.

Correction factors are considered for the following measurements: the 
inclusive jet multiplicity, $\ptjet$ and $|\rapjet|$ distributions; the $\ptjet$ and $|\rapjet|$ 
distributions for the leading- and second-leading jets in events with at least one and two jets, respectively; and
the invariant mass and angular separation distributions in the inclusive dijet sample. 
Typical correction factors are about $1.40$ for the electron channel and about $1.15$ for the muon channel 
(see below), where the difference is mainly attributed to the identification of the Z boson candidate 
in the final state. 

The measured  differential cross sections are defined as functions of a given $\xi$: 
\begin{equation}
\frac{\mathrm{d} \sigma}{\mathrm{d} \xi} = \frac{1}{\mathcal{L}} \ \frac{1}{\Delta \xi} (N_{\rm data} - N_{\rm backg}) \times U(\xi) 
\end{equation}

\noindent
where, for each bin in $\xi$, $N_{\rm data}$ and  $N_{\rm backg}$ denote the number of entries (events or jets) observed in data and the 
background prediction, respectively,  $\Delta \xi$ is the bin width, $U(\xi)$ is the correction factor, and $\mathcal{L}$ is the total integrated luminosity. 
The bin widths were chosen to be commensurate with the resolution, with typical correct-bin purities 
above $70\%$, and   
the cross section measurements 
are limited to bins in $\xi$ 
that contain at least ten entries in the data.


\subsection{Correction factors in the $\zees$ channel}

In the case of the inclusive jet multiplicity, the correction factors vary with the number of jets and are between 1.40 and 1.50.
The correction factors for the 
inclusive jet $\ptjet$ distribution and the $\ptjet$ distribution for the leading jet 
vary from  1.45 at $\ptjet$ around 30~GeV and 1.50 at $\ptjet$ about 60~GeV to 
1.42 at very large $\ptjet$. 
The corresponding factors for the $\ptjet$ distribution of the second-leading
jet increase from about 1.40 to 1.55 
with increasing $\ptjet$. 

The correction factors for the inclusive $|\rapjet|$ distribution and the $|\rapjet|$ 
distribution of the leading jet vary from 1.40 for central jets to about 1.60 for very forward jets. 
The correction factors for the $|\rapjet|$ distribution of the second-leading jets are about 1.45 and 
show a mild rapidity dependence. 

The correction factors for the $\Delta y$, $\Delta \phi$, and $\Delta R$ distributions between 
the two leading jets increase from 1.30 to 1.50
as the jet separation increases. 
Finally, the correction factor for the dijet invariant mass distribution varies between 1.40 and 1.55 
as $\mjj$ increases from 60~GeV to 300~GeV. At very low $\mjj$, the correction factors are about 0.90 and  
reflect a  large sensitivity to the $\ptjet$ thresholds 
applied in the analysis. Therefore, the cross section as a function of $\mjj$ is only reported  
for $\mjj > 60$~GeV.


\subsection{Correction factors in the $\zmms$ channel}

The correction factors for the inclusive jet multiplicity decrease from 1.15 to 1.08 with increasing $\njet$. 
The correction factors for the different $\ptjet$ distributions 
increase from 1.10 to 1.20 as $\ptjet$ increases from 30~GeV to 50~GeV and present a mild $\ptjet$ dependence 
for $\ptjet > 50$~GeV. Similarly, the corresponding factors for the different jet $|\rapjet|$ distributions 
vary between 1.15 for central jets and 1.20 for forward jets.  

The correction factors for the $\Delta y$,  $\Delta \phi$, and   
$\Delta R$ distributions, for
the two leading jets in events with at least two jets in the final state, 
vary between 1.10 and 1.20 as the jet separation increases.
The correction factors for the $\mjj$  
distribution vary between 1.10 and 1.20 as $\mjj$ increases. As in the electron case, the cross section as a function
of $\mjj$ is limited to the region $\mjj > 60$~GeV.

%
%

\section{Study of systematic uncertainties}
\label{sec:sys}

\begin{figure}
\begin{center}
\mbox{
\includegraphics[width=0.495\textwidth]{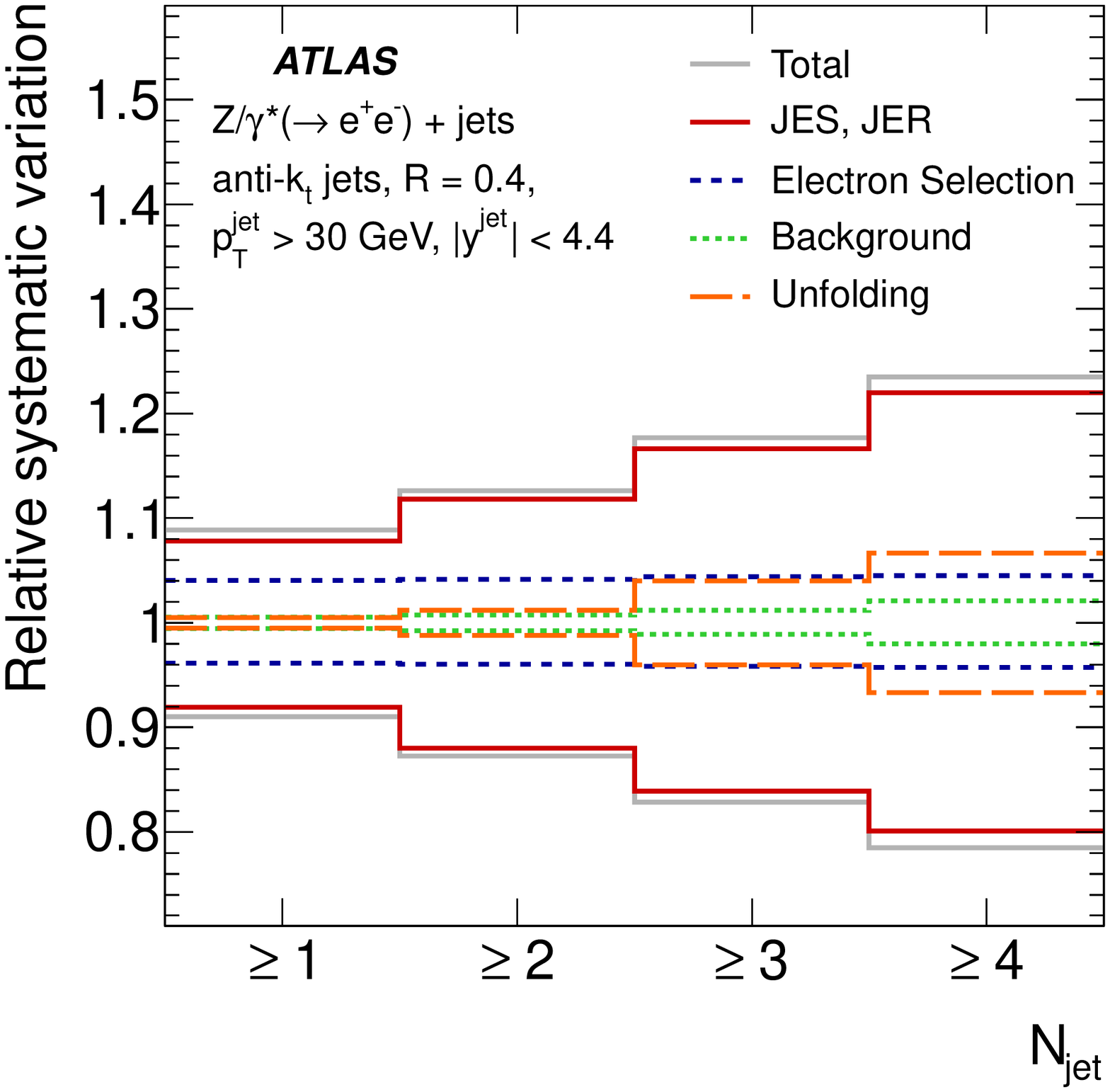}
\includegraphics[width=0.495\textwidth]{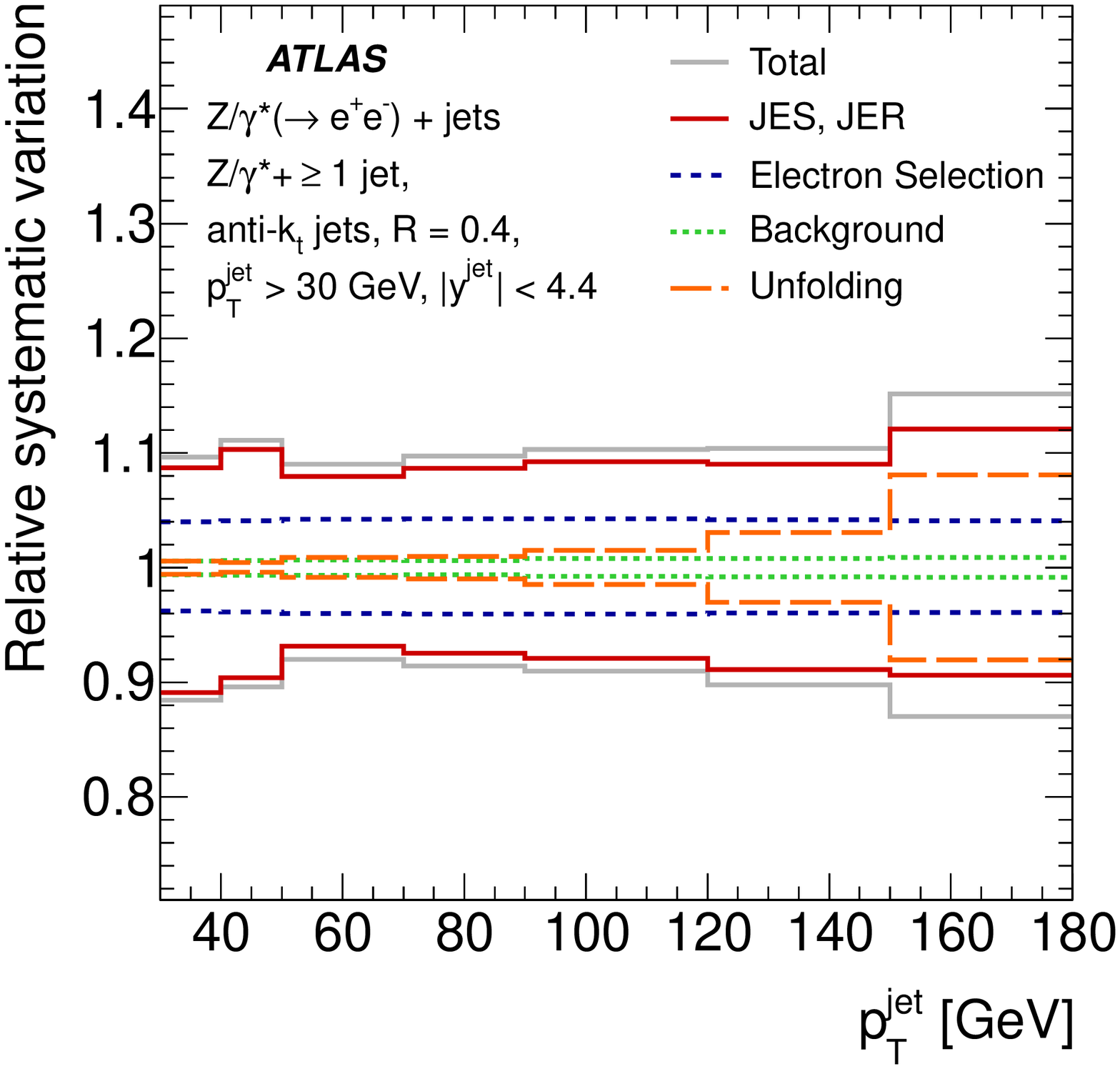}
}\\
\mbox{
\includegraphics[width=0.495\textwidth]{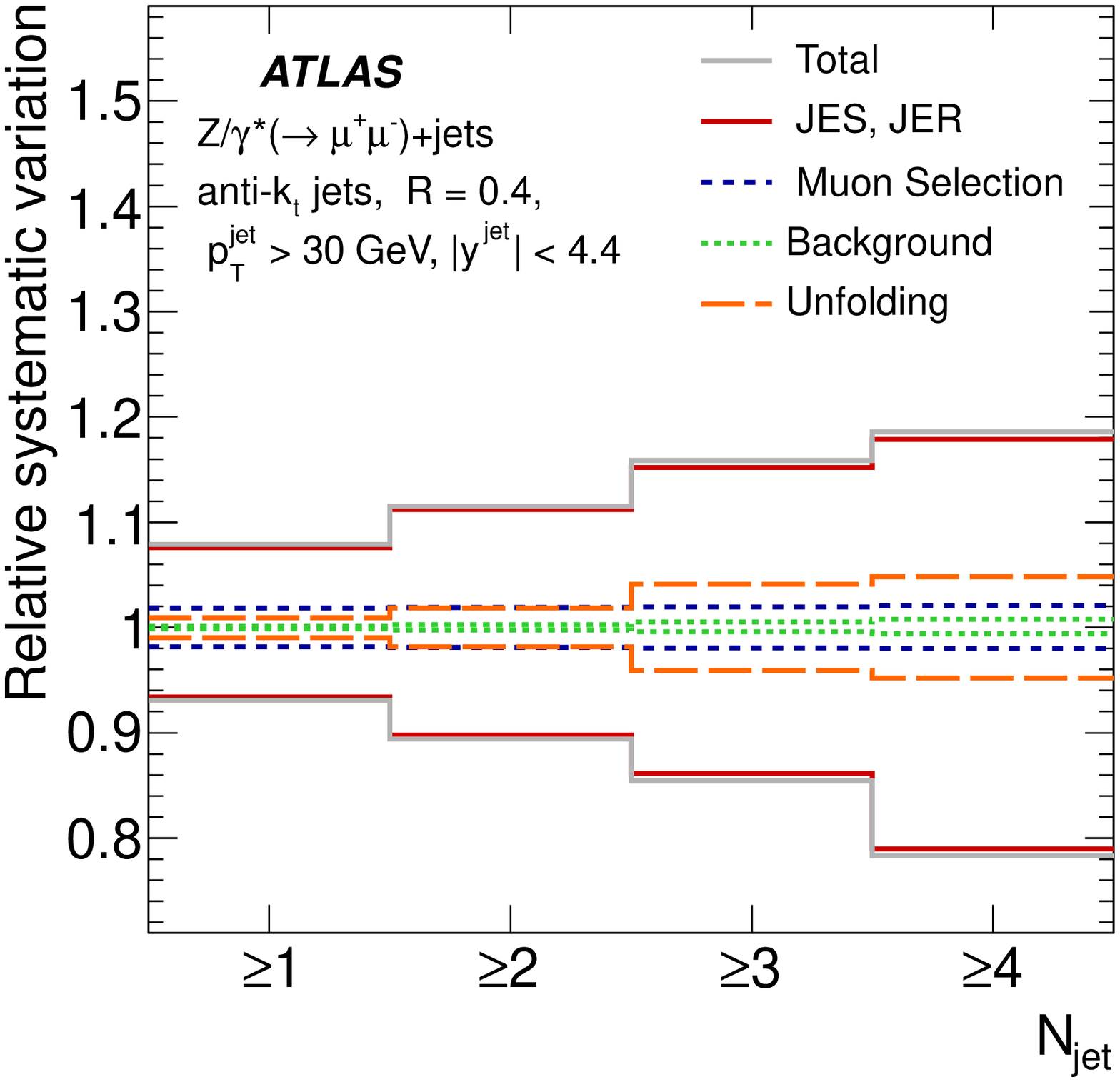}
\includegraphics[width=0.495\textwidth]{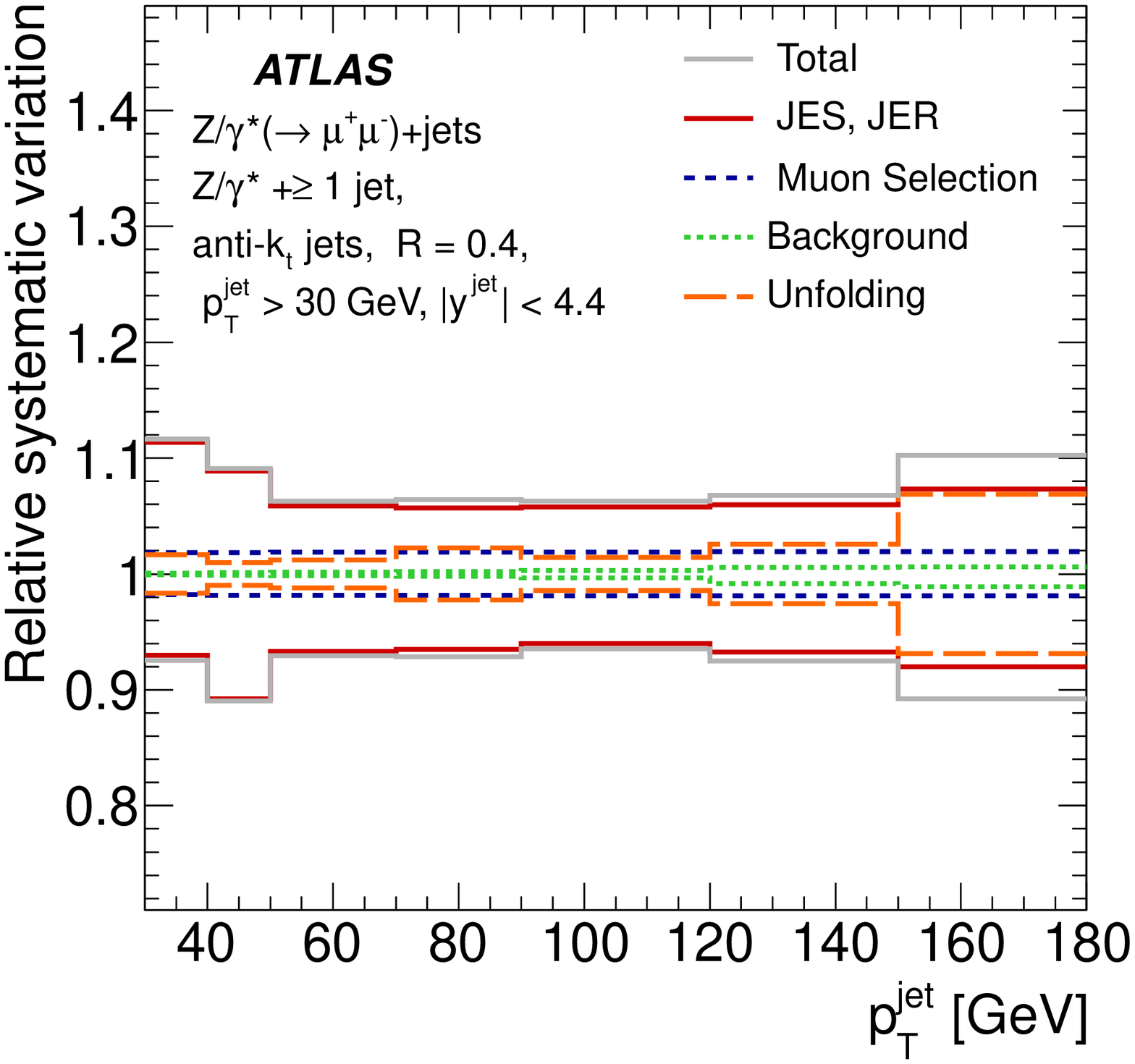}
}
\end{center}
\caption{\small 
Relative systematic uncertainties from different sources 
in the (top) $\zee$+jets and (bottom) $\zmm$+jets analyses for  the measured cross section as a function of 
inclusive jet multiplicity, and the inclusive differential cross sections as a function of 
$\ptjet$, for events with at least one jet with $\ptjet >30$~GeV and $|\rapjet|<4.4$ in the final state
 (see Section~\ref{sec:sys}). The total systematic uncertainty is obtained by summing all contributions 
in quadrature.
}
\label{fig:sys_e_m_1}
\end{figure}

A detailed study of systematic uncertainties is carried out.
In the following, 
a complete description is given for two of the observables: 
 the  inclusive cross 
section as a function of $\njet$ and the inclusive jet cross section  as a function of $\ptjet$, in 
events with at least one jet in the final state (see Fig.~\ref{fig:sys_e_m_1}). The same 
sources of systematic uncertainty are considered for the rest of the observables.
 

\begin{itemize}

\item  The measured jet energies are increased and decreased by factors between 3$\%$ and 10$\%$, depending on $\ptjet$ and 
$\etajet$,  to account for the absolute jet energy scale (JES) uncertainty, 
as determined in inclusive jet studies~\cite{jet_pro_atlas}. For a given jet $|\etajet|$, the 
jet energy uncertainty tends to decrease with increasing $\ptjet$, while the uncertainties increase
with increasing $|\etajet|$. An additional 0.1$\%$ to 1.5$\%$ uncertainty on the jet energy, 
depending on $\ptjet$ and $|\etajet|$, is considered for each additional  reconstructed primary vertex in the event to 
account for the uncertainty on the pileup offset subtraction, where the uncertainty 
decreases (increases) with increasing $\ptjet$ ($|\etajet|$). Additional uncertainties are 
included to account for  the different quark- and gluon-jet relative 
population in multi-jets and $Z/\gamma^*$+jets processes and the presence of close-by jets in the final state,
leading to a different average calorimeter response.
These effects added in quadrature result in an uncertainty on the measured cross sections that increases
from 7$\%$ to $22 \%$ as $\njet$ increases and from 8$\%$ to 12$\%$ as $\ptjet$ increases, 
and constitutes the dominant source of 
systematic uncertainty for each of the measured distributions. The uncertainty on the jet energy resolution (JER)~\cite{jet_pro_atlas} translates into 
a 1$\%$ uncertainty on the cross section as a function of $\njet$ and into a 1$\%$ to 3$\%$ uncertainty 
on the measured cross sections with increasing jet $\ptjet$ and $|\rapjet|$.\\

 
\item The uncertainty on the estimated multi-jets background in the electron channel translates 
into an uncertainty on the measured cross sections which rises from 
$0.6\%$ to $2\%$ as $\njet$ and $\ptjet$ increase.
In addition, the background contributions from top quark, $W$+jets, $\ztt$+jets, and diboson 
production processes are varied by ${}^{+7}_{-9.6}\%$, $5\%$, $5\%$, and $5\%$, respectively, 
to account for the uncertainty on the 
absolute normalization of the different MC samples. This translates into a less than $1\%$ uncertainty
in the measured cross sections. In the $\zmm$+jets measurements, the impact from 
the background uncertainties is negligible.    


\item The correction factors are re-computed using Sherpa instead of ALPGEN  
to account for possible dependencies
on the parton shower, underlying event and fragmentation models, and the PDF sets
used in the MC samples. 
This introduces an uncertainty on the measured cross sections that increases from 
0.4$\%$ to 4.5$\%$ with increasing $\njet$ and $\ptjet$.
In addition, a Bayesian iterative method~\cite{bayes} 
is used to unfold the data,  which accounts for the full migration matrix across bins for a given 
observable.  The ALPGEN MC samples are used to construct the input migration matrices for the different 
measured distributions and up to three iterations are considered, as optimized separately for 
each observable using the simulation.
The 
differences with respect to the nominal bin-by-bin 
correction factors are less than 1$\%$ except at very large $\ptjet$ where they vary between 3$\%$ and 6$\%$, and are included as an additional source of  systematic 
uncertainty.
Altogether, this introduces an uncertainty on the measured cross sections that increases from 
0.7$\%$ to 7$\%$ with increasing $\njet$ and $\ptjet$.


\item The uncertainty on the electron selection is taken into account. 
It includes uncertainties on the electron absolute energy scale and  
energy resolution, the uncertainty on the electron identification efficiency, and the
uncertainties on the electron reconstruction scale factors applied to the MC simulation. 
This translates into a 4$\%$ uncertainty in the measured $\zee$+jets cross sections, approximately 
independent of $\njet$, and jet $\ptjet$ and $\etajet$.
The uncertainty on the measured cross sections due to the determination of 
the electron trigger efficiency is negligible. 


\item The uncertainty on the muon reconstruction efficiency, the muon momentum scale, and the muon momentum resolution 
translate into a conservative $2\%$ uncertainty 
in the measured $\zmm$+jets cross sections, approximately independent of $\njet$, and jet $\ptjet$ and $\etajet$. The uncertainty on the muon trigger efficiency introduces a less than $1\%$ uncertainty on the measured cross sections.

\end{itemize}

\noindent
For each channel, the different sources of systematic uncertainty are added in
quadrature to the statistical uncertainty to obtain the total uncertainty.
 The total systematic uncertainty increases from 9$\%$ to $23\%$ 
as $\njet$ increases; and from 10$\%$ at low $\ptjet$ to $13\%$ at very high $\ptjet$. 
Finally, the additional 3.4$\%$ uncertainty on the 
total integrated luminosity~\cite{lumi} is also taken into account.       

\section{Next-to-leading order pQCD predictions}
\label{sec:nlo}

NLO pQCD predictions for $\zee$+jets and $\zmm$+jets production  
are computed using the BlackHat program~\cite{black}. CTEQ6.6 PDFs~\cite{cteq66} are employed and   
renormalization and factorization scales are set to $\mu  =  H_{\rm T}/2$, where $H_{\rm T}$ is defined event-by-event 
as the scalar sum of the $p_{\rm T}$ of all particles and partons in the final state. The   
$\akt$ algorithm with $R=0.4$ is used to reconstruct jets at the parton level. 

Systematic uncertainties on the 
predictions related to PDF uncertainties are computed using the Hessian method~\cite{hessian} and are
defined as 90$\%$ confidence level uncertainties.
For the total cross sections, they increase from 2$\%$ to 5$\%$ with increasing $\njet$.
Additional changes in the PDFs due to the variation of the input value 
for $\alpha_s(M_Z)$ by $\pm 0.002$ around its nominal 
value $\alpha_s(M_Z)=0.118$ introduce uncertainties on the measured cross sections that increase 
from 2$\%$ to 7$\%$ with increasing $\njet$. These  
are added in quadrature to the PDF uncertainties. 
Variations of the  renormalization and factorization scales by a factor of two (half) 
reduce (increase) the predicted cross sections by 4$\%$ to 14$\%$  as $\njet$ increases.

The theoretical predictions are corrected for QED radiation effects. The 
correction factors $\delta^{\rm QED}$ are determined using ALPGEN MC samples 
with and without photon radiation in the final state, defined by
the lepton four-momentum and 
photons within a cone of radius 0.1 around the lepton direction. 
The correction factors are about $2\%$ for the electron and muon channels, 
and do not present a significant $\njet$ dependence.

The theoretical predictions include  parton-to-hadron  
correction  factors $\delta^{\rm had}$ that approximately account for 
non-perturbative contributions from  the underlying event
and fragmentation into particles. 
In each measurement, the correction factor is estimated using HERWIG+JIMMY  
MC samples, as the ratio at the particle level between the 
nominal distribution and the one obtained 
by turning off both the interactions between proton remnants and the 
cluster fragmentation in the MC samples.  
 The non-perturbative correction factors for the inclusive $\njet$ and $\ptjet$ distributions 
are about 0.99 and exhibit a moderate $\njet$ and $\ptjet$ dependence. 
However, for very forward jets $\delta^{\rm had}$
is about  0.9. The non-perturbative corrections are
also computed using PYTHIA-AMBT1 MC 
samples with different parton shower, fragmentation model,  and UE settings.
The uncertainty on $\delta^{\rm had}$, defined as the difference between the results obtained with 
HERWIG/JIMMY-AUET1 and PYTHIA-AMBT1, varies between 2$\%$ and 5$\%$.


\section{Results}
\label{sec:results}

As mentioned in Section~\ref{sec:unfold}, the measured cross sections refer to particle level jets identified using the 
$\akt$ algorithm with $R=0.4$, for jets with $\ptjet >30$~GeV and $|\rapjet|<4.4$, and the 
results are defined in a limited kinematic range for the $Z/\gamma^*$ decay products.
The data are compared to the predictions from the different MC event generators implementing 
$\zee$+jets and $\zmm$+jets production, as 
discussed in Section~\ref{sec:sim}, as well as  to NLO pQCD predictions, as discussed in Section~\ref{sec:nlo}.
Tabulated values
of the results are available in Tables~\ref{tab:comb_njet} to \ref{tab:comb_rjj} and in the Durham HEP database~\cite{tab}. 

\subsection{Inclusive jet multiplicity}
\label{sec:results_n}
\begin{figure}[h]
\begin{center}
\mbox{
\includegraphics[width=0.495\textwidth]{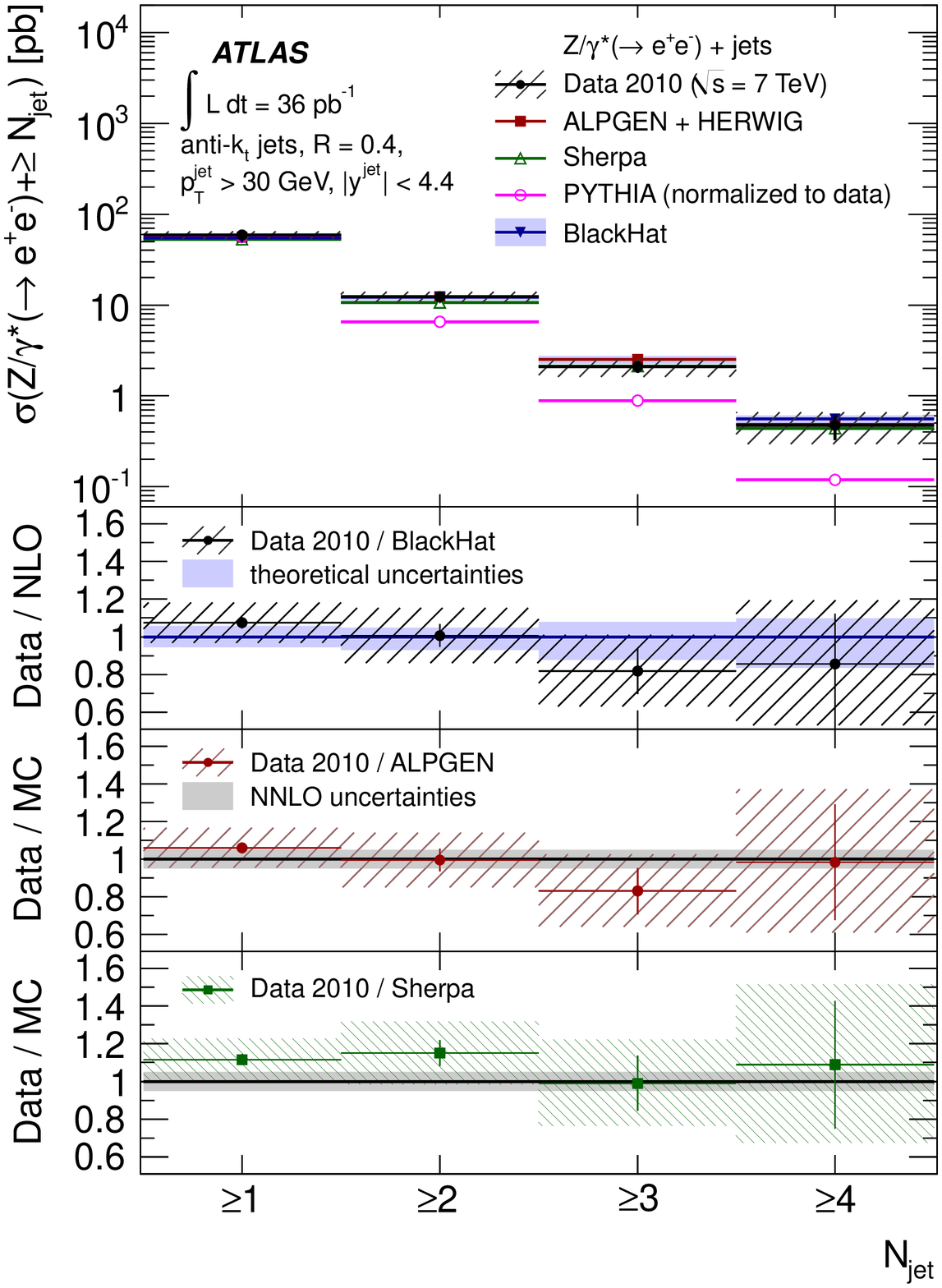}
\includegraphics[width=0.495\textwidth]{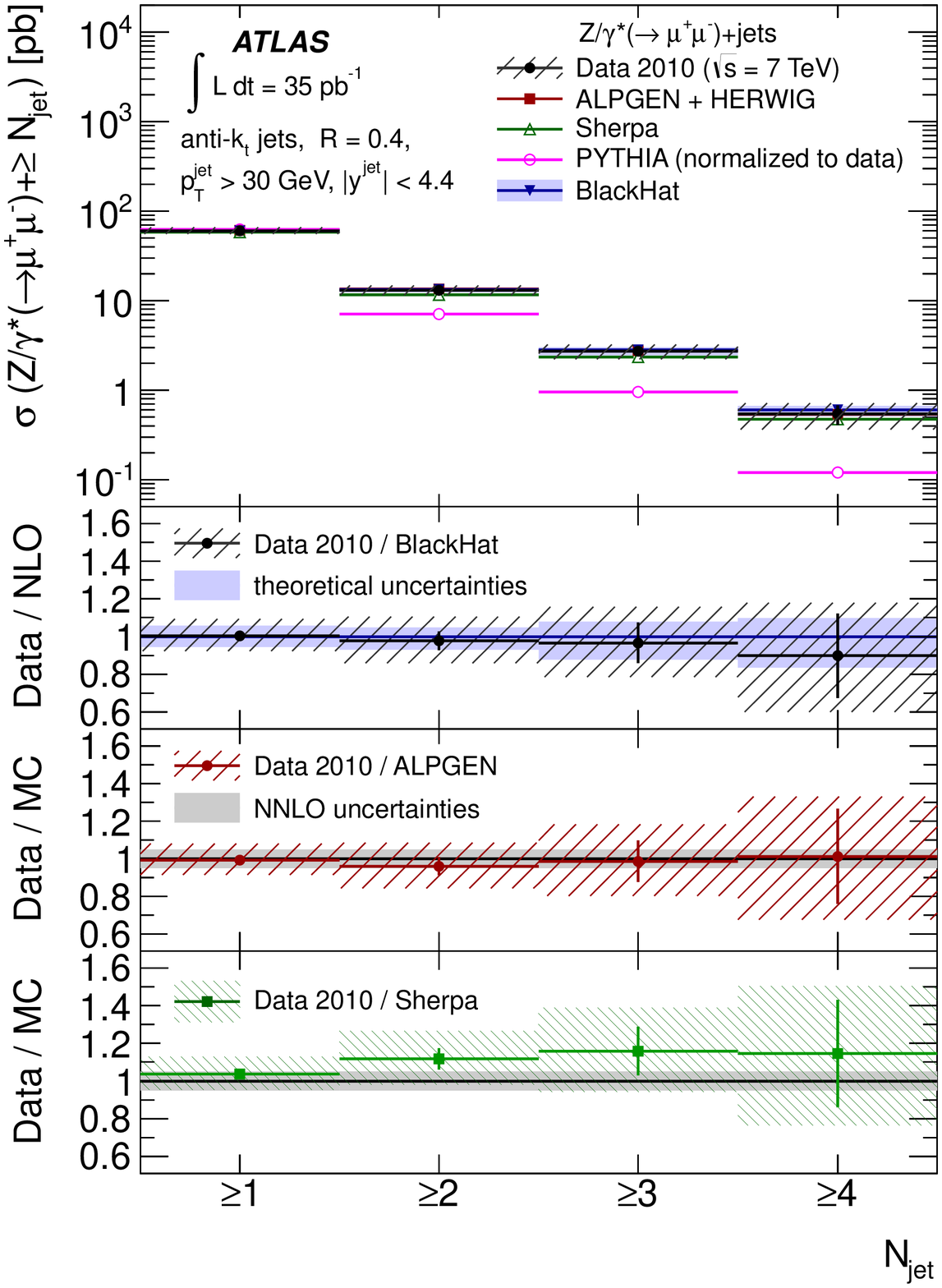}
}
\end{center}
\caption{\small 
Measured cross section $\sigma_{\njet}$ (black dots) for (left) $\zee$+jets and (right) $\zmm$+jets production
as a function of the inclusive jet multiplicity, for events with at least one jet with 
$\ptjet >30$~GeV and $|\rapjet|<4.4$ in the final state. 
In this and subsequent figures~\ref{fig:ratio} - \ref{fig:dr}  
the error bars indicate the statistical uncertainty
and the dashed  areas the statistical and systematic uncertainties added in quadrature. 
The measurements are compared to NLO pQCD predictions from BlackHat, 
 as well as the predictions from ALPGEN and Sherpa (both normalized to the FEWZ value for the total 
cross section), and PYTHIA (normalized to the data as discussed in Section~\ref{sec:results}). 
}
\label{fig:njet}
\end{figure}

Figure~\ref{fig:njet} presents the measured cross sections as functions of the inclusive jet multiplicity
($\geq \njet$) for $\zees$  and $\zmms$ interactions, in events with up to at least four jets 
in the final state. 
The data are well described by the predictions from ALPGEN and Sherpa, and 
BlackHat NLO pQCD. 
ALPGEN and Sherpa predictions include a $5\%$ uncertainty  
from the NNLO pQCD normalization, as discussed in Section~\ref{sec:sim}, and the 
systematic uncertainty on the BlackHat NLO pQCD predictions  
is discussed in Section~\ref{sec:nlo}. 
In the case of PYTHIA, the LO pQCD ($q\bar{q} \to Z/\gamma^* g$ and $qg \to Z/\gamma^* q$ processes) MC predictions are multiplied by 
a factor 1.19, as determined from data and extracted from 
the average of electron and muon results in the $\geq 1$ jet bin in Fig.~\ref{fig:njet}. This brings the PYTHIA  
predictions close to the data. However, for larger $\njet$, and despite the 
additional normalization applied, PYTHIA predictions underestimate the measured cross sections.

\begin{figure}[h]
\begin{center}
\mbox{
\includegraphics[width=0.495\textwidth]{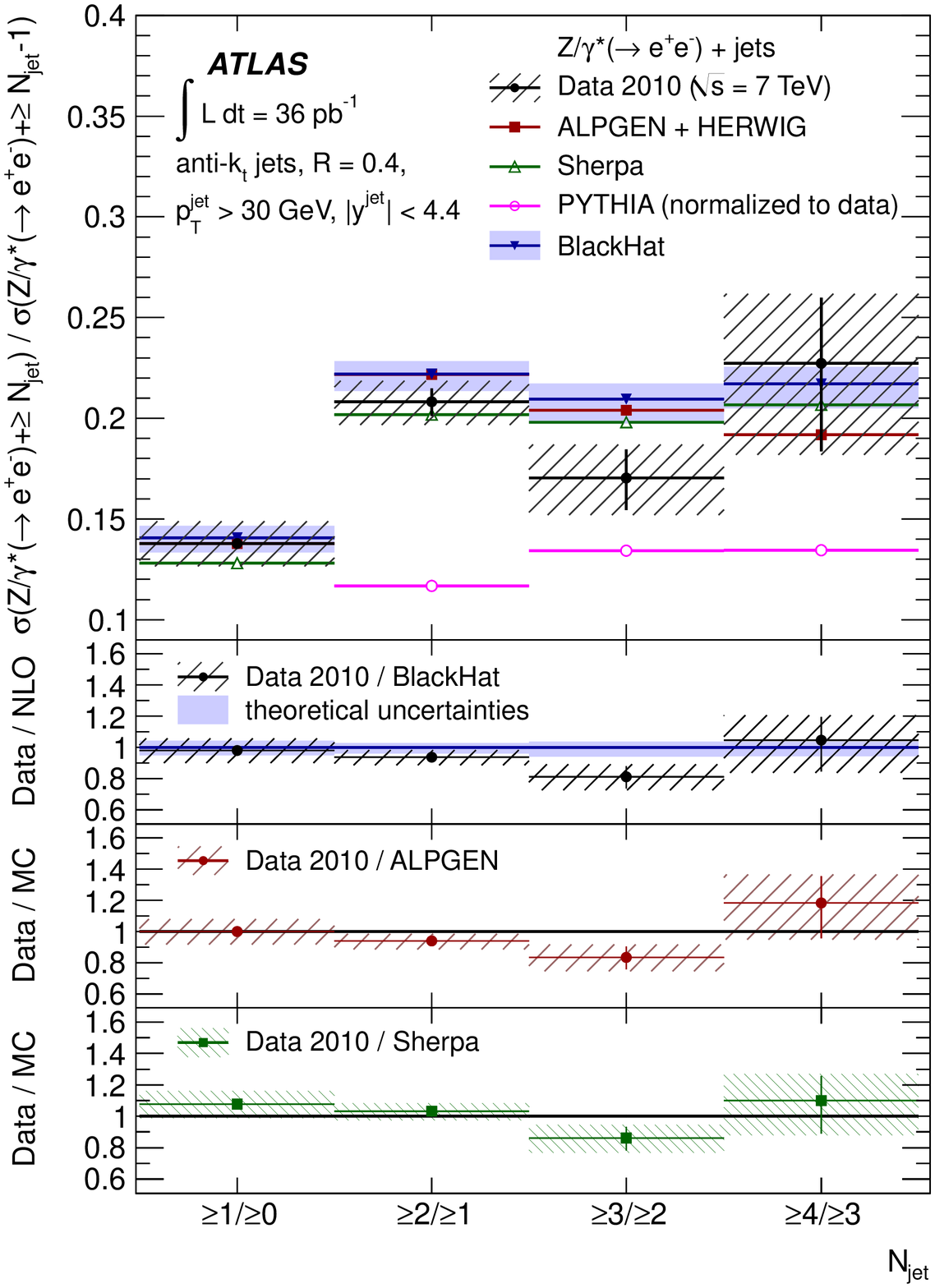}
\includegraphics[width=0.495\textwidth]{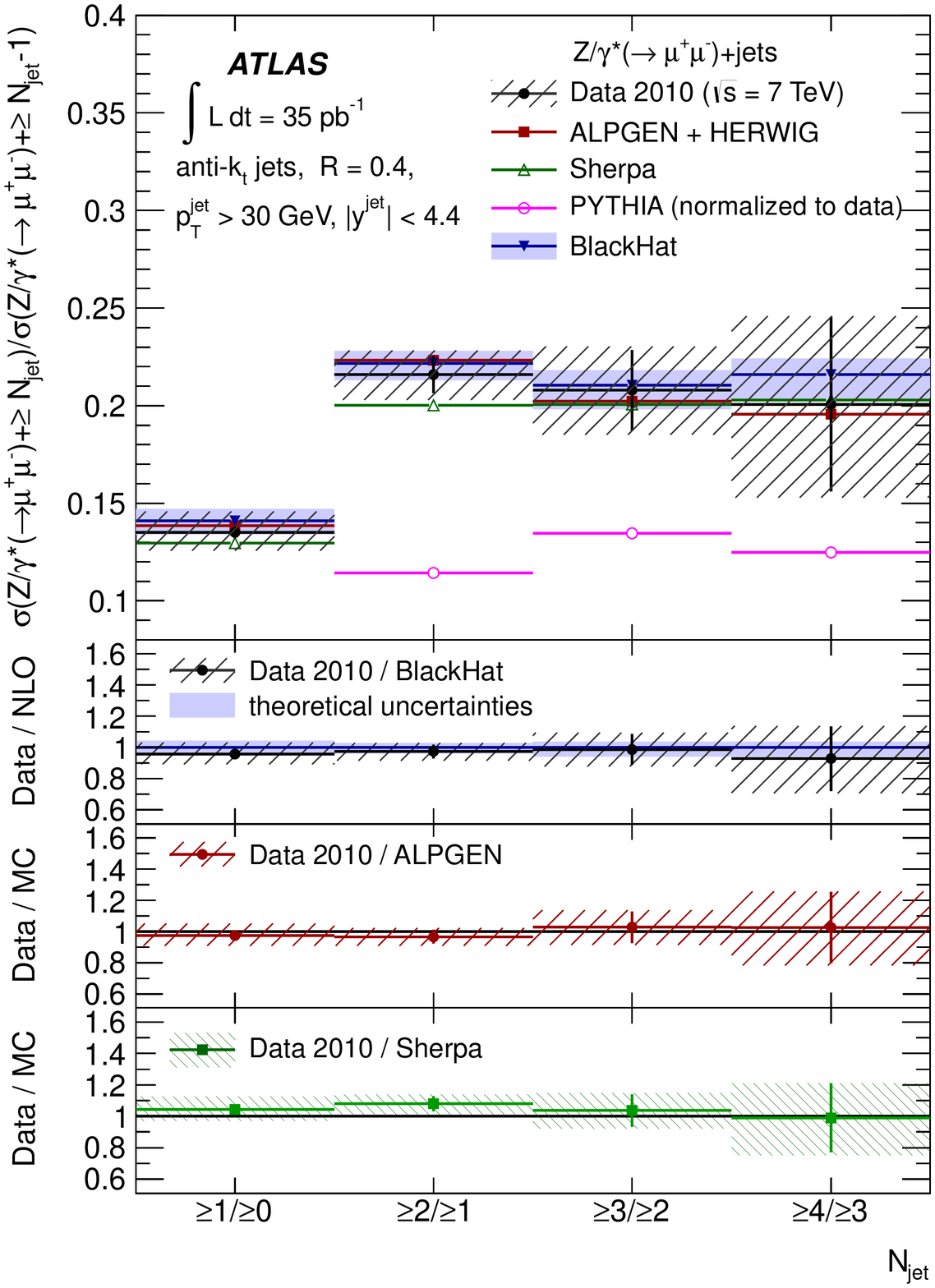}
}
\end{center}
\caption{\small 
Measured ratio of cross sections ($\sigma_{\njet} / \sigma_{\njet - 1}$) (black dots) for 
(left) $\zee$+jets and (right) $\zmm$+jets production
as a function of the inclusive jet multiplicity, for events with at least one jet with 
$\ptjet >30$~GeV and $|\rapjet|<4.4$ in the final state. 
}
\label{fig:ratio}
\end{figure}

The measured ratio of cross sections for $\njet$ and $\njet - 1$
is shown in Fig.~\ref{fig:ratio}, compared to the different theoretical predictions.  
This observable cancels part of the systematic uncertainty and constitutes 
an improved test of the SM. The ratio is sensitive to the value of the strong 
coupling, and to the details of the implementation of higher-order matrix elements and 
soft-gluon radiation contributions in the theoretical predictions. 
The data indicate that
the cross sections decrease by a factor of five with the requirement of each additional jet
in the final state. The electron and muon measurements are well described by ALPGEN and Sherpa, and the BlackHat 
NLO pQCD predictions. PYTHIA predictions underestimate the measured ratios.


\subsection{$\mathrm{d}\sigma /\mathrm{d} \ptjet$ and $\mathrm{d}\sigma / \mathrm{d} |\rapjet|$}
\label{sec:results_pt}

\begin{figure}[h] 
\begin{center}
\mbox{
\includegraphics[width=0.495\textwidth]{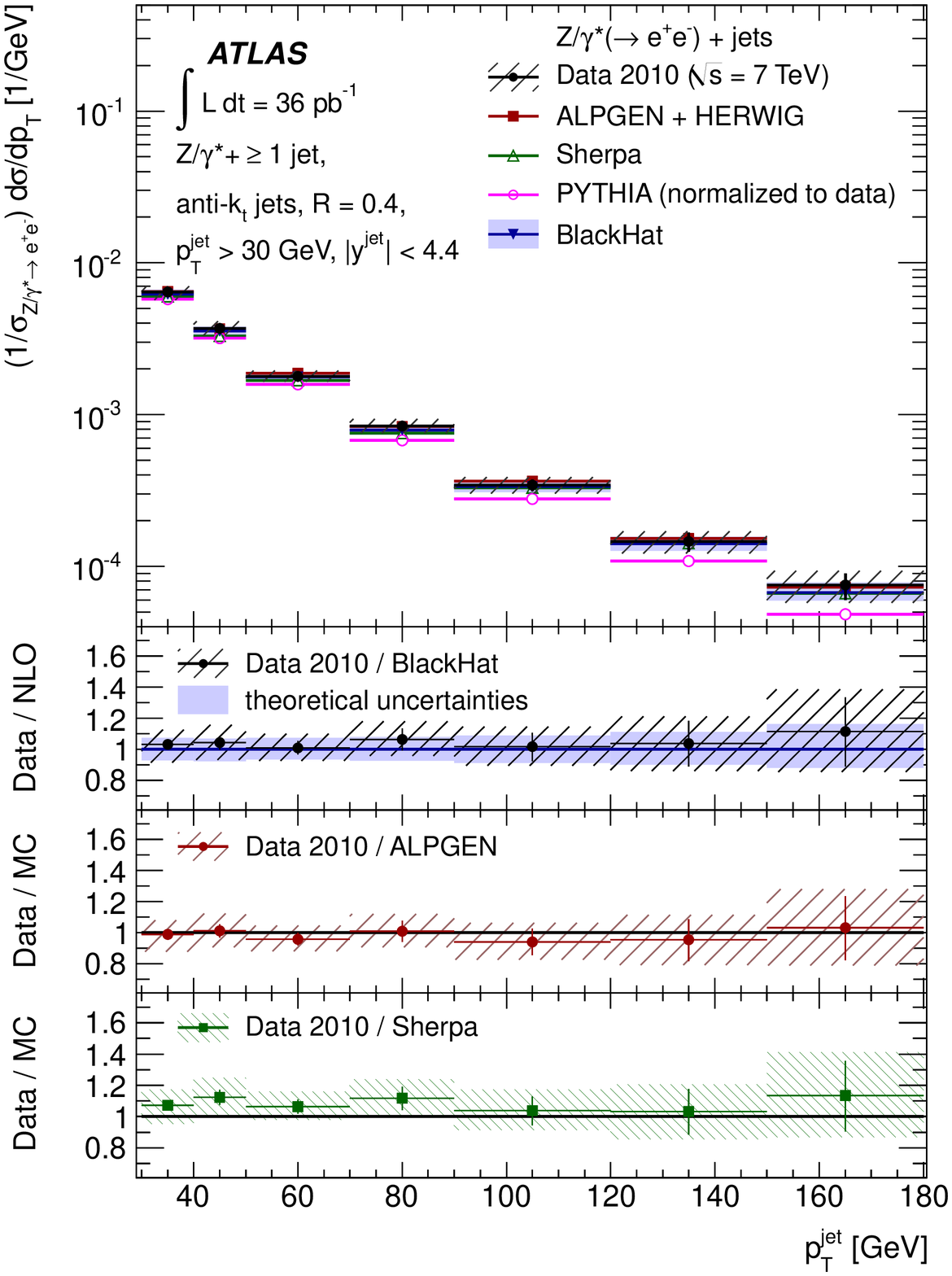}
\includegraphics[width=0.495\textwidth]{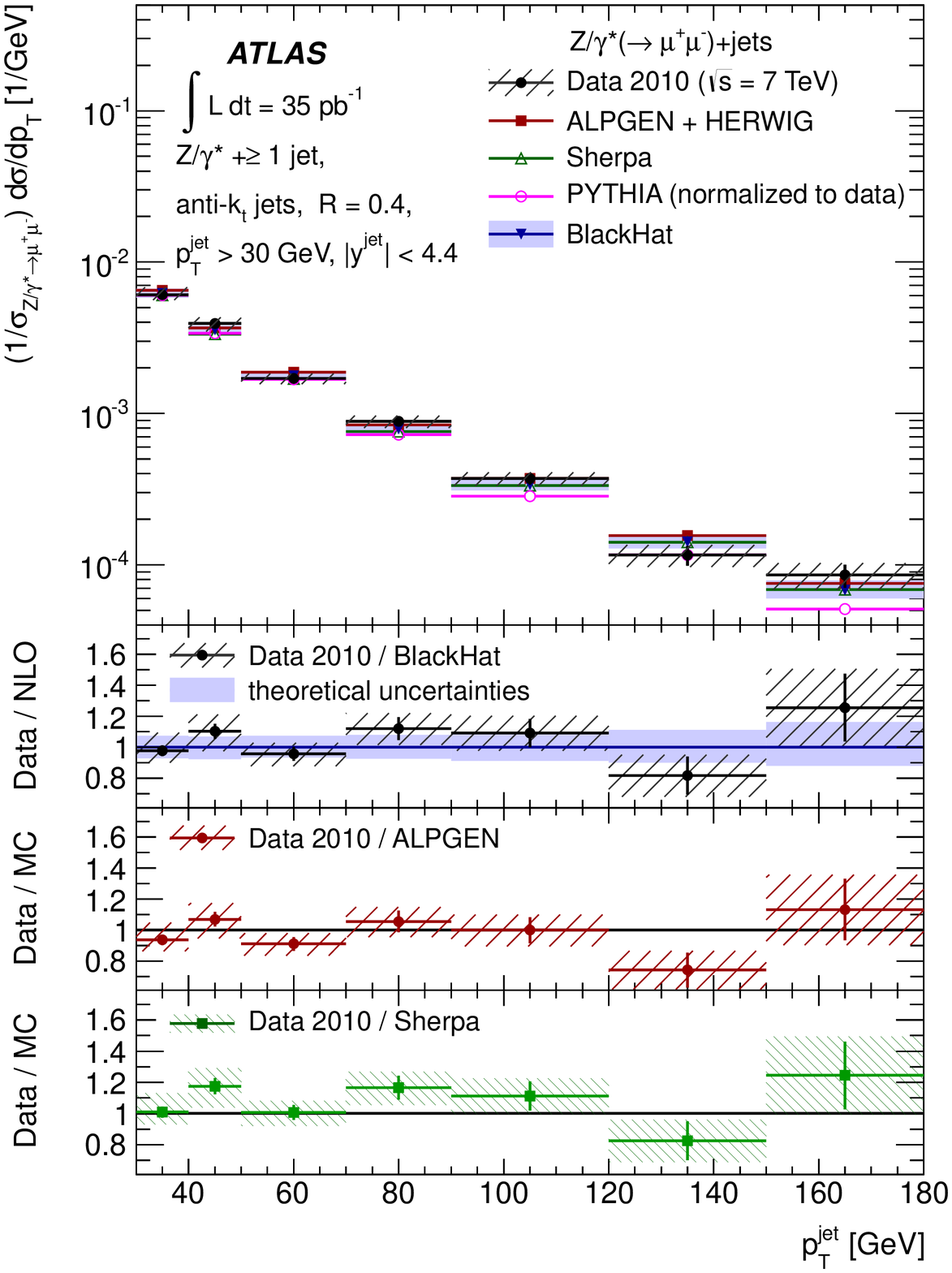}
}
\end{center}
\caption{\small 
Measured normalized inclusive jet cross section $(1/\sigma_{\zlls}) \mathrm{d}\sigma / \mathrm{d}\ptjet$ (black dots) in 
 (left) $\zee$+jets and (right) $\zmm$+jets production
as a function of $\ptjet$, in events with at least one jet with 
$\ptjet >30$~GeV and $|\rapjet|<4.4$ in the final state, and normalized by 
$\sigma_{\zees}$ and $\sigma_{\zmms}$ Drell-Yan cross sections, respectively.
}
\label{fig:ptincl}
\end{figure}

\begin{figure}[h] 
\begin{center}
\mbox{
\includegraphics[width=0.495\textwidth]{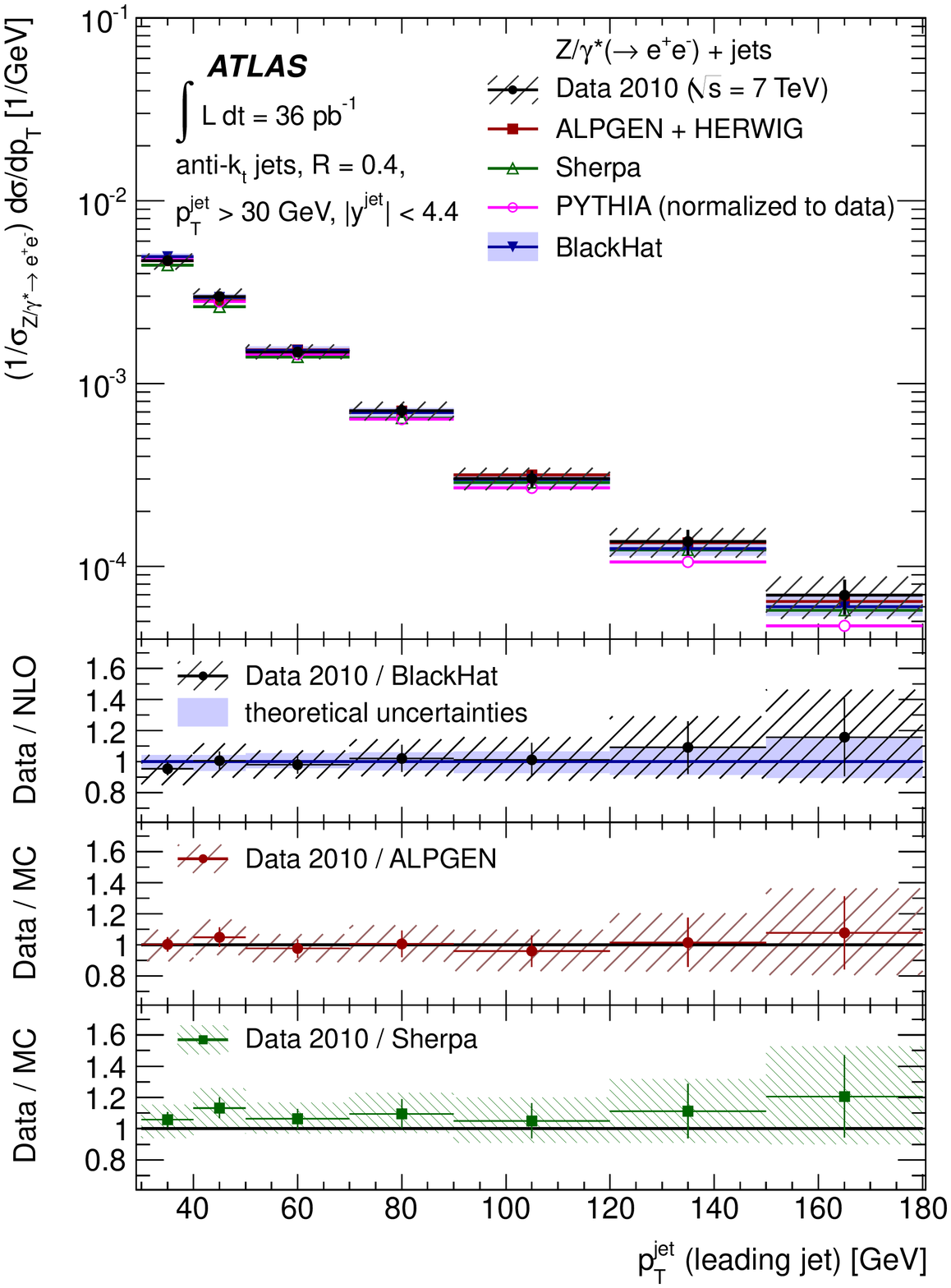}
\includegraphics[width=0.495\textwidth]{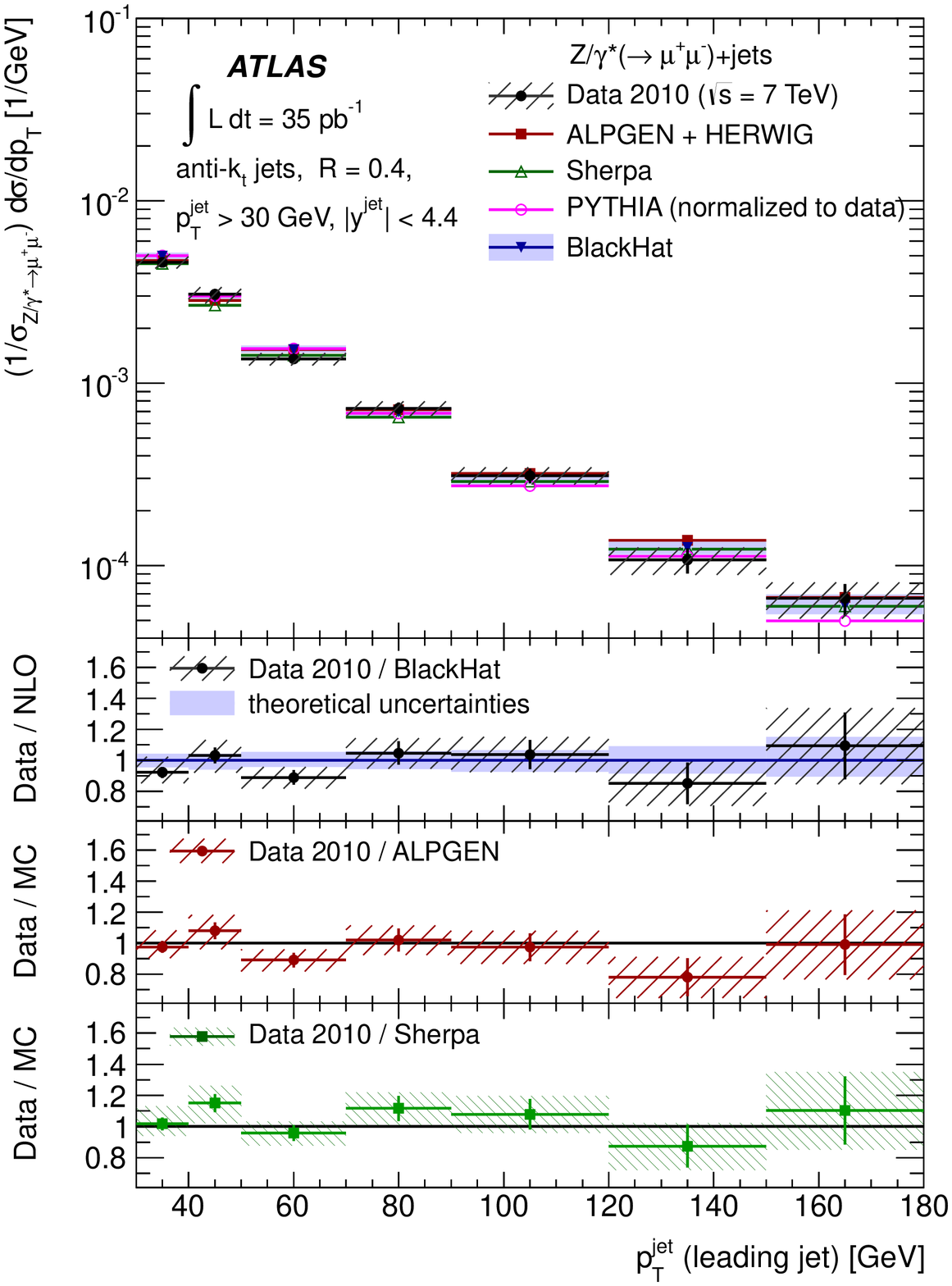}
}
\end{center}
\caption{\small 
Measured normalized jet cross section $(1/\sigma_{\zlls}) \mathrm{d}\sigma / \mathrm{d}\ptjet$ (black dots) in 
 (left) $\zee$+jets and (right) $\zmm$+jets production
as a function of the leading jet $\ptjet$, in events with at least one jet with 
$\ptjet >30$~GeV and $|\rapjet|<4.4$ in the final state, and normalized by 
$\sigma_{\zees}$ and $\sigma_{\zmms}$ Drell-Yan cross sections, respectively.
}
\label{fig:pt1}
\end{figure}

The inclusive jet differential cross section $\mathrm{d}\sigma/\mathrm{d}\ptjet$  as a function of $\ptjet$ 
is presented in Fig.~\ref{fig:ptincl}, for both electron and muon analyses, in 
events with at least one jet in the final state. The cross sections
are divided by  the corresponding  inclusive $Z/\gamma^*$ cross section times branching ratio $\sigma_{\zlls} \ (\ell = e,\mu)$, separately for 
$\zees $ and $\zmms $, measured in the same kinematic region for the leptons and consistent with 
the results in Ref.~\cite{lepton_scales}, with the aim of cancelling 
systematic uncertainties related to lepton identification and the luminosity.   The measured 
differential cross sections decrease by more than two orders of magnitude as $\ptjet$ 
increases between 30~GeV and 180~GeV.
The data are well described by ALPGEN and Sherpa, and the BlackHat NLO pQCD predictions. 
PYTHIA predictions include the multiplicative factor 1.19 (as described above) and are then divided 
by the measured $\sigma_{\zlls}$ cross sections in this analysis. This results  in total  
normalization factors $(\times 0.0028$~pb${}^{-1}$) and $(\times 0.0027$~pb${}^{-1}$) for the electron and muon channels, respectively.
PYTHIA  shows a slightly softer jet 
$\ptjet$ spectrum than the data.    
Similar conclusions are extracted  from Fig.~\ref{fig:pt1}, 
where the differential cross sections are presented as a function of the leading-jet $\ptjet$.

\begin{figure}[h] 
\begin{center}
\mbox{
\includegraphics[width=0.495\textwidth]{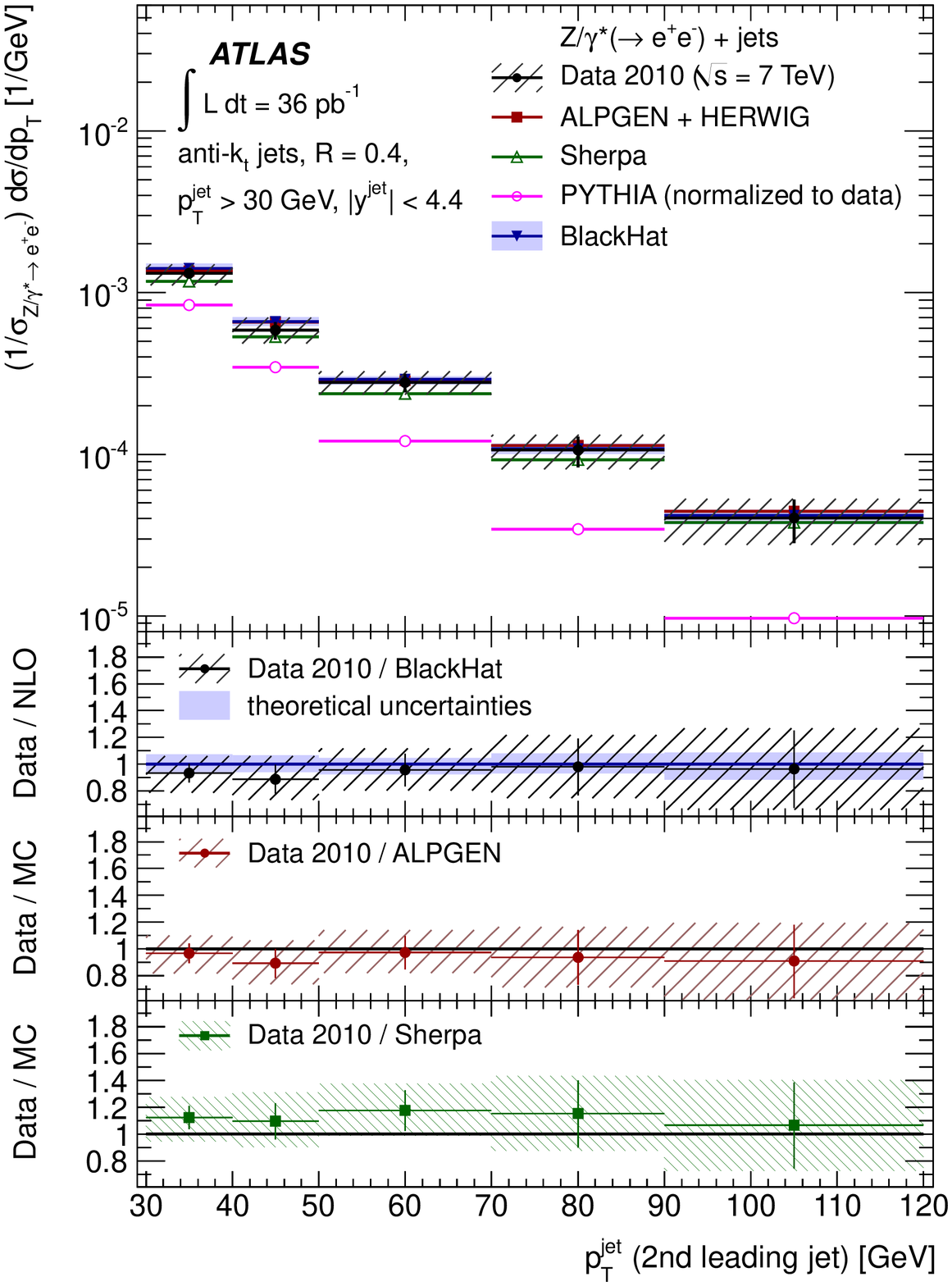}
\includegraphics[width=0.495\textwidth]{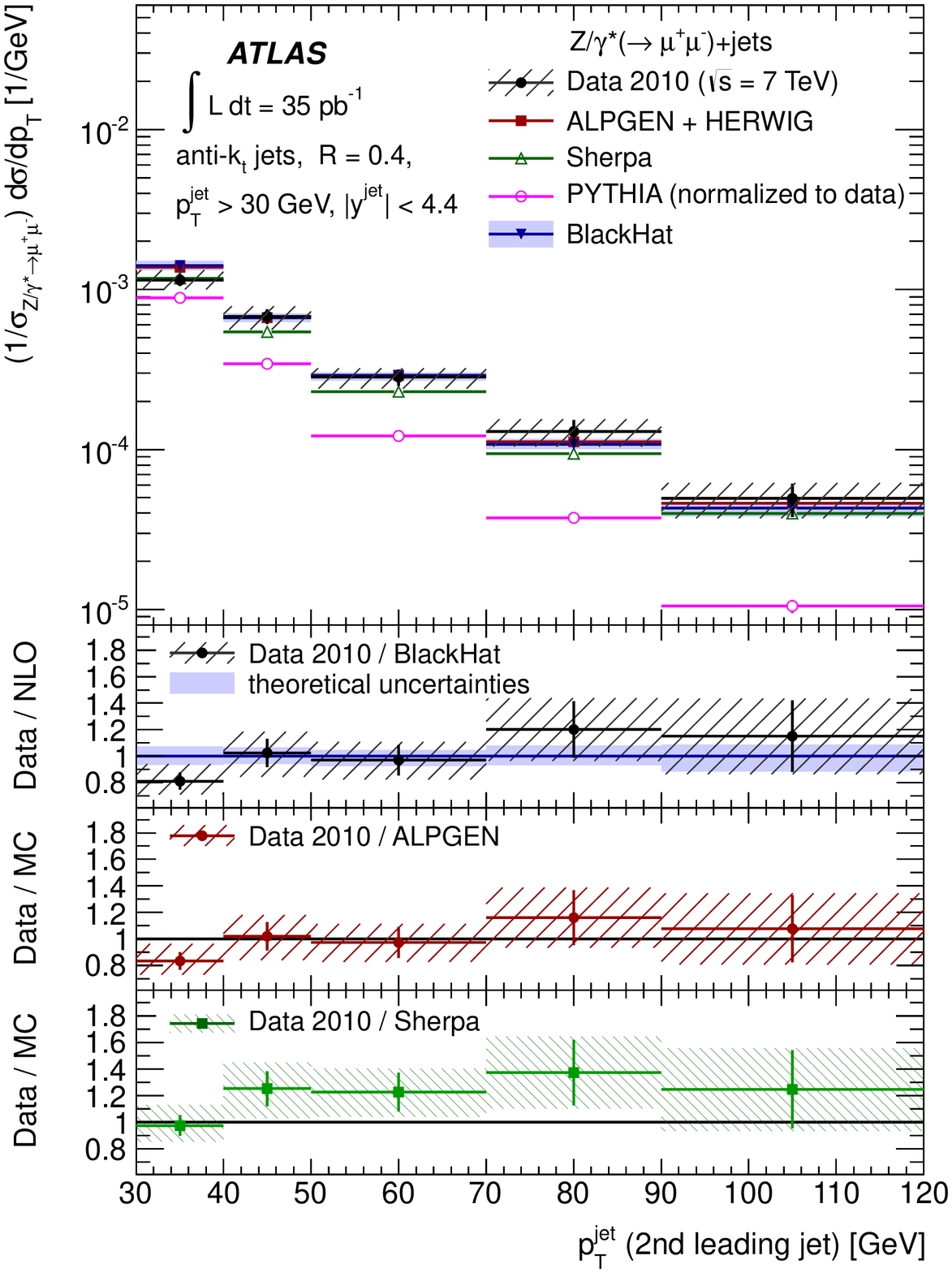}
}
\end{center}
\caption{\small 
Measured normalized jet cross section $(1/\sigma_{\zlls}) \mathrm{d}\sigma / \mathrm{d}\ptjet$ (black dots) in 
 (left) $\zee$+jets and (right) $\zmm$+jets production
as a function of the second-leading jet $\ptjet$, in events with at least two jets with 
$\ptjet >30$~GeV and $|\rapjet|<4.4$ in the final state, and normalized by 
$\sigma_{\zees}$ and $\sigma_{\zmms}$ Drell-Yan cross sections, respectively.
}
\label{fig:pt2}
\end{figure}

Figure~\ref{fig:pt2} shows the measured 
differential cross sections  $(1/\sigma_{\zlls})\mathrm{d}\sigma / \mathrm{d}\ptjet$, for electron and muon channels, as a function 
of $\ptjet$ of the second leading jet for jets with $30$~GeV $< \ptjet < 120$~GeV, 
in events with at least two jets in the final state.  
The measured cross sections decrease with increasing $\ptjet$, and 
are again well described by ALPGEN and Sherpa, and the BlackHat NLO pQCD predictions, 
while PYTHIA 
does not describe the data. This is expected since PYTHIA only implements  pQCD matrix elements for 
$Z/\gamma^*$+1 jet production, with the additional parton radiation produced 
via parton shower.

\begin{figure}[h] 
\begin{center}
\mbox{
\includegraphics[width=0.495\textwidth]{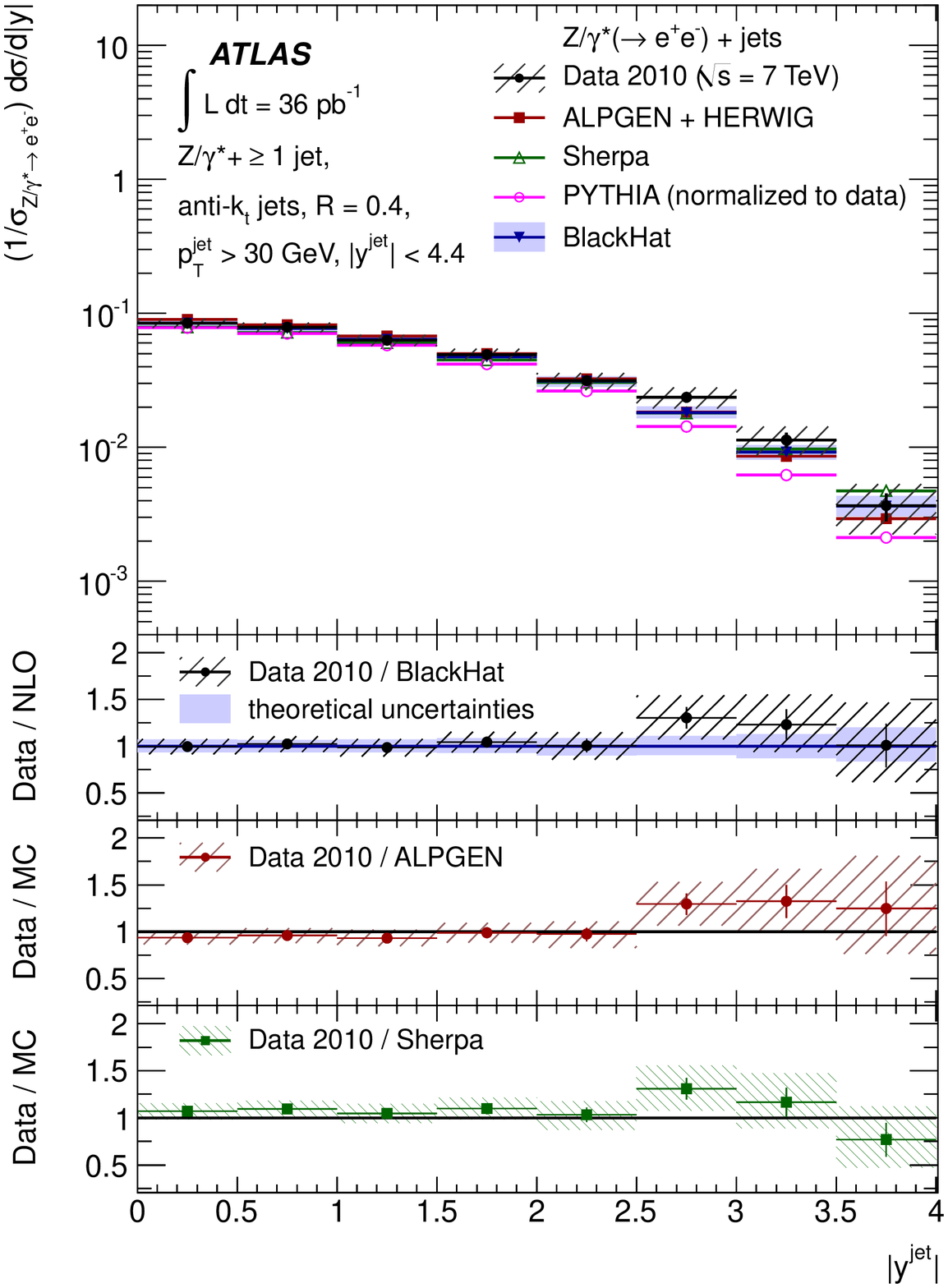}
\includegraphics[width=0.495\textwidth]{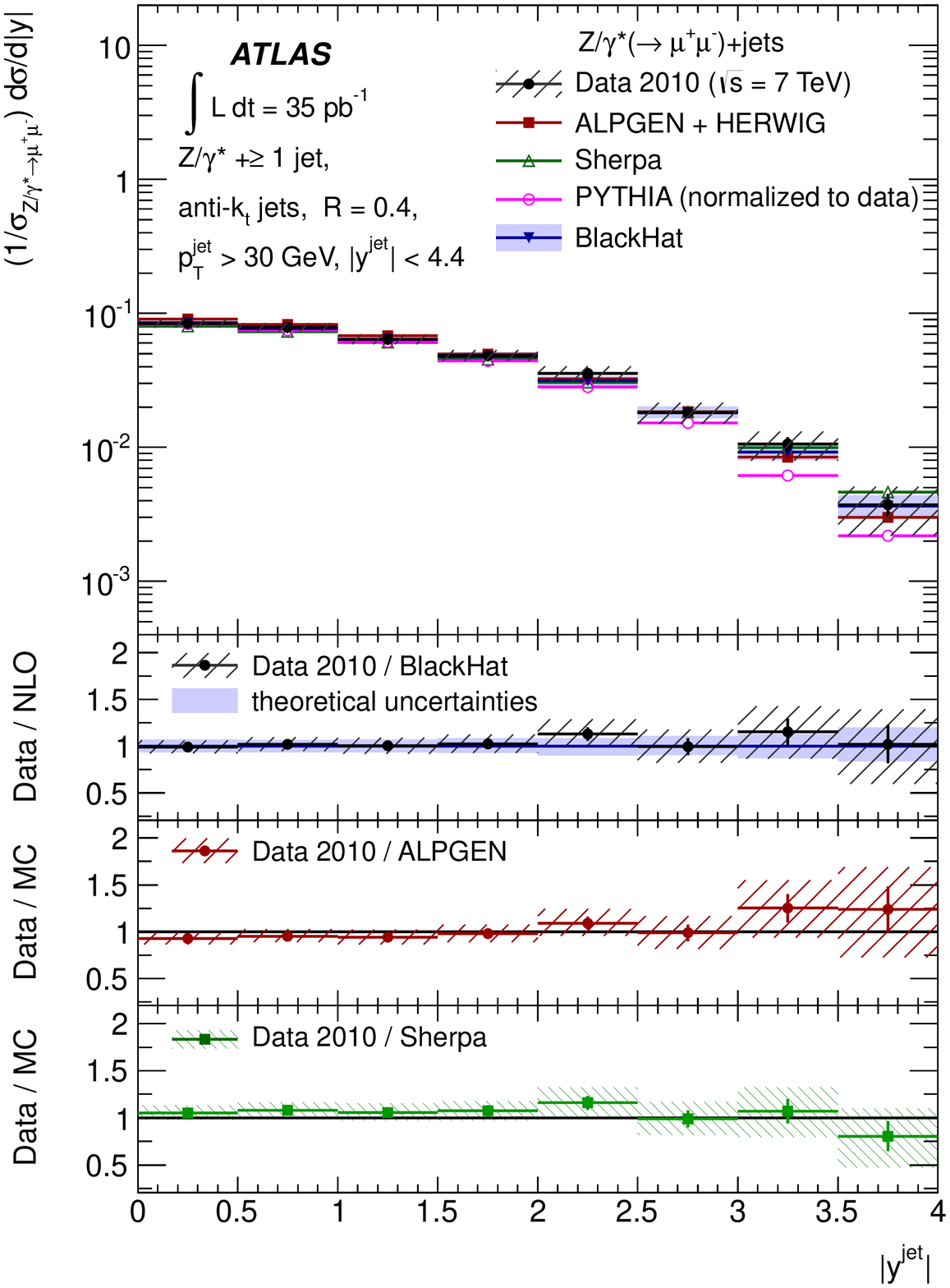}
}
\end{center}
\caption{\small 
Measured normalized inclusive jet cross section $(1/\sigma_{\zlls})  \mathrm{d}\sigma / \mathrm{d}|\rapjet|$ (black dots) in 
 (left) $\zee$+jets and (right) $\zmm$+jets production
as a function of $|\rapjet|$, in events with at least one jet with 
$\ptjet >30$~GeV and $|\rapjet|<4.4$ in the final state, and normalized by 
$\sigma_{\zees}$ and $\sigma_{\zmms}$ Drell-Yan cross sections, respectively.
}
\label{fig:rapincl}
\end{figure}
\begin{figure}[h] 
\begin{center}
\mbox{
\includegraphics[width=0.495\textwidth]{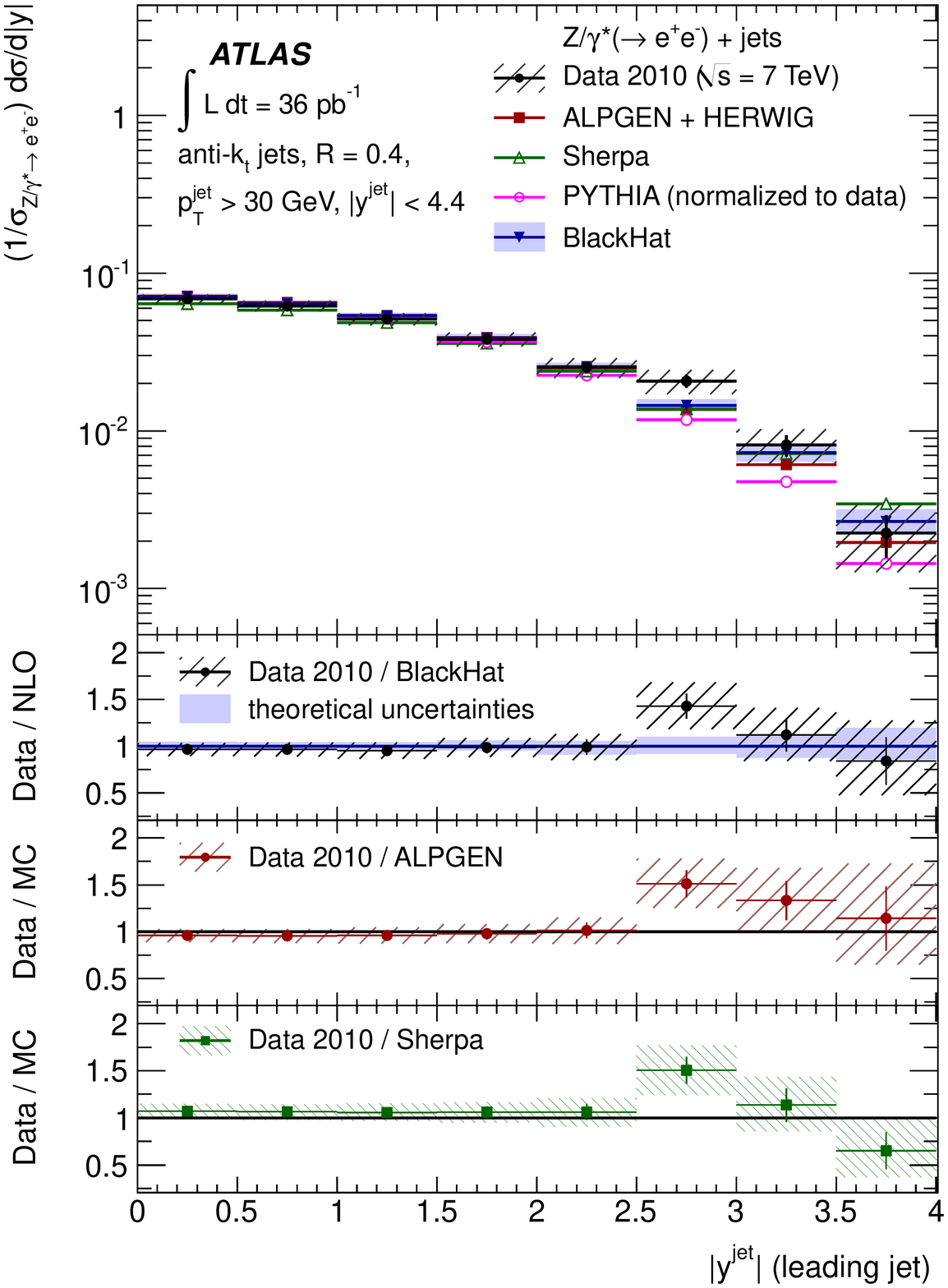}
\includegraphics[width=0.495\textwidth]{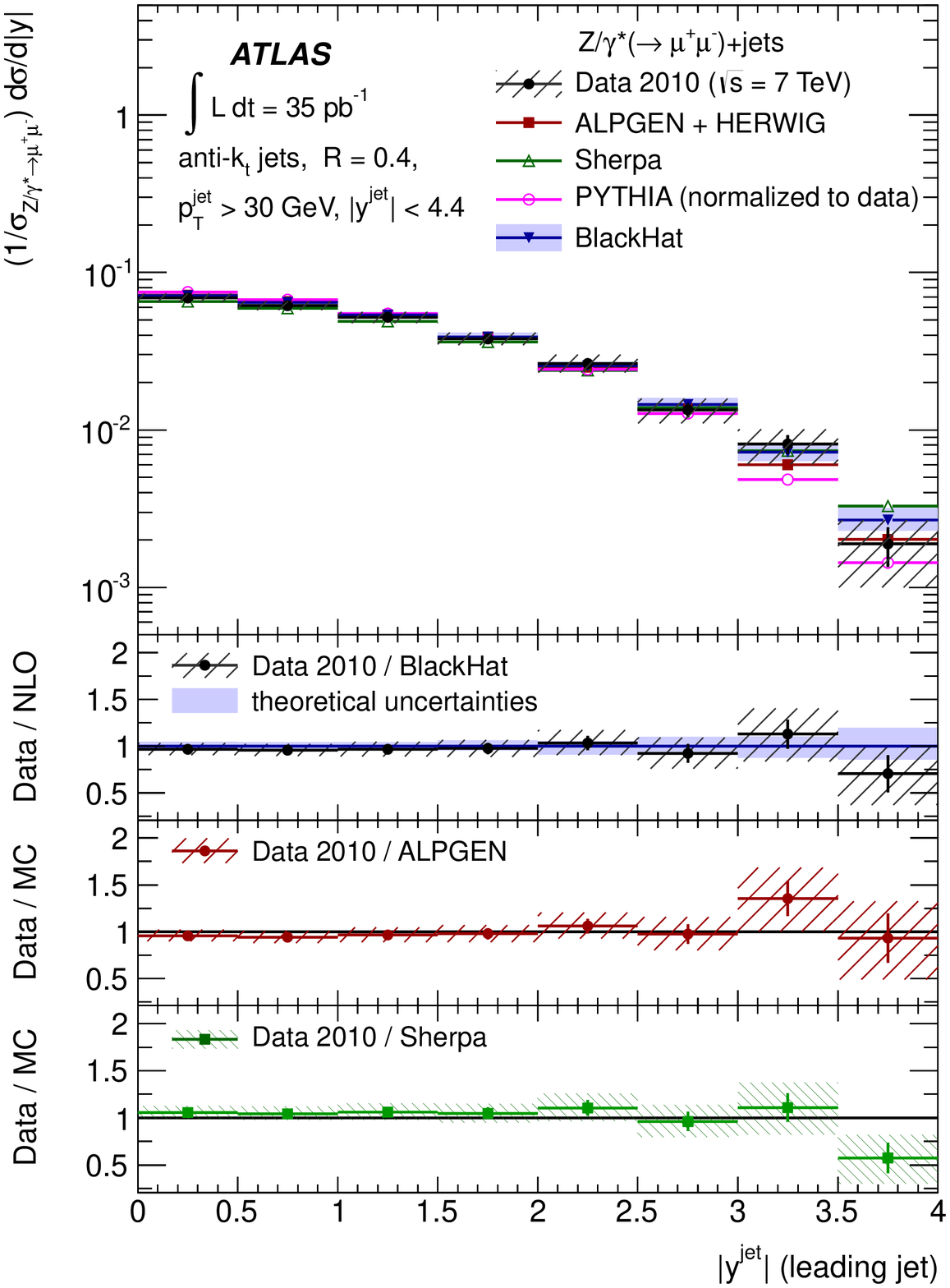}
}
\end{center}
\caption{\small 
Measured normalized jet cross section $(1/\sigma_{\zlls})  \mathrm{d}\sigma / \mathrm{d}|{\rapjet}|$ (black dots) in 
 (left) $\zee$+jets and (right) $\zmm$+jets production
as a function of the leading jet $|{\rapjet}|$, in events with at least one jet with 
$\ptjet >30$~GeV and $|\rapjet|<4.4$ in the final state, and normalized by 
$\sigma_{\zees}$ and $\sigma_{\zmms}$ Drell-Yan cross sections, respectively.
}
\label{fig:rap1}
\end{figure}
\begin{figure}[h] 
\begin{center}
\mbox{
\includegraphics[width=0.495\textwidth]{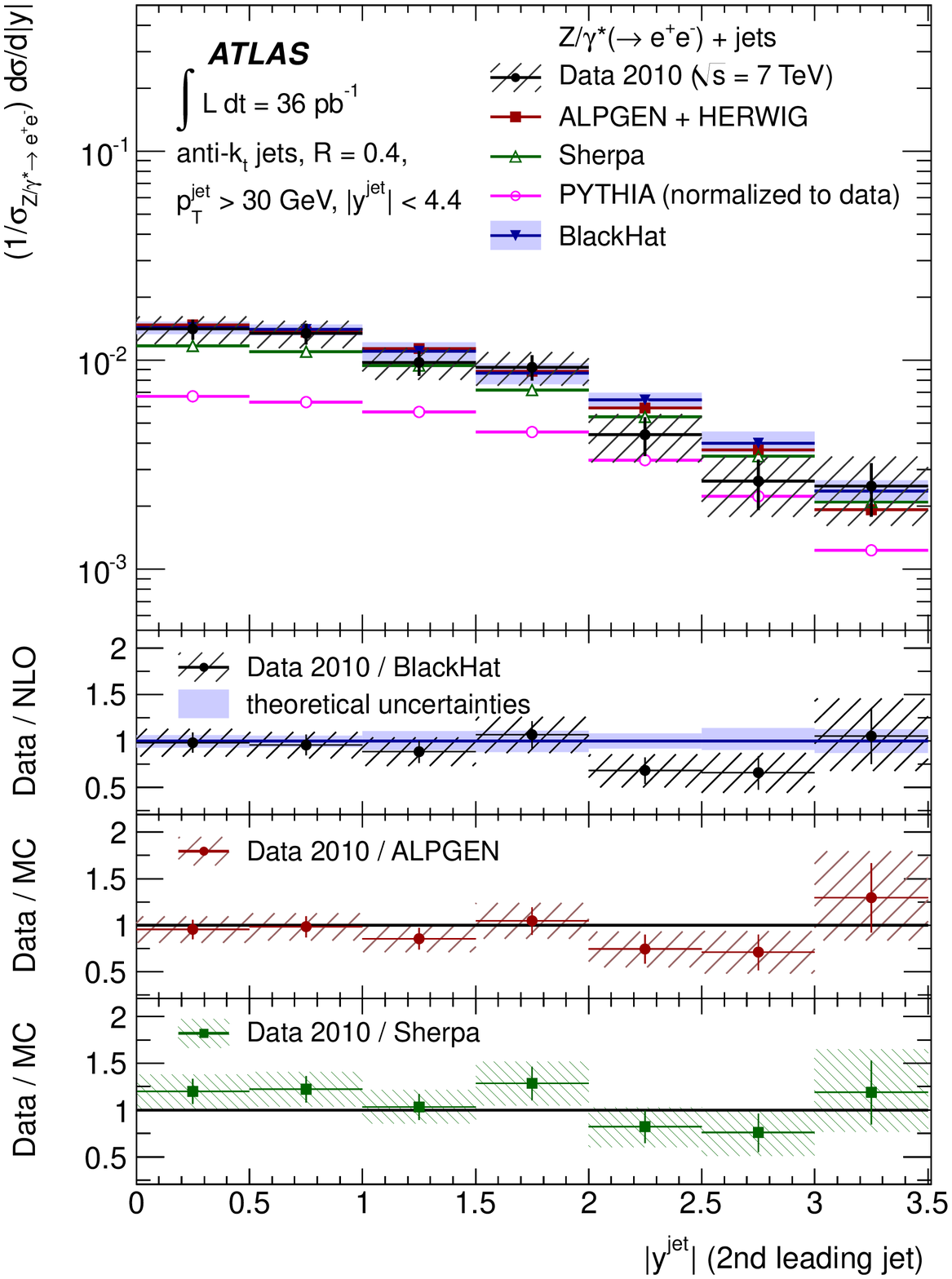}
\includegraphics[width=0.495\textwidth]{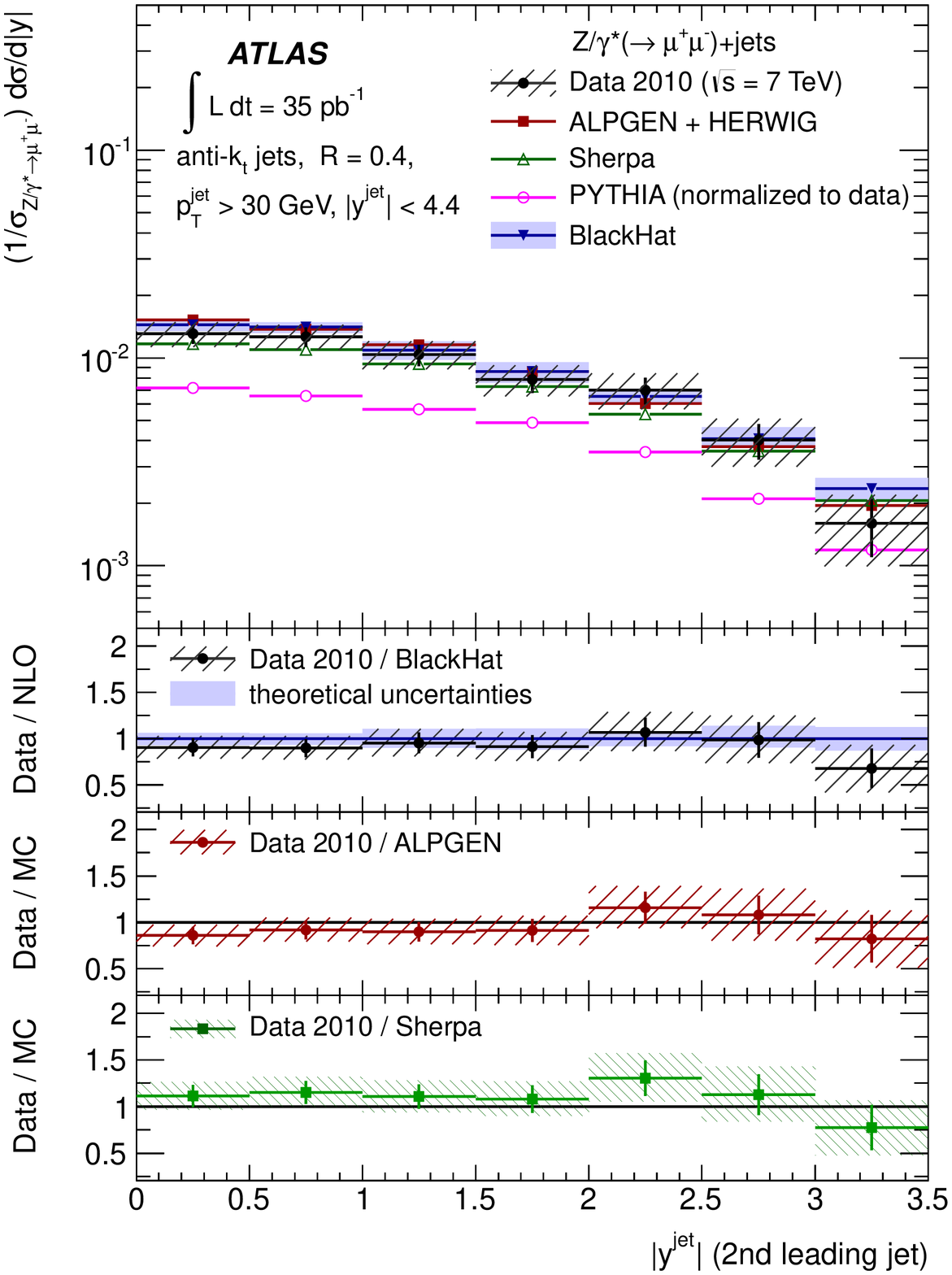}
}
\end{center}
\caption{\small 
Measured normalized jet cross section $(1/\sigma_{\zlls})  \mathrm{d}\sigma / \mathrm{d}|{\rapjet}|$ (black dots) in 
 (left) $\zee$+jets and (right) $\zmm$+jets production
as a function of the second-leading jet $|{\rapjet}|$, in events with at least two jets with 
$\ptjet >30$~GeV and $|\rapjet|<4.4$ in the final state, and normalized by 
$\sigma_{\zees}$ and $\sigma_{\zmms}$ Drell-Yan cross sections, respectively.
}
\label{fig:rap2}
\end{figure}

Inclusive jet differential cross sections $(1/\sigma_{\zlls}) \mathrm{d}\sigma / \mathrm{d}|\rapjet|$ 
as a function of $|\rapjet|$ for jets with $\ptjet > 30$~GeV are presented in Fig.~\ref{fig:rapincl}, while 
Fig.~\ref{fig:rap1} shows the jet measurements as a function of the rapidity of the 
leading jet.  
The measured cross sections decrease with increasing $|\rapjet|$ and are well described by 
ALPGEN and the BlackHat NLO pQCD predictions. Sherpa provides a good description of the data in the region 
$|\rapjet| < 3.5$ but predicts a slightly larger cross section than observed in data for very forward jets.  
PYTHIA provides a good description of the shape of the measured cross sections in the region
$|\rapjet| < 2.5$ but  predicts a smaller cross section than the data in the forward region. 
In Fig.~\ref{fig:rap2}, the measured differential cross sections are presented as functions of the $|\rapjet|$ of 
the second leading jet, for events with at least two jets in the final state. 
The data are described by the  predictions from ALPGEN and Sherpa, and BlackHat NLO pQCD, while again PYTHIA 
does not describe the data. 

\subsection{$\mathrm{d}\sigma/\mathrm{d}\mjj$}
\label{sec:results_mass}

\begin{figure}[h] 
\begin{center}
\mbox{
\includegraphics[width=0.495\textwidth]{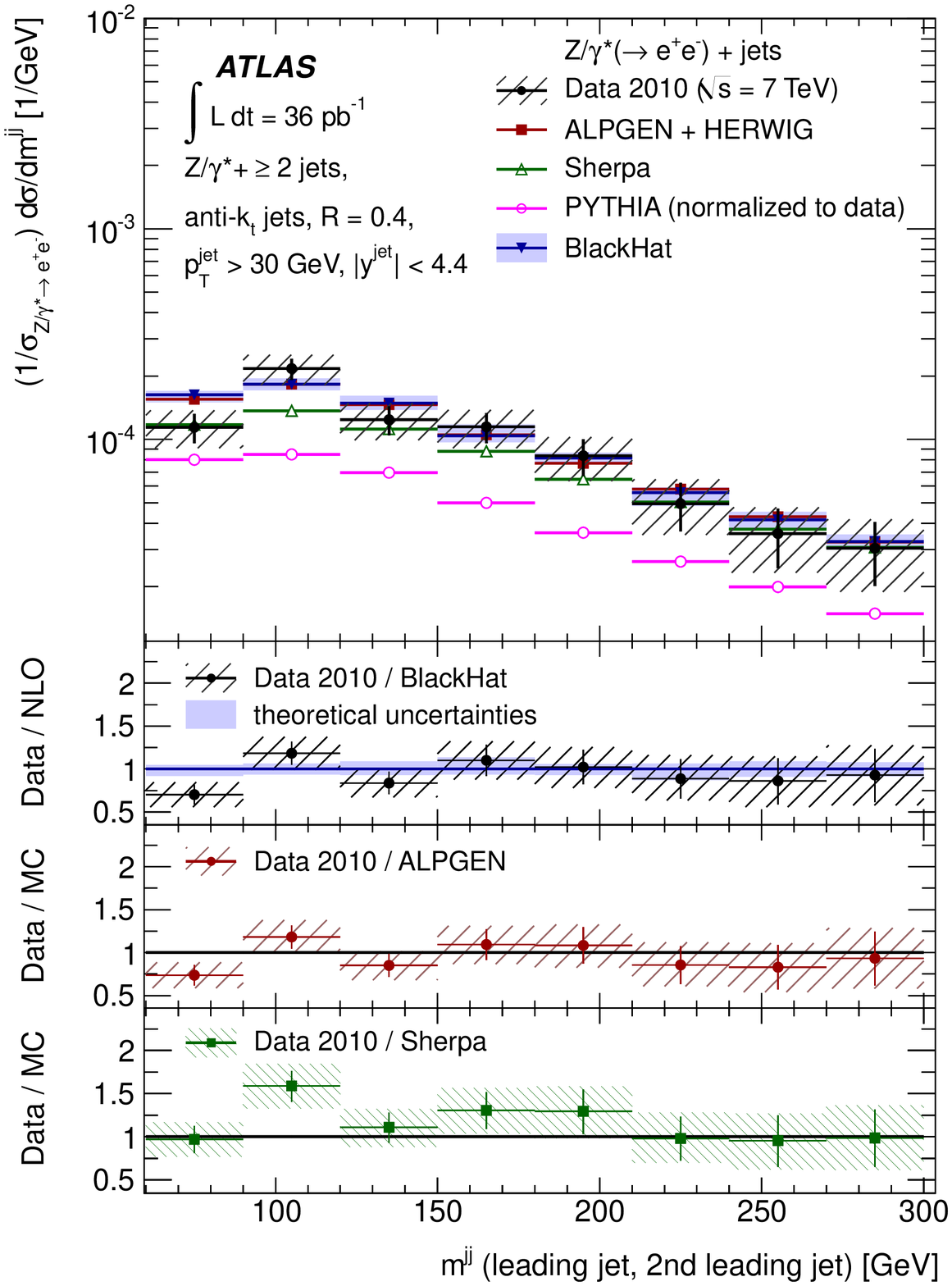}
\includegraphics[width=0.495\textwidth]{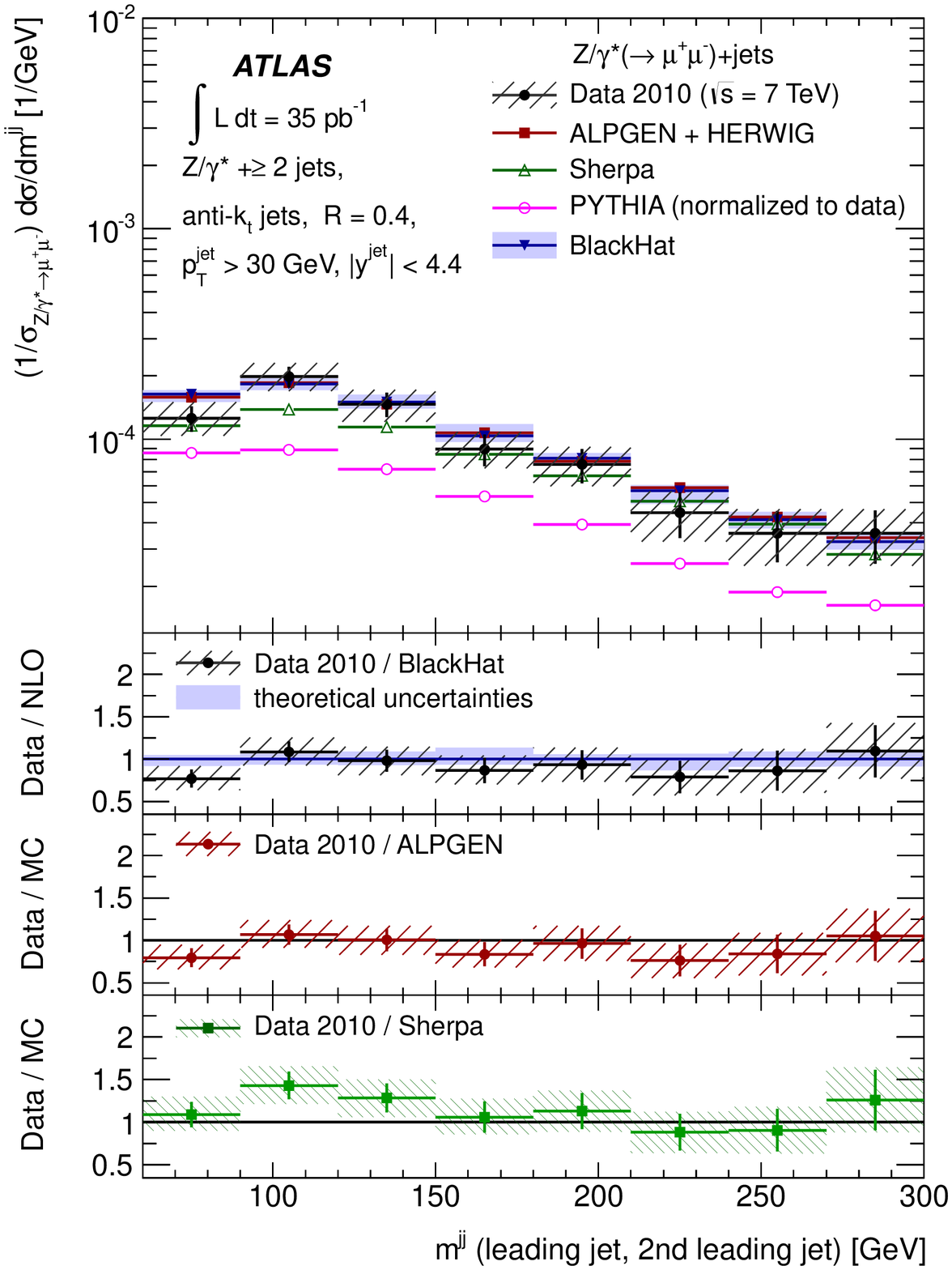}
}
\end{center}
\caption{\small 
Measured normalized dijet cross section $ (1/\sigma_{\zlls}) \mathrm{d}\sigma /\mathrm{d}\mjj$ (black dots) in 
 (left) $\zee$+jets and (right) $\zmm$+jets production
as a function of the invariant mas of the two leading jets $\mjj$, in events with at least two jets with 
$\ptjet >30$~GeV and $|\rapjet|<4.4$ in the final state, and normalized by 
$\sigma_{\zees}$ and $\sigma_{\zmms}$ Drell-Yan cross sections, respectively.
}
\label{fig:mass}
\end{figure}

The measured differential cross sections  $(1/\sigma_{\zlls}) \mathrm{d}\sigma / \mathrm{d}\mjj$ as a function of the invariant mass
of the two leading jets in the event for $60$~GeV $ < \mjj < 300$~GeV are presented in Fig.~\ref{fig:mass} for both
electron and muon channels.  The shape of the measured cross section at low $\mjj$ 
is affected by the jet $\ptjet$ threshold in the cross section definition. For $\mjj > 100$~GeV, the measured cross sections 
decrease with increasing $\mjj$. The measurements are well described by ALPGEN and Sherpa, and the BlackHat NLO pQCD
predictions. PYTHIA approximately reproduces the shape of the measured distribution  
but underestimates the measured cross sections.

\subsection{$\mathrm{d}\sigma/\mathrm{d} |\rapjj|$, $\mathrm{d}\sigma/\mathrm{d} |\phijj|$, and $\mathrm{d}\sigma/\mathrm{d} \rjj$}
\label{sec:results_dijet}

\begin{figure}[h] 
\begin{center}
\mbox{
\includegraphics[width=0.495\textwidth]{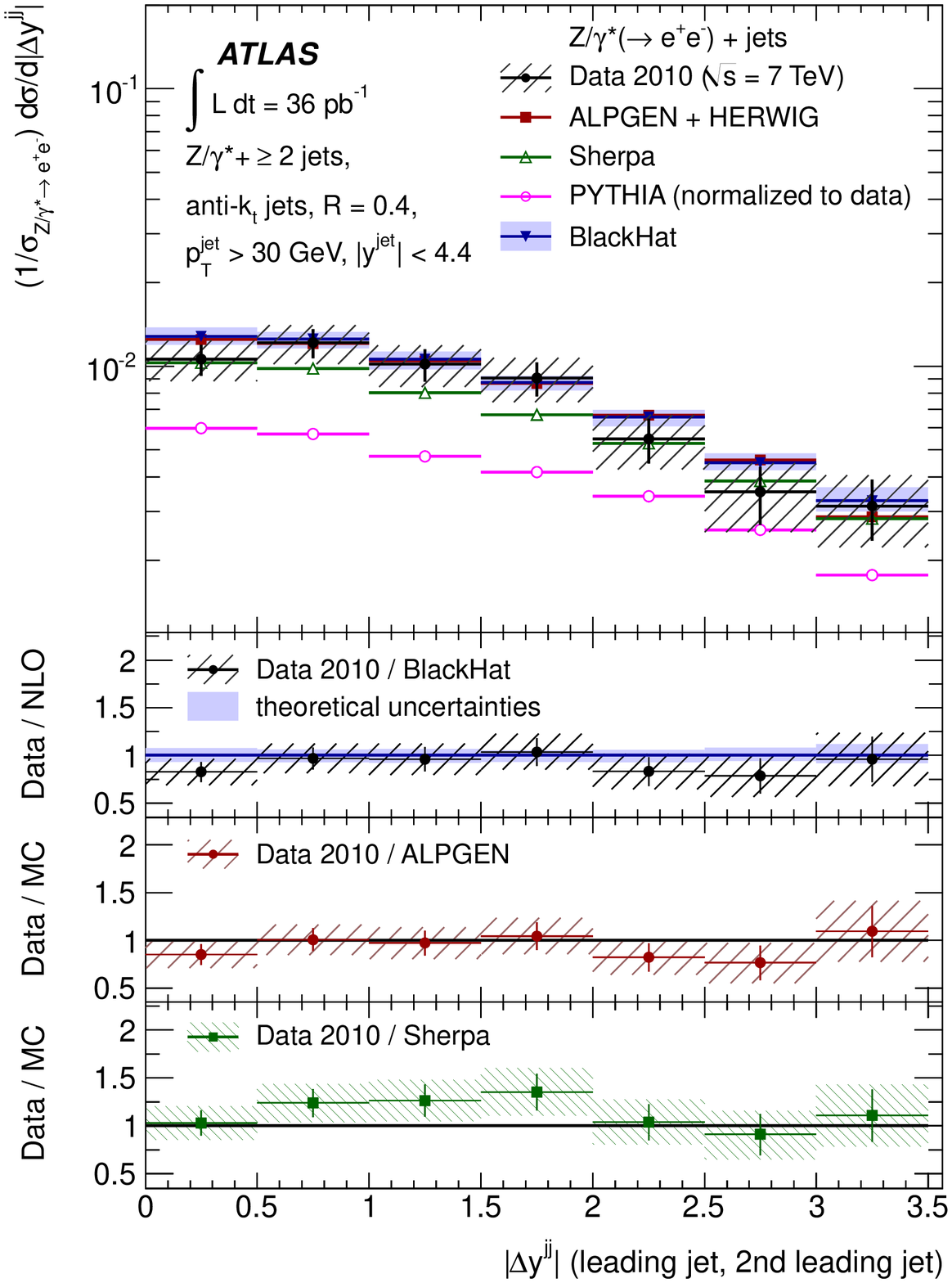}
\includegraphics[width=0.495\textwidth]{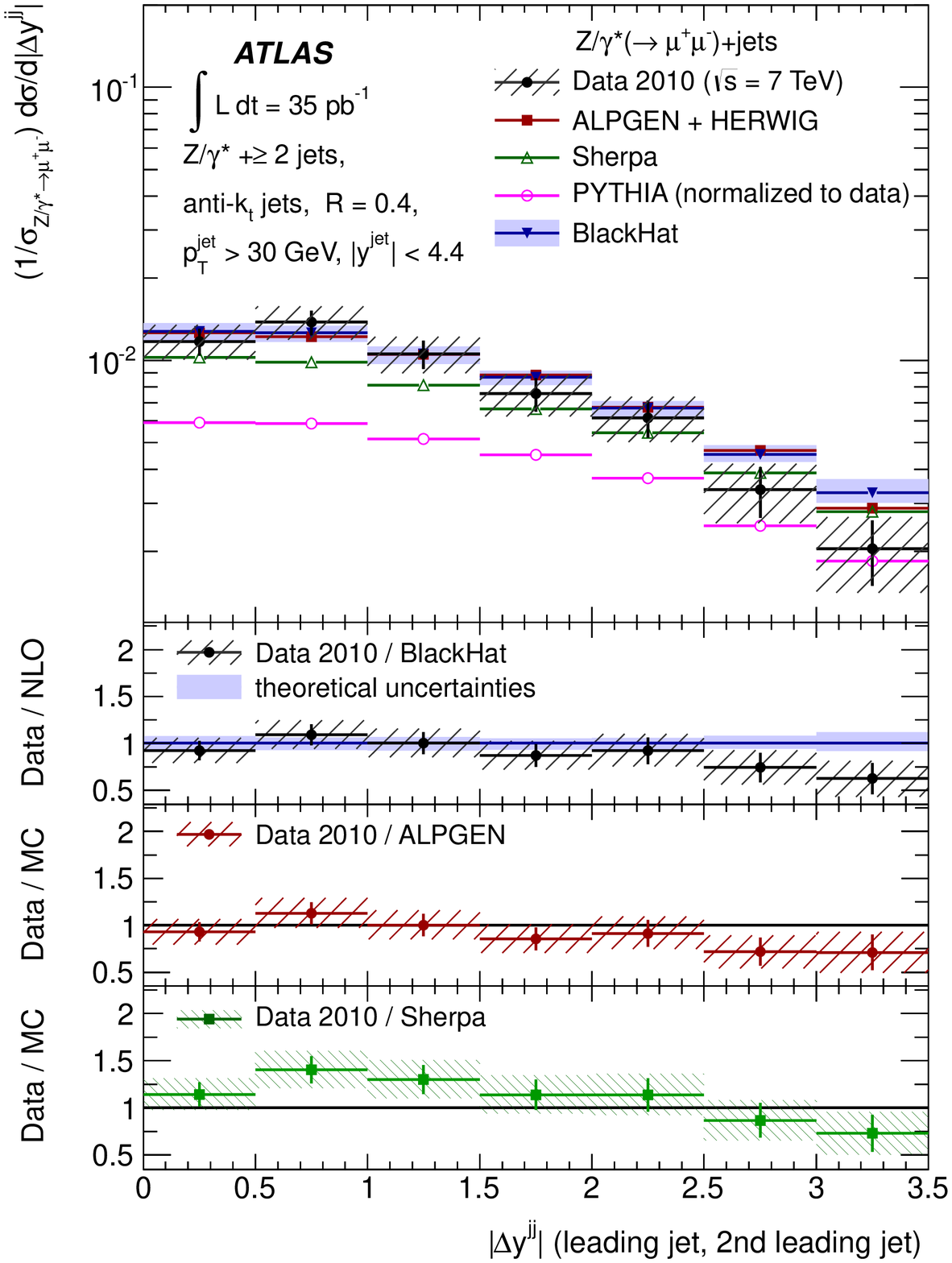}
}
\end{center}
\caption{\small 
Measured normalized dijet cross section $(1/\sigma_{\zlls}) \mathrm{d}\sigma / \mathrm{d}|\rapjj|$ (black dots) in 
 (left) $\zee$+jets and (right) $\zmm$+jets production
as a function of the rapidity separation  of the two leading jets $|\rapjj|$, in events with at least two jets with 
$\ptjet >30$~GeV and $|\rapjet|<4.4$ in the final state, and normalized by 
$\sigma_{\zees}$ and $\sigma_{\zmms}$ Drell-Yan cross sections, respectively.
}
\label{fig:dy}
\end{figure}
\begin{figure}[h] 
\begin{center}
\mbox{
\includegraphics[width=0.495\textwidth]{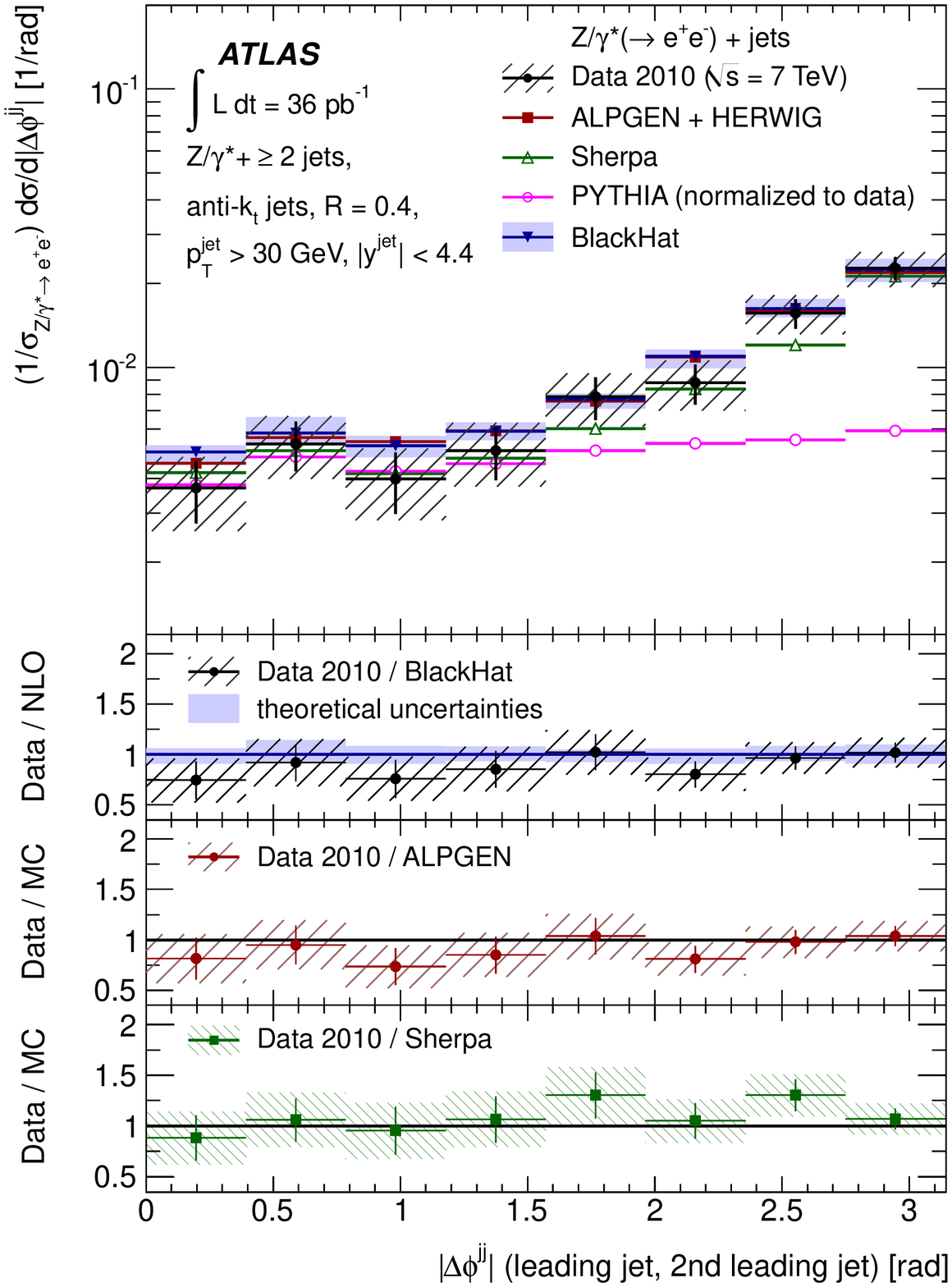}
\includegraphics[width=0.495\textwidth]{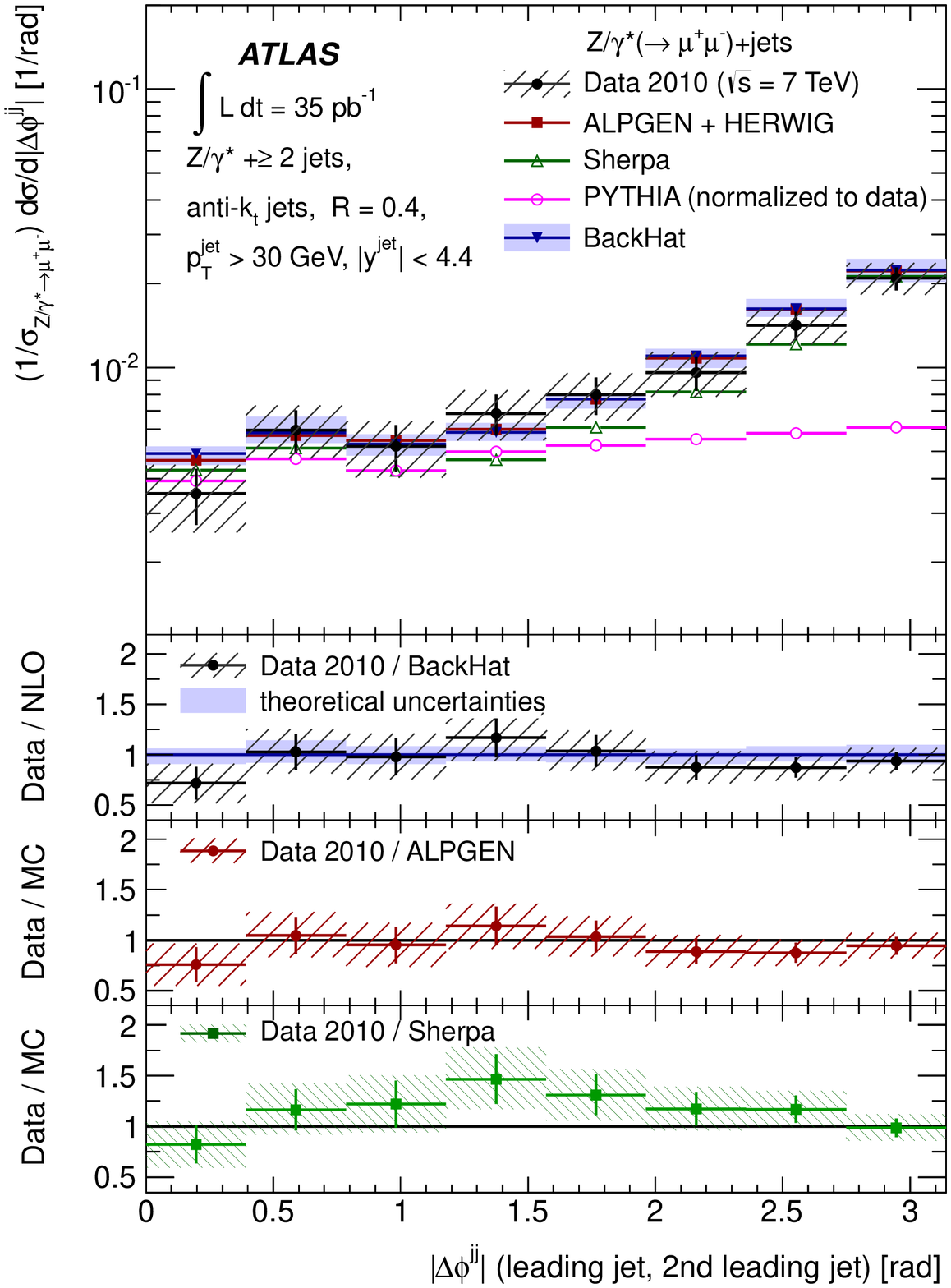}
}
\end{center}
\caption{\small 
Measured normalized dijet cross section $(1/\sigma_{\zlls}) \mathrm{d}\sigma / \mathrm{d}|\phijj|$ (black dots) in 
 (left) $\zee$+jets and (right) $\zmm$+jets production
as a function of the azimuthal separation  of the two leading jets $|\phijj|$, in events with at least two jets with 
$\ptjet >30$~GeV and $|\rapjet|<4.4$ in the final state, and normalized by 
$\sigma_{\zees}$ and $\sigma_{\zmms}$ Drell-Yan cross sections, respectively.
}
\label{fig:dphi}
\end{figure}
\begin{figure}[h] 
\begin{center}
\mbox{
\includegraphics[width=0.495\textwidth]{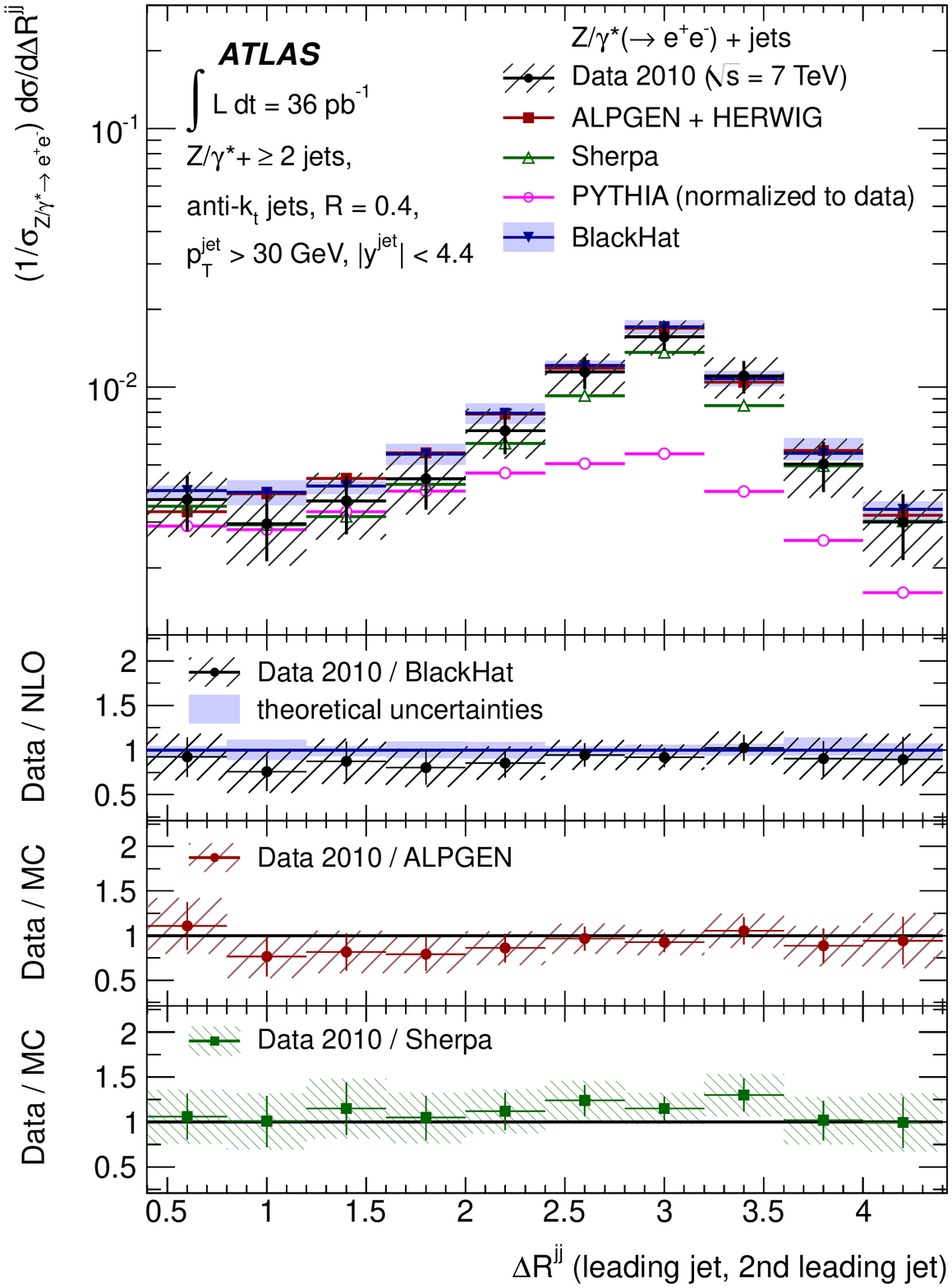}
\includegraphics[width=0.495\textwidth]{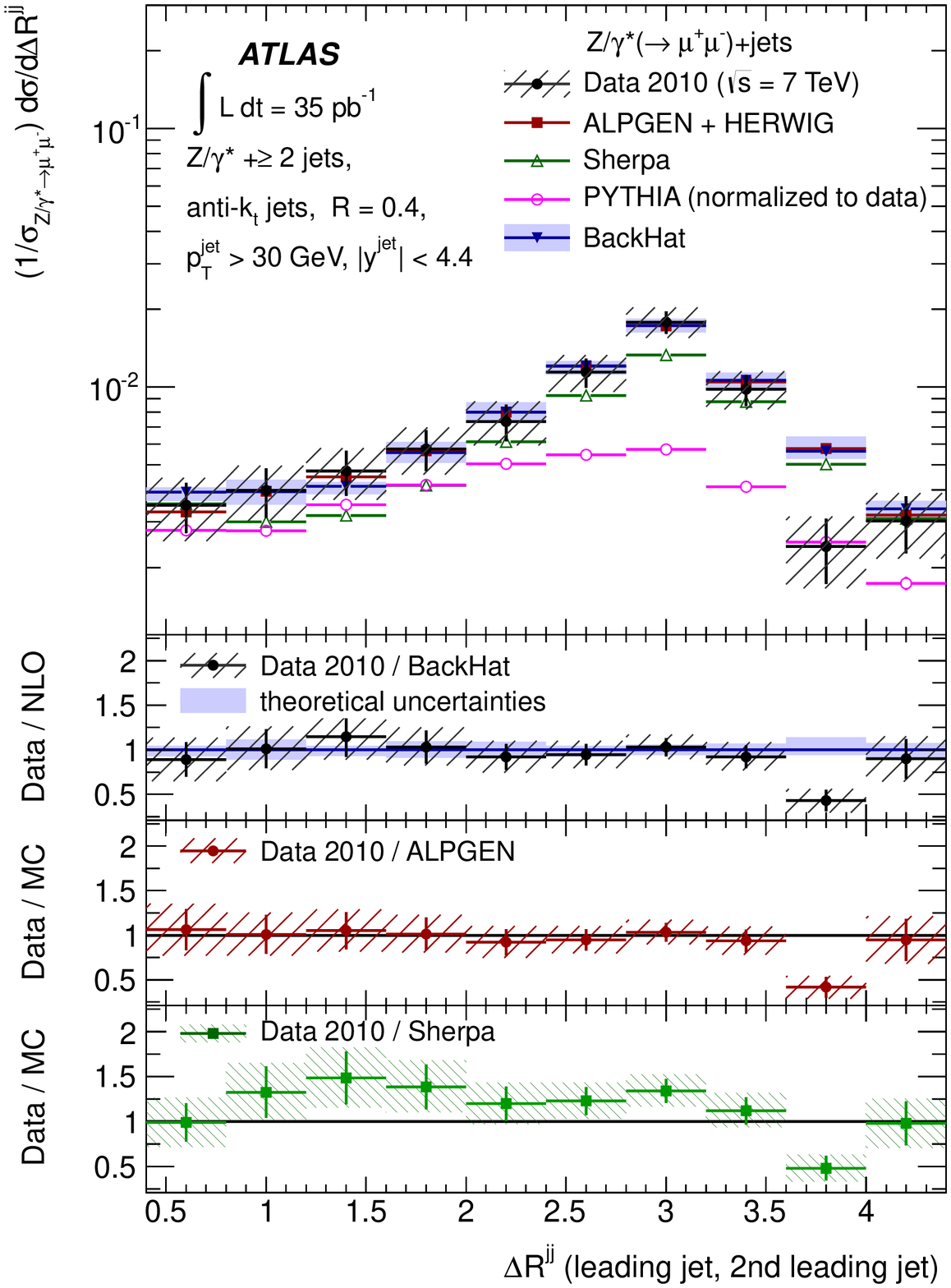}
}
\end{center}
\caption{\small 
Measured normalized dijet cross section $(1/\sigma_{\zlls}) \mathrm{d}\sigma / \mathrm{d} \rjj$ (black dots) in 
 (left) $\zee$+jets and (right) $\zmm$+jets production
as a function of the angular separation ($y-\phi$ space)  
of the two leading jets $\rjj$, in events with at least two jets with 
$\ptjet >30$~GeV and $|\rapjet|<4.4$ in the final state, and normalized by 
$\sigma_{\zees}$ and $\sigma_{\zmms}$ Drell-Yan cross sections, respectively.
}
\label{fig:dr}
\end{figure}

Inclusive dijet cross sections are also measured as a function of the spatial separation of the two leading jets 
in the final state. Figure~\ref{fig:dy} shows the measured differential cross section as a function of the rapidity 
separation of the jets $(1/\sigma_{\zlls}) \mathrm{d}\sigma / \mathrm{d} |\rapjj|$, for both the electron and muon analysis, compared to the different 
predictions.  The measured differential cross sections as a function of the 
azimuthal separation between jets $(1/\sigma_{\zlls}) \mathrm{d}\sigma / \mathrm{d} |\phijj|$ are presented in Fig.~\ref{fig:dphi}, and Fig.~\ref{fig:dr} shows the measured 
differential cross sections $(1/\sigma_{\zlls})$ $\mathrm{d}\sigma / \mathrm{d} \rjj$ as a function of the angular separation $\rjj$ 
between the two leading jets in the event.
The measurements are well described by ALPGEN and Sherpa, and the BlackHat NLO pQCD predictions, while 
PYTHIA underestimates the measured cross sections. In particular, PYTHIA underestimates the data for large
$|\phijj|$ values and for those topologies corresponding to well-separated jets.

\subsection{Combination of electron and muon results}
\label{sec:results_comb}
The measured cross section distributions for the $\zee$+jets and $\zmm$+jets analyses are combined. 
In this case, the results are not 
normalized by the inclusive $Z/\gamma^*$ cross section after the combination, with the aim to 
present also precise absolute jet cross section measurements.

As already discussed, the electron and muon measurements  
are performed in different fiducial regions for the rapidity of the leptons in the 
final state. In addition, the QED radiation effects are different in both channels. 
For each measured distribution, bin-by-bin correction factors, as extracted from ALPGEN  $\zee$+jets and  $\zmm$+jets MC samples,  
are used to extrapolate the measurements to the region   
$\ptjet > 20$~GeV and  $|\eta| <2.5$ for the leptons, where the   
lepton kinematics are defined at the decay vertex of the $Z$ boson.
The increased acceptance in the lepton rapidities translates into about a 
14$\%$ and  a 5$\%$ increase of the measured cross sections in the electron and muon channels, respectively. 
As already mentioned in Section~\ref{sec:nlo}, the correction for QED effects increases the cross sections by 
about 2$\%$. The uncertainties on the acceptance corrections are at the per mille level, as determined by using 
Sherpa instead of ALPGEN, and by considering different PDFs among the CTEQ6.6 and MSTW sets.     
A $\chi^2$ test is performed for each observable to quantify the agreement between the electron and muon results before they are combined, where the statistical and uncorrelated uncertainties are taken into account. The statistical tests lead to 
probabilities larger than 60$\%$  for the electron and muon measurements to be compatible with each other, 
consistent with slightly conservative systematic uncertainties.    

\begin{figure}[h] 
\begin{center}
\mbox{
\includegraphics[width=0.80\textwidth]{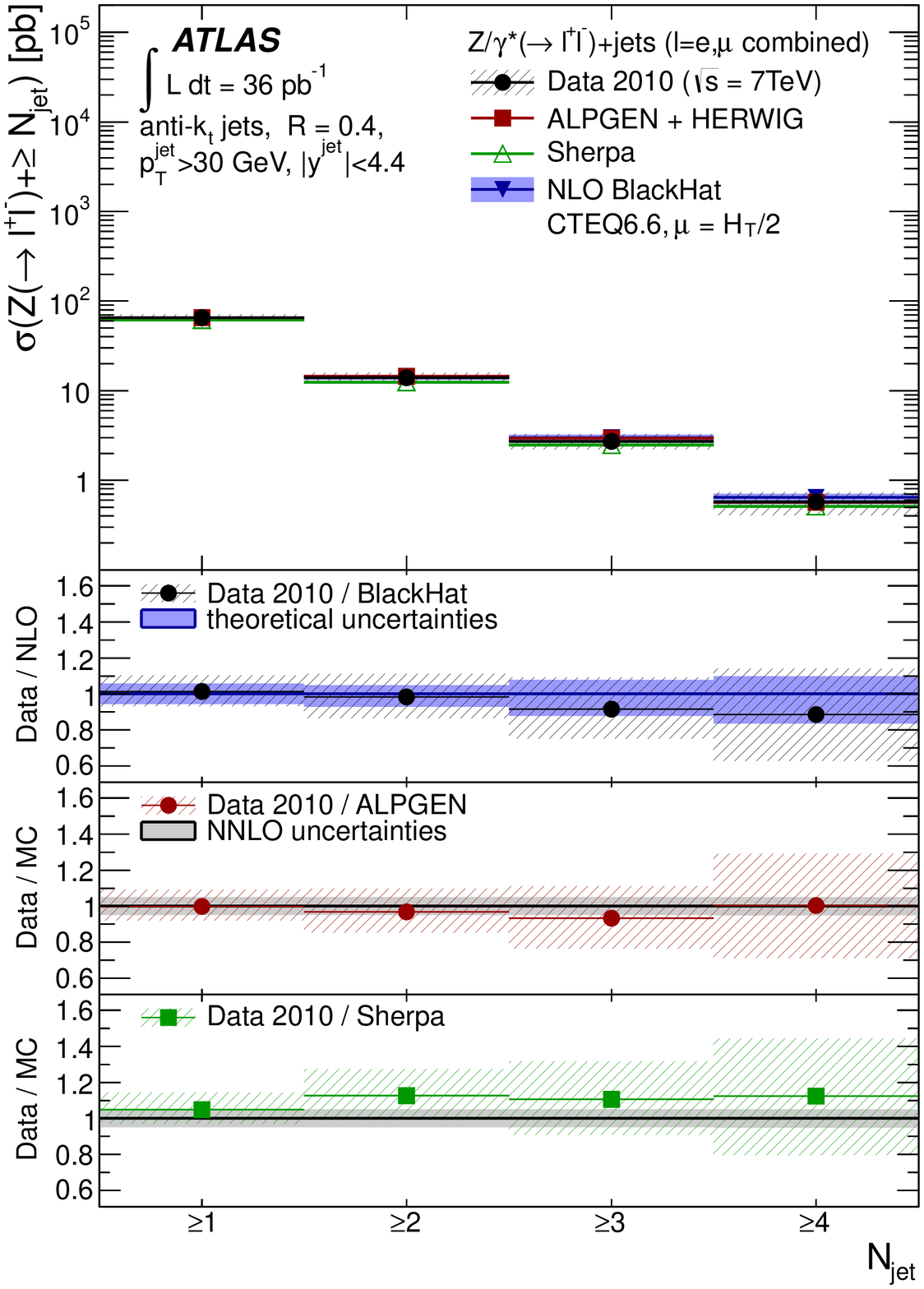}
}
\end{center}
\caption{\small 
Measured cross section $\sigma_{\njet}$ (black dots) in $\zll$+jets production
as a function of the inclusive jet multiplicity, for events with at least one jet with 
$\ptjet >30$~GeV and $|\rapjet|<4.4$ in the final state. 
In this and subsequent figures~\ref{fig:comb_ratio} - \ref{fig:comb_dr} the error bands indicate the total uncertainty from the  combination of electron and muon results.
The measurements are compared to NLO pQCD predictions from BlackHat, as well as the predictions from ALPGEN and Sherpa (both normalized to the FEWZ value for the total cross section).
}
\label{fig:comb_njet}
\end{figure}

\begin{figure}[h] 
\begin{center}
\mbox{
\includegraphics[width=0.80\textwidth]{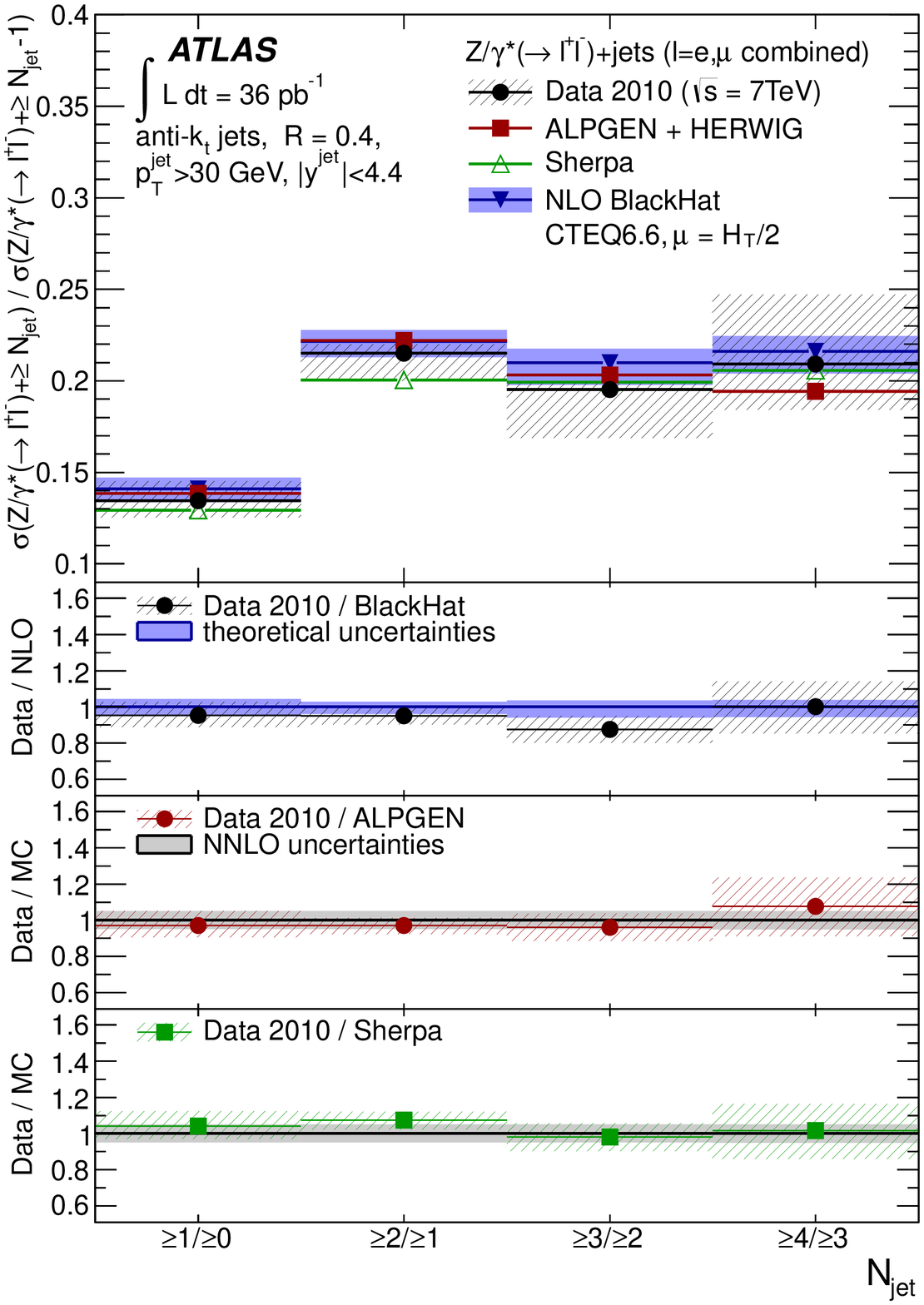}
}
\end{center}
\caption{\small 
Measured ratio of cross sections ($\sigma_{\njet} / \sigma_{\njet - 1}$) (black dots)  
in $\zll$+jets production
as a function of the inclusive jet multiplicity, for events with at least one jet with 
$\ptjet >30$~GeV and $|\rapjet|<4.4$ in the final state. 
}
\label{fig:comb_ratio}
\end{figure}

\begin{figure}[h] 
\begin{center}
\mbox{
\includegraphics[width=0.80\textwidth]{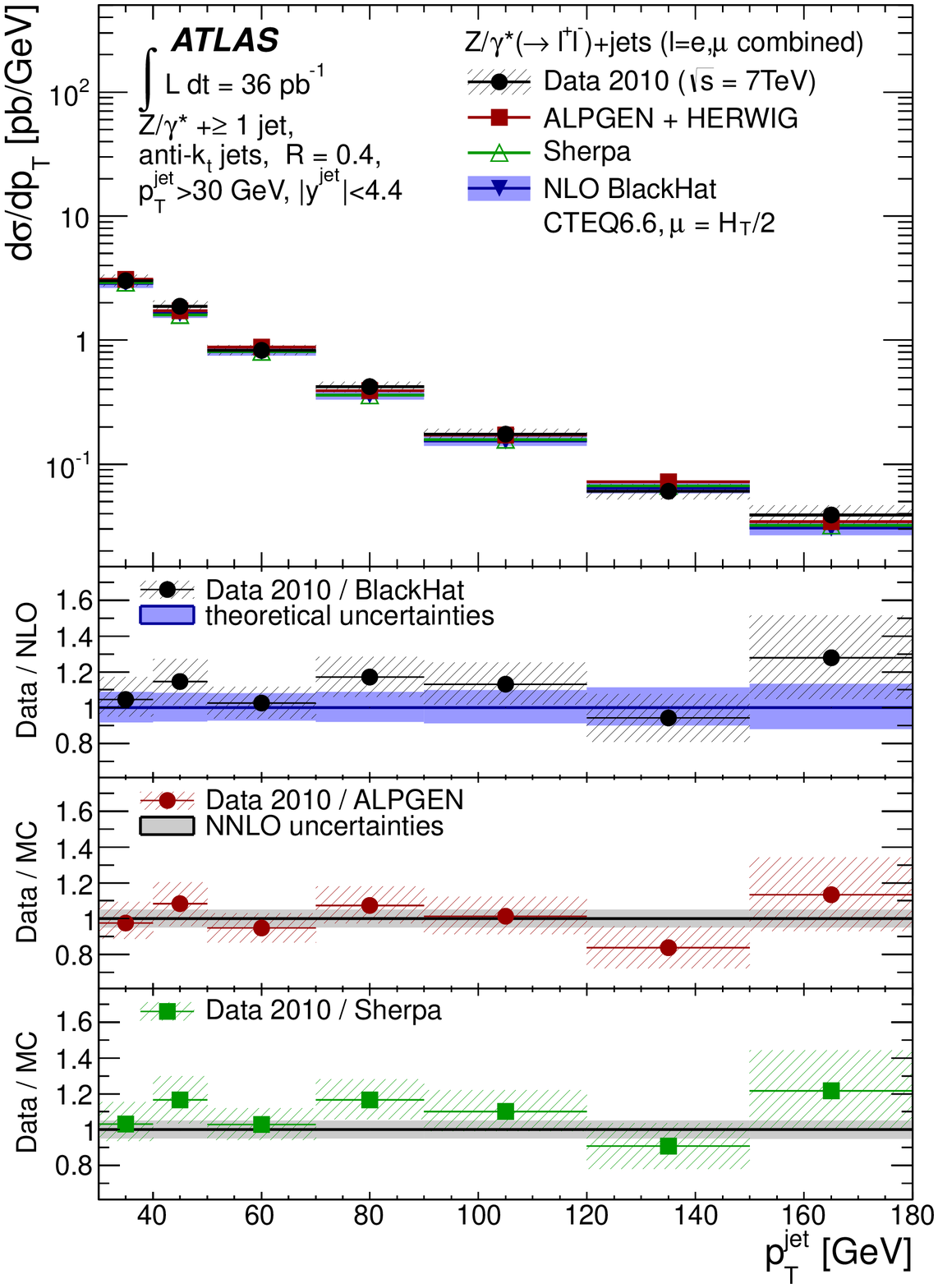}
}
\end{center}
\caption{\small 
Measured inclusive jet cross section $\mathrm{d}\sigma / \mathrm{d} \ptjet$ (black dots) in 
$\zll$+jets production
as a function of $\ptjet$, in events with at least one jet with 
$\ptjet >30$~GeV and $|\rapjet|<4.4$ in the final state.
}
\label{fig:comb_pt}
\end{figure}

\begin{figure}[h] 
\begin{center}
\mbox{
\includegraphics[width=0.80\textwidth]{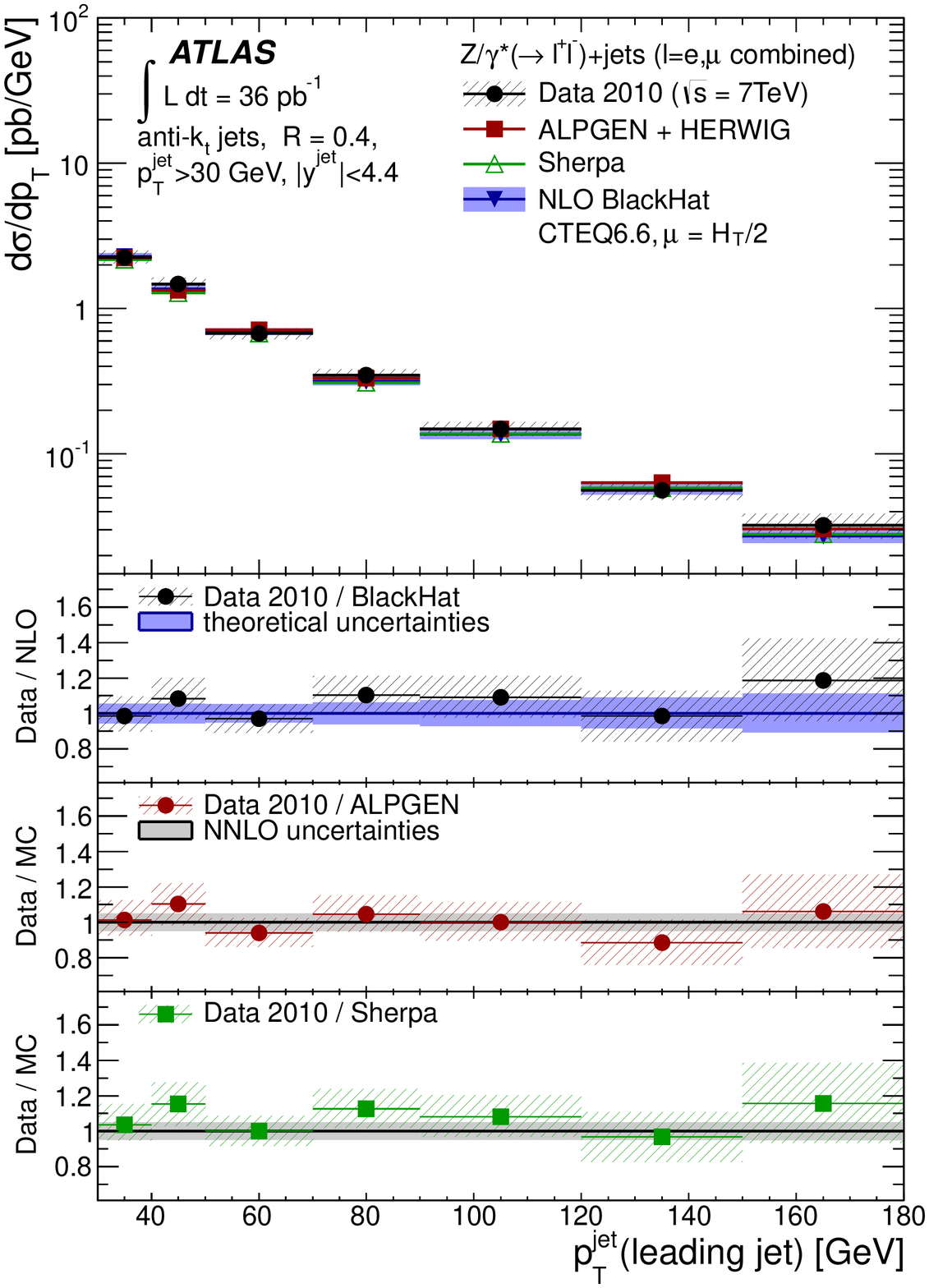}
}
\end{center}
\caption{\small 
Measured jet cross section $\mathrm{d}\sigma / \mathrm{d} \ptjet$ (black dots) in 
$\zll$+jets production
as a function of the leading jet $\ptjet$, in events with at least one jet with 
$\ptjet >30$~GeV and $|\rapjet|<4.4$ in the final state.
}
\label{fig:comb_pt1}
\end{figure}

\begin{figure}[h] 
\begin{center}
\mbox{
\includegraphics[width=0.80\textwidth]{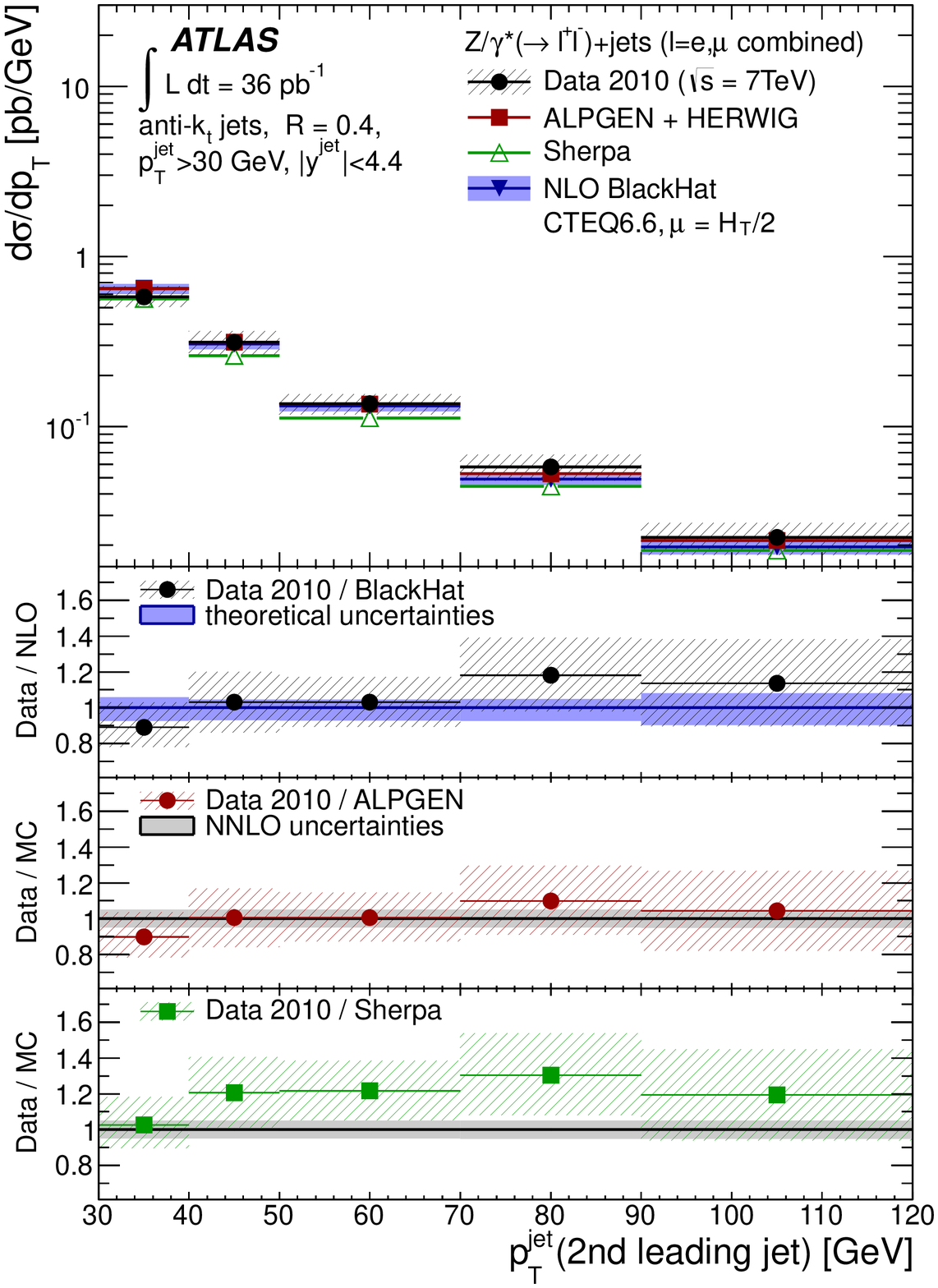}
}
\end{center}
\caption{\small 
Measured jet cross section $\mathrm{d}\sigma / \mathrm{d} \ptjet$ (black dots) in 
$\zll$+jets production
as a function of the second-leading jet $\ptjet$, in events with at least two jets with 
$\ptjet >30$~GeV and $|\rapjet|<4.4$ in the final state.
}
\label{fig:comb_pt2}
\end{figure}

\begin{figure}[h] 
\begin{center}
\mbox{
\includegraphics[width=0.80\textwidth]{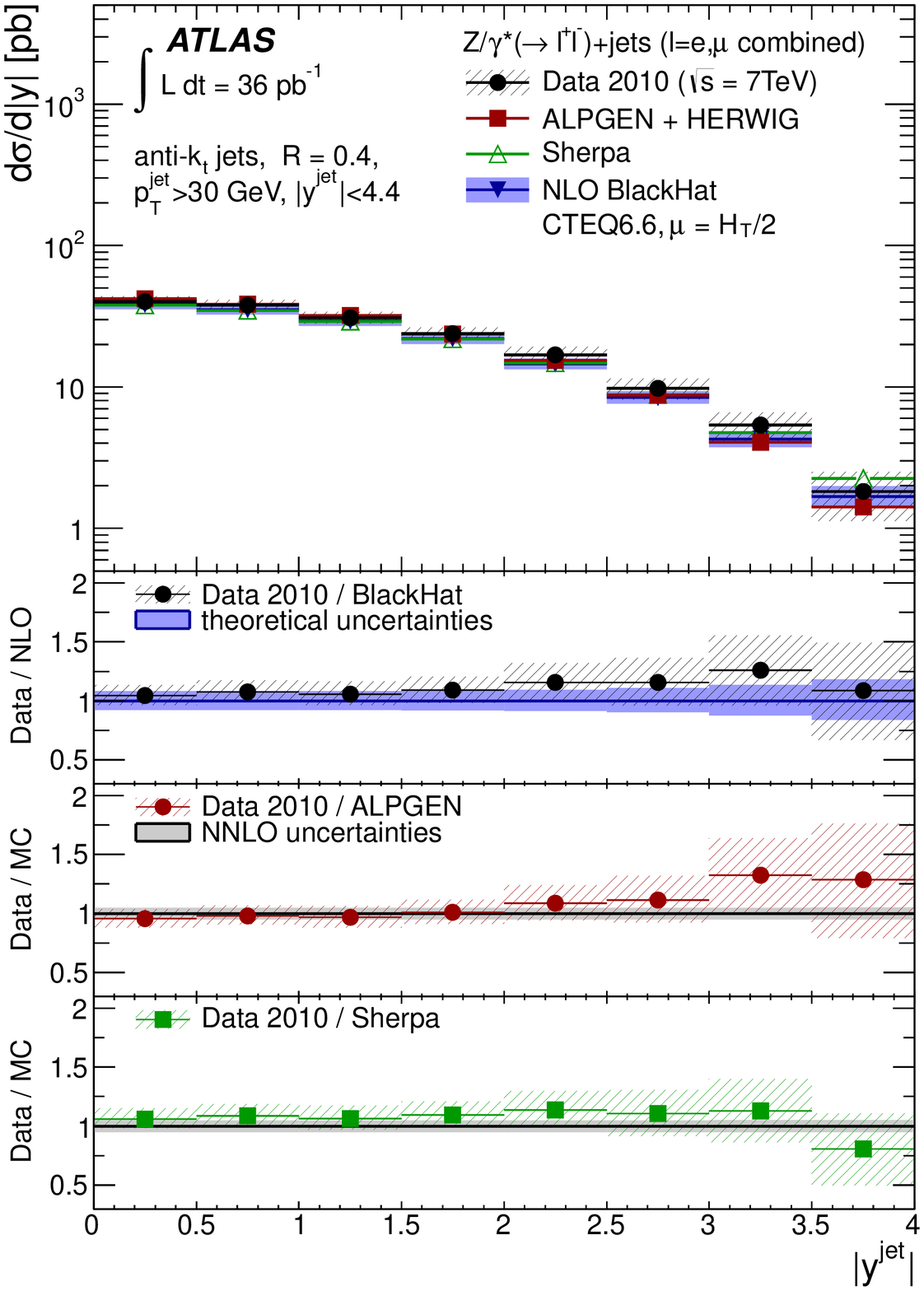}
}
\end{center}
\caption{\small 
Measured inclusive jet cross section $\mathrm{d}\sigma / \mathrm{d}|\rapjet|$ (black dots) in 
$\zll$+jets  production
as a function of $|\rapjet|$, in events with at least one jet with 
$\ptjet >30$~GeV and $|\rapjet|<4.4$ in the final state.
}
\label{fig:comb_rap}
\end{figure}

\begin{figure}[h] 
\begin{center}
\mbox{
\includegraphics[width=0.80\textwidth]{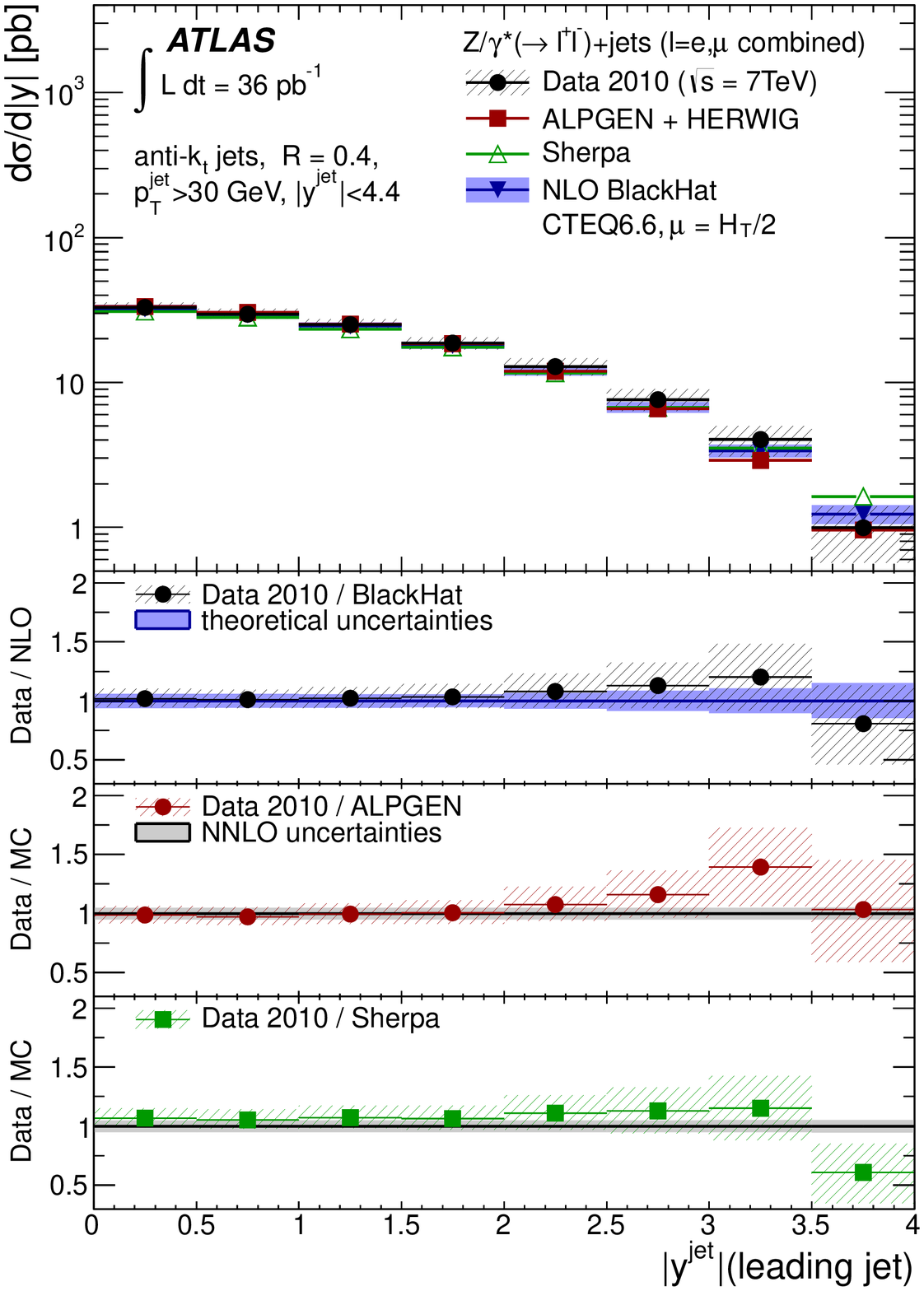}
}
\end{center}
\caption{\small 
Measured jet cross section $\mathrm{d}\sigma / \mathrm{d}|\rapjet|$ (black dots) in 
$\zll$+jets  production
as a function of the leading jet $|\rapjet|$, in events with at least one jet with 
$\ptjet >30$~GeV and $|\rapjet|<4.4$ in the final state.
}
\label{fig:comb_rap1}
\end{figure}

\begin{figure}[h] 
\begin{center}
\mbox{
\includegraphics[width=0.80\textwidth]{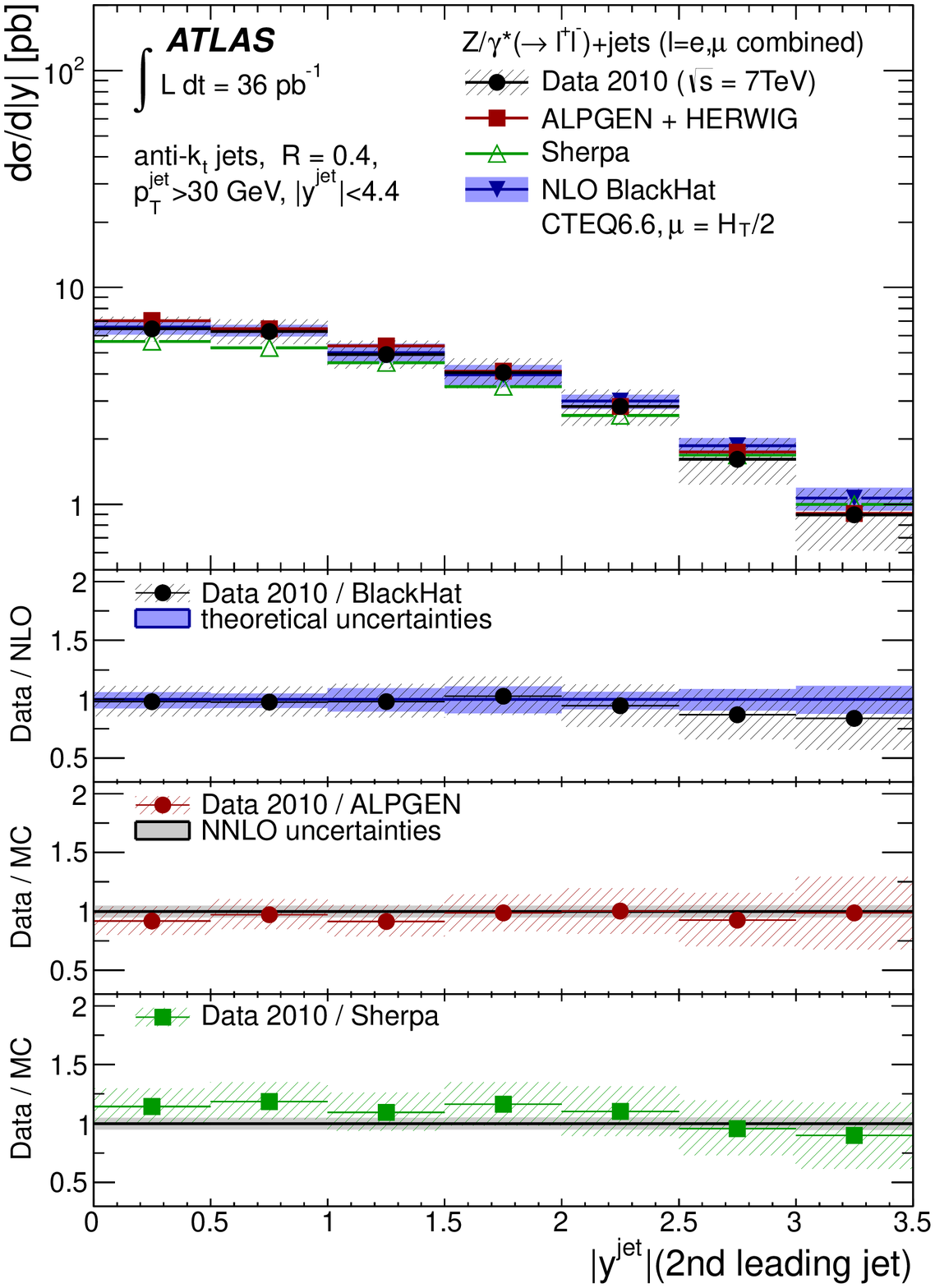}
}
\end{center}
\caption{\small 
Measured jet cross section $\mathrm{d}\sigma / \mathrm{d}|\rapjet|$ (black dots) in 
$\zll$+jets  production
as a function of the second-leading jet $|\rapjet|$, in events with at least two jets with 
$\ptjet >30$~GeV and $|\rapjet|<4.4$ in the final state.
}
\label{fig:comb_rap2}
\end{figure}

\begin{figure}[h] 
\begin{center}
\mbox{
\includegraphics[width=0.8\textwidth]{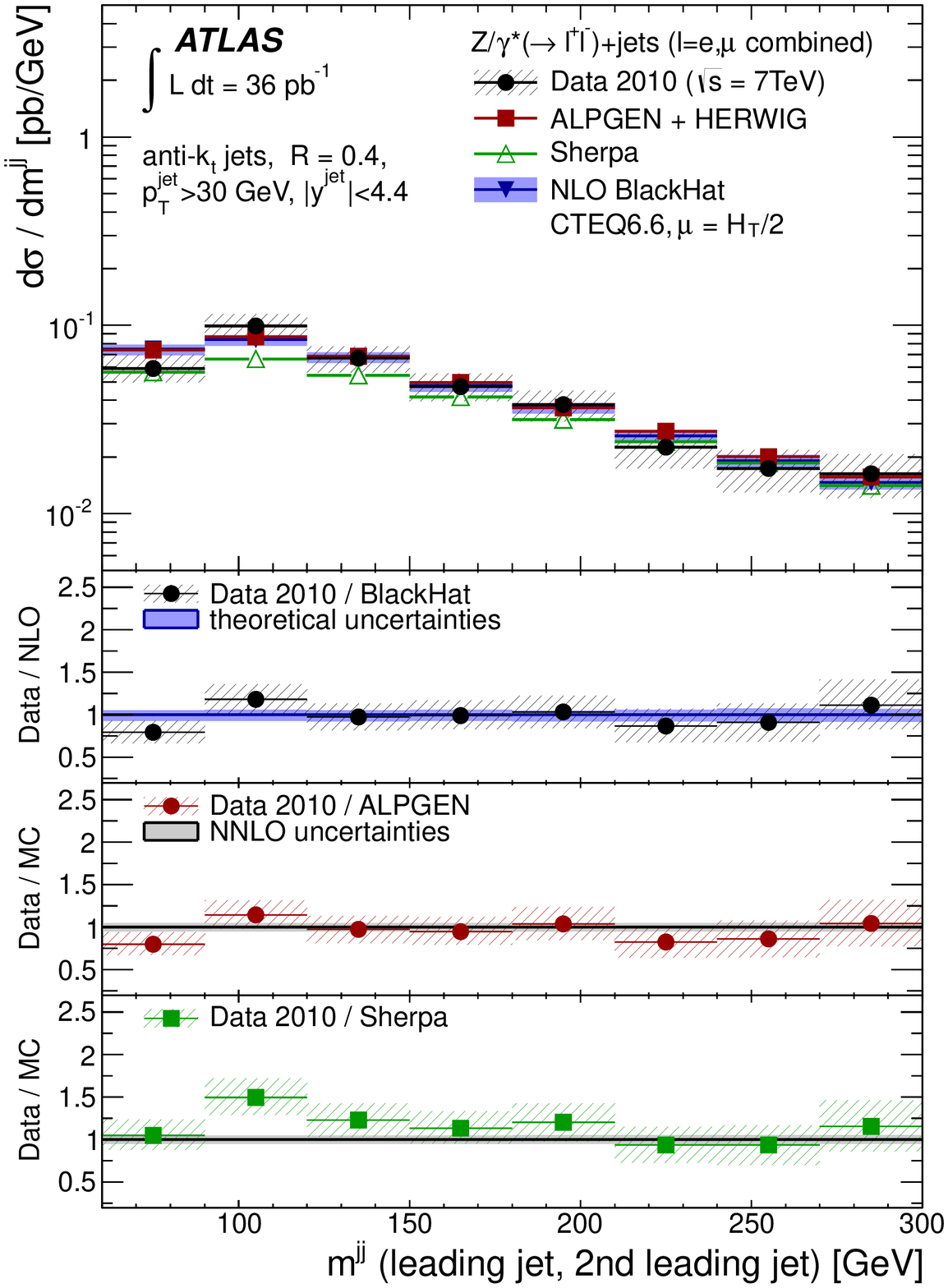}
}
\end{center}
\caption{\small 
Measured dijet cross section $ \mathrm{d}\sigma / \mathrm{d} \mjj$ (black dots) in 
$\zll$+jets  production
as a function of the invariant mass  of the two leading jets $\mjj$, in events with at least two jets with 
$\ptjet >30$~GeV and $|\rapjet|<4.4$ in the final state.
}
\label{fig:comb_mjj}
\end{figure}

\begin{figure}[h] 
\begin{center}
\mbox{
\includegraphics[width=0.8\textwidth]{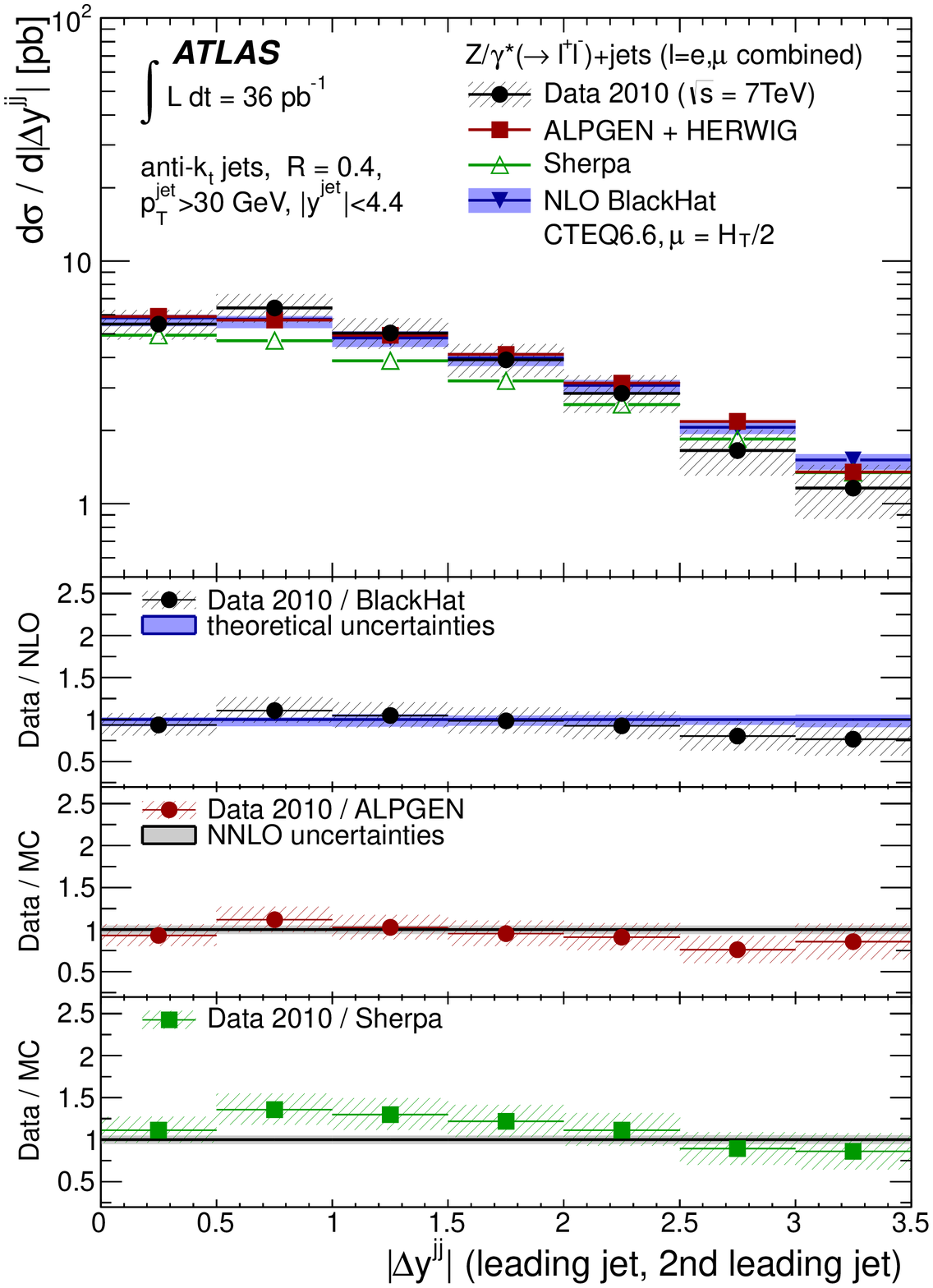}
}
\end{center}
\caption{\small 
Measured dijet cross section $ \mathrm{d}\sigma / \mathrm{d}|\rapjj|$ (black dots) in 
$\zll$+jets  production
as a function of the rapidity separation  of the two leading jets $|\rapjj|$, in events with at least two jets with 
$\ptjet >30$~GeV and $|\rapjet|<4.4$ in the final state.
}
\label{fig:comb_dy}
\end{figure}

\begin{figure}[h] 
\begin{center}
\mbox{
\includegraphics[width=0.8\textwidth]{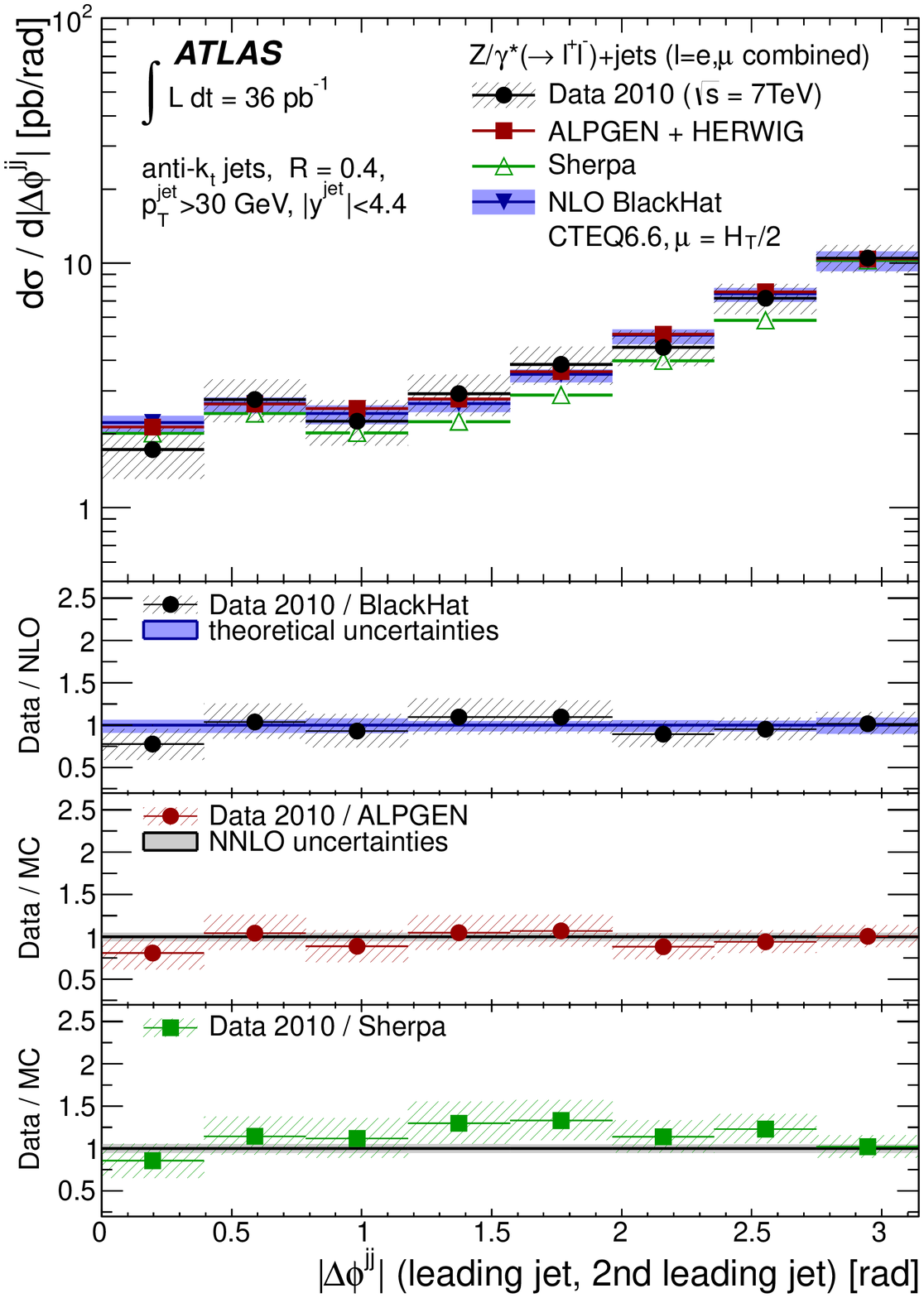}
}
\end{center}
\caption{\small 
Measured dijet cross section $ \mathrm{d}\sigma / \mathrm{d}|\phijj|$ (black dots) in 
$\zll$+jets  production
as a function of the azimuthal separation  of the two leading jets $|\phijj|$, in events with at least two jets with 
$\ptjet >30$~GeV and $|\rapjet|<4.4$ in the final state.
}
\label{fig:comb_dphi}
\end{figure}

\begin{figure}[h] 
\begin{center}
\mbox{
\includegraphics[width=0.8\textwidth]{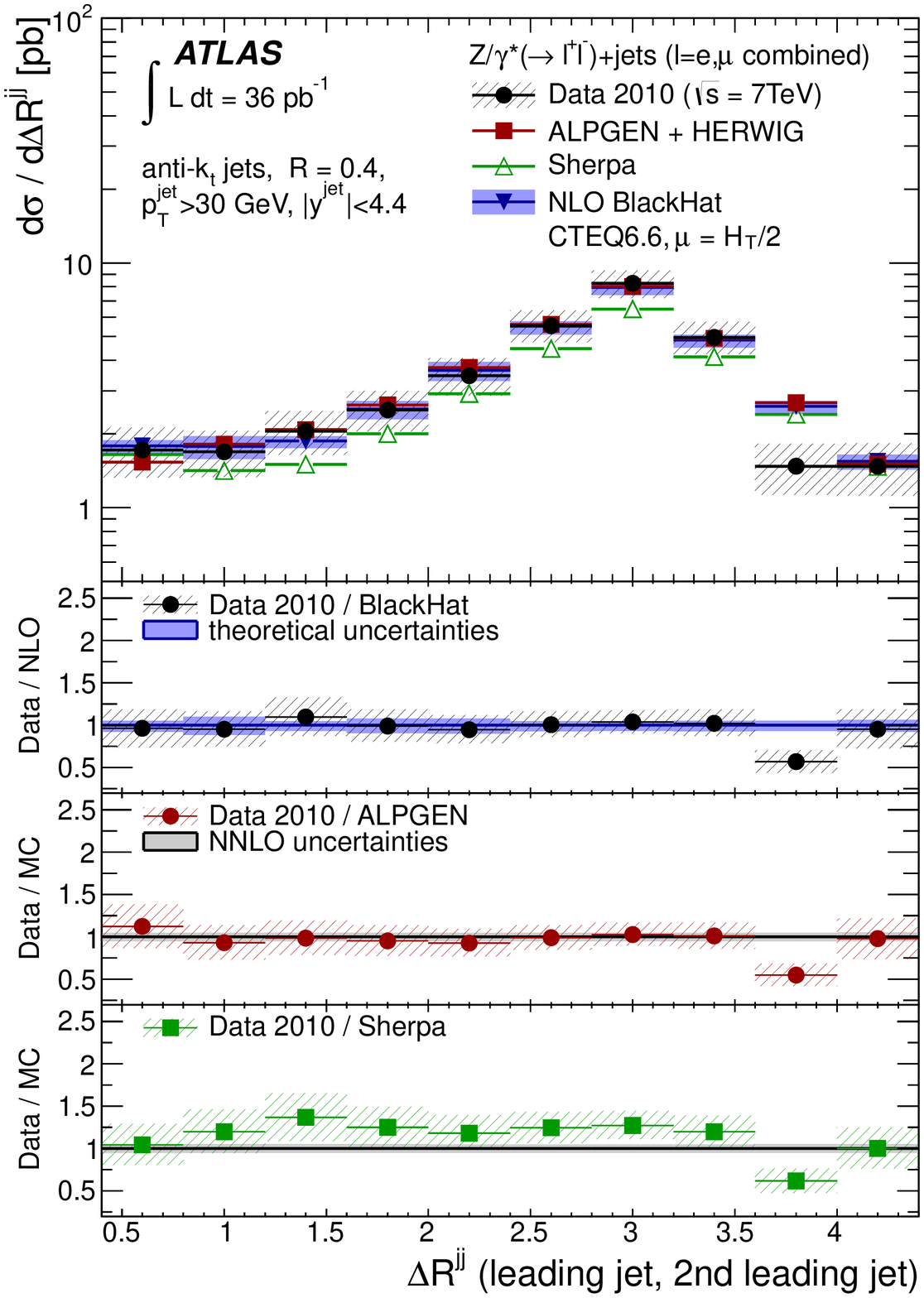}
}
\end{center}
\caption{\small 
Measured dijet cross section $ \mathrm{d}\sigma / \mathrm{d} \rjj$ (black dots) in 
$\zll$+jets  production
as a function of the angular separation ($y - \phi$ space) of the two leading jets $\rjj$, in events with at least two jets with 
$\ptjet >30$~GeV and $|\rapjet|<4.4$ in the final state.
}
\label{fig:comb_dr}
\end{figure}

The electron and muon results are combined using the BLUE (Best Linear Unbiased Estimate)~\cite{blue} method,
 which considers the correlations between the systematic uncertainties in the two channels. The uncertainties 
related to the trigger, the lepton reconstruction, and the multi-jets background estimation are considered uncorrelated 
between the two channels, while the rest of the systematic uncertainties are treated as fully correlated. 
Figures~\ref{fig:comb_njet} to \ref{fig:comb_dr} show the combined results, and  
Tables~\ref{tab:comb_njet} to \ref{tab:comb_rjj} collect the final measurements for the electron and muon channels and 
their combination, together with the multiplicative parton-to-hadron correction factors 
$\delta^{\rm had}$ applied to the BlackHat NLO pQCD predictions (see Section~\ref{sec:nlo}). 
The measurements are well described by the BlackHat NLO pQCD predictions, and by the predictions 
from ALPGEN and Sherpa. 
The corresponding $\chi^2$ tests relative to the different predictions,
 performed separately in each channel and for each observable, 
lead to $\chi^2$ per degree of freedom values in the range between 0.05 and 2.70.
Further details of the combination and the $\chi^2$ tests are presented in the Appendix.


\section{Summary}
\label{sec:sum}

In summary, results are reported for inclusive jet production in $\zees$ and $\zmms$
events in  proton-proton collisions at $\sqrt{s}=7$~TeV. The analysis considers the  data collected by 
the ATLAS detector in 2010 corresponding
to a total integrated luminosity of  about 
36 pb${}^{-1}$. Jets are defined using the $\akt$
algorithm with $R=0.4$ and the measurements are performed for jets in the region 
$\ptjet >30$~GeV and $|\rapjet|<4.4$. Cross sections are measured as a function of the
inclusive jet multiplicity, and the transverse momentum and rapidity of the jets in the final state. 
Measurements are also performed as a function of the dijet invariant mass and the 
angular separation between the two leading jets in events with at least two jets in the final state. 
The measured cross sections are well described by NLO pQCD predictions
including non-perturbative corrections, as well as by predictions of LO matrix elements of up to $2 \to 5$ 
parton scatters, supplemented by parton showers, as implemented in the ALPGEN and Sherpa MC generators.


\input{acknowledgements.tex}


\clearpage


%
%


\clearpage



\clearpage


\clearpage


\clearpage


\input{tables}

\clearpage
\appendix
\section*{Appendix - Combined Results}

The results for the electron and muon channels are  
extrapolated to a common acceptance region $\ptjet > 20$~GeV and $|\eta|<2.5$
for the kinematics of the leptons,  defined at the decay vertex of the Z boson before QED radiation.
For each bin in a given observable $\xi$, the measured cross section $\sigma^{\rm fiducial}_{\xi}$  in each channel
is corrected  according to
\begin{equation}
\sigma^{\rm extrapolated}_{\xi} = \sigma^{\rm fiducial}_{\xi} \times \delta^{\rm QED} \times \mathcal{A},  
\end{equation}
where  $\delta^{\rm QED}$ corrects  for QED radiation effects  back to the Born level and 
$\mathcal{A}$ extrapolates the result to the new lepton acceptance region.  
Tables~\ref{tab:accept1} to \ref{tab:accept4} present
the correction factors applied to the measured cross sections, 
separately for the electron and muon analyses.

The  results are then combined using the BLUE~\cite{blue} method  
that takes into account the correlations between systematic uncertainties in the two channels. The method
assumes Gaussian $\chi^2$ distributions and is not directly able to treat the asymmetric 
systematic uncertainties present in 
the measured cross sections.  Therefore, a modified asymmetric iterative BLUE method is employed.

Three separate BLUE combinations are computed, using 
as an input the upper, the lower, and the average of the upper and lower uncertainties in the electron and muon 
channels, leading to three different results here denoted as $\sigma^{\rm up}_{\xi} \pm  \Delta\sigma^{\rm up}_{\xi}$,  
$\sigma^{\rm low}_{\xi} \pm \Delta\sigma^{\rm low}_{\xi}$, and 
$\sigma^{\rm ave}_{\xi} \pm \Delta\sigma^{\rm ave}_{\xi}$, respectively. The central value for the 
combined cross section $\sigma_{\xi}$, and its upper and lower 
uncertainties,  ${\Delta^{+}\sigma_{\xi}}$ and ${\Delta^{-}\sigma_{\xi}}$ respectively, are 
given by the expressions

\begin{equation}
\sigma_{\xi} = \sigma^{\rm ave}_{\xi}, \\
\end{equation} 
\begin{equation}
\Delta^{+}\sigma_{\xi} = 2 \times R \times \Delta\sigma^{\rm ave}_{\xi}, \ {\rm and} \\
\end{equation} 
\begin{equation}
\Delta^{-}\sigma_{\xi} = 2 \times (1 - R) \times \Delta\sigma^{\rm ave}_{\xi},  \\
\end{equation} 

\noindent
with
\begin{equation}
R = \frac{\Delta\sigma^{\rm up}_{\xi}}{\Delta\sigma^{\rm up}_{\xi} + \Delta\sigma^{\rm low}_{\xi}}.
\end{equation}

\noindent
The BLUE method provides uncertainties on the combined measurement that include 
both statistical and systematic uncertainties.

Finally, $\chi^2$ tests to the  data points in each measured cross section before and after extrapolation 
are performed  with respect to the NLO pQCD, ALPGEN, and Sherpa predictions, according to 
\begin{equation}
\chi^2 = \sum_{j=1}^{bins} \frac{[{  d}_j - {  t}_j(\bar{s})]^2}{[\delta{  d}_j]^2 + [\delta{  t}_j(\bar{s})]^2
 } 
+ \sum_{i=1}^{7} [s_i]^2 \ , 
\end{equation}
\noindent
where ${  d}_j$ is the measured data point $j$,   ${  t}_j(\bar{s})$ is 
the corresponding prediction, and  $\bar{s}$  
denotes the vector of  standard deviations, $s_i$, for the different independent sources of 
systematic uncertainty in data and theory, which are considered fully correlated across bins. 
For each measurement considered,  the sums above  run over the total number of data points 
and seven independent sources of systematic uncertainty, 
and the correlations among systematic uncertainties  
are taken into account in ${  t}_j(\bar{s})$.
The 
average of the  upper and lower uncertainties in data and theory are employed, and the 
$\chi^2$ is minimized with respect to $\bar{s}$. 
The results of the $\chi^2$ tests are tabulated in 
Tables~XVIII to~XX.


\input{extrapol}

\input{chi2}


\clearpage
\input{atlas_authlist}

\end{document}

%% file: acknowledgements.tex

\section{Acknowledgements}

We thank CERN for the very successful operation of the LHC, as well as the
support staff from our institutions without whom ATLAS could not be
operated efficiently.

We acknowledge the support of ANPCyT, Argentina; YerPhI, Armenia; ARC,
Australia; BMWF, Austria; ANAS, Azerbaijan; SSTC, Belarus; CNPq and FAPESP,
Brazil; NSERC, NRC and CFI, Canada; CERN; CONICYT, Chile; CAS, MOST and
NSFC, China; COLCIENCIAS, Colombia; MSMT CR, MPO CR and VSC CR, Czech
Republic; DNRF, DNSRC and Lundbeck Foundation, Denmark; ARTEMIS, European
Union; IN2P3-CNRS, CEA-DSM/IRFU, France; GNAS, Georgia; BMBF, DFG, HGF, MPG
and AvH Foundation, Germany; GSRT, Greece; ISF, MINERVA, GIF, DIP and
Benoziyo Center, Israel; INFN, Italy; MEXT and JSPS, Japan; CNRST, Morocco;
FOM and NWO, Netherlands; RCN, Norway; MNiSW, Poland; GRICES and FCT,
Portugal; MERYS (MECTS), Romania; MES of Russia and ROSATOM, Russian
Federation; JINR; MSTD, Serbia; MSSR, Slovakia; ARRS and MVZT, Slovenia;
DST/NRF, South Africa; MICINN, Spain; SRC and Wallenberg Foundation,
Sweden; SER, SNSF and Cantons of Bern and Geneva, Switzerland; NSC, Taiwan;
TAEK, Turkey; STFC, the Royal Society and Leverhulme Trust, United Kingdom;
DOE and NSF, United States of America.

The crucial computing support from all WLCG partners is acknowledged
gratefully, in particular from CERN and the ATLAS Tier-1 facilities at
TRIUMF (Canada), NDGF (Denmark, Norway, Sweden), CC-IN2P3 (France),
KIT/GridKA (Germany), INFN-CNAF (Italy), NL-T1 (Netherlands), PIC (Spain),
ASGC (Taiwan), RAL (UK) and BNL (USA) and in the Tier-2 facilities
worldwide.

%% file: tables.tex


\begin{table}[htbp]
\begin{center}
\begin{footnotesize}
\begin{tabular}{|c| c| c| c| c|} \hline\hline
\multicolumn{5}{|c|}{\small{$\sigma_{\njet} \ [\rm{pb}]$}} \\ \hline\hline 
 $\njet$ & $\zee$ & $\zmm$ & $\zll$ & $\delta^{\rm had} \pm ({\rm{total \ unc.}})$\\ 
  & $\sigma \pm {\rm{(stat.)}} \pm {\rm{(syst.)}} $ & $\sigma \pm {\rm{(stat.)}} \pm {\rm{(syst.)}}$ & $\sigma \pm {\rm{(total \ unc.)}}$ & parton $\to$ hadron \\ \hline 
$\geq$ 1 jet &  $  69 \pm 2 \pm 7 $ &  $  65 \pm 2 {}^{+ 6 }_{- 5 }$ & $ 65 {}^{+ 6 }_{- 5 }$ & $ 0.99 \pm  0.02 $ \\
$\geq$ 2 jets &  $  14.3 \pm 0.9 \pm 1.9 $ &  $  13.9 \pm 0.7 {}^{+ 1.7 }_{- 1.6 }$ & $ 14.0 \pm 1.8  $ & $ 0.98 \pm  0.03 $ \\
$\geq$ 3 jets &  $  2.4 \pm 0.4 \pm 0.4 $ &  $  2.9 \pm 0.3 {}^{+ 0.5 }_{- 0.4 }$ & $ 2.7 \pm 0.5 $ & $0.98 \pm 0.05 $  \\
$\geq$ 4 jets &  $  0.6 \pm 0.2 \pm 0.1 $ &  $  0.6 \pm 0.2 \pm 0.1 $ & $ 0.6 \pm 0.2$ & $1.03 \pm 0.05 $ \\ \hline\hline
\end{tabular}
\end{footnotesize}
\caption{\small
Measured cross section $\sigma_{\njet}$ as a function of the inclusive jet multiplicity, for events with at least one jet with 
$\ptjet >30$~GeV and $|\rapjet|<4.4$ in the final state. 
In this and subsequent tables~\ref{tab:comb_njetminus} - \ref{tab:comb_rjj}
the results are presented for the $\zee$ and $\zmm$ analyses separately, as extrapolated to the Born level in the common acceptance region $\ptjet > 20$~GeV and $|\eta| < 2.5$ for the lepton kinematics,  and their combination. 
The multiplicative parton-to-hadron correction factors $\delta^{\rm had}$ are applied to the NLO pQCD predictions.
}
\label{tab:comb_njet}
\end{center}
\end{table}


\begin{table}[htbp]
\begin{center}
\begin{footnotesize}
\begin{tabular}{|c| c| c| c| c|} \hline\hline
\multicolumn{5}{|c|}{\small{$\sigma_{\njet}/\sigma_{\njet - 1}$}} \\ \hline\hline 
 $\njet$ & $\zee$ & $\zmm$ & $\zll$ & $\delta^{\rm had} \pm ({\rm{total \ unc.}})$\\ 
  & ratio $\pm {\rm{(stat.)}} \pm {\rm{(syst.)}} $ & ratio $\pm {\rm{(stat.)}} \pm {\rm{(syst.)}}$ & ratio $\pm {\rm{(total \ unc.)}}$ & parton $\to$ hadron \\ \hline 
$\geq$ 1 jet &  $0.139  \pm 0.002 \pm 0.011$ & $0.135 \pm 0.003 {}^{+0.010}_{-0.009}$& $0.135 {}^{ +0.011}_{-0.009}$ & $0.99 \pm 0.03 $\\
$\geq$ 2 jets & $0.208  \pm 0.007 {}^{+0.008}_{-0.009}$ & $0.215 \pm 0.010 {}^{+0.008}_{-0.009}$& $0.215 {}^{ +0.010}_{-0.011}$ & $0.99 \pm 0.01 $\\
$\geq$ 3 jets & $0.17  \pm 0.02 \pm 0.01$ & $0.21 \pm 0.02 \pm 0.01$& $0.20 \pm 0.02$ & $1.00 \pm 0.02 $\\
$\geq$ 4 jets & $0.23  \pm 0.04 \pm 0.01$ & $0.20 \pm 0.05 {}^{+0.01}_{-0.02}$& $0.21 \pm 0.03$ & $1.05 \pm 0.03 $\\ \hline\hline
\end{tabular}
\end{footnotesize}
\caption{\small
Measured cross section ratio $\sigma_{\njet}/\sigma_{\njet - 1}$ as a function of the inclusive jet multiplicity, for events with at least one jet with 
$\ptjet >30$~GeV and $|\rapjet|<4.4$ in the final state. 
}
\label{tab:comb_njetminus}
\end{center}
\end{table}



\begin{table}[htbp]
\begin{center}
\begin{footnotesize}
\begin{tabular}{|c| c| c| c| c|} \hline\hline
\multicolumn{5}{|c|}{\small{$\mathrm{d}\sigma / \mathrm{d}\ptjet \ [\rm{pb}/\rm{GeV}]$ \ (inclusive) }} \\ \hline\hline 
 $\ptjet$ & $\zee$ & $\zmm$ & $\zll$ & $\delta^{\rm had} \pm ({\rm{total \ unc.}})$\\ 
 $[\rm{GeV}]$ & $\sigma \pm {\rm{(stat.)}} \pm {\rm{(syst.)}}$ & $\sigma \pm {\rm{(stat.)}} \pm {\rm{(syst.)}}$ & $\sigma \pm {\rm{(total \ unc.)}}$ & parton $\to$ hadron  \\ \hline
30-40 &  $  3.2 \pm 0.1 {}^{+ 0.3 }_{- 0.4 }$ &  $  2.9 \pm 0.1 {}^{+ 0.4 }_{- 0.2 }$ & $ 3.0 {}^{+ 0.4 }_{- 0.3 }$ & $ 1.00 \pm  0.04 $ \\
40-50 &  $  1.9 \pm 0.1 \pm 0.2$ &  $  1.9 \pm 0.1 \pm 0.2$ & $ 1.9 \pm 0.2$ & $ 0.99 \pm  0.02 $ \\
50-70 &  $  0.89 \pm 0.05 {}^{+ 0.09 }_{- 0.08 }$ &  $  0.81 \pm 0.04 \pm 0.06 $ & $ 0.83 \pm 0.07$ & $ 0.99 \pm  0.02 $ \\
70-90 &  $  0.42 \pm 0.03 \pm 0.04$ &  $  0.42 \pm 0.03 \pm 0.03$ & $ 0.42 \pm 0.04$ & $ 0.98 \pm  0.01 $ \\
90-120 &  $  0.17 \pm 0.02 \pm 0.02$ &  $  0.18 \pm 0.02 \pm 0.01$ & $ 0.17 \pm 0.02$ & $ 0.98 \pm  0.01 $ \\
120-150 &  $  0.073 \pm 0.011 \pm 0.008$ &  $  0.055 \pm 0.008 {}^{+ 0.004 }_{- 0.005 }$ & $ 0.061 {}^{+ 0.009 }_{- 0.008 }$ & $ 1.00 \pm  0.02 $ \\
150-180 &  $  0.037 \pm 0.008 {}^{+ 0.006 }_{- 0.005 }$ &  $  0.040 \pm 0.007 {}^{+ 0.004 }_{- 0.005 }$ & $ 0.039 \pm 0.007$ & $ 1.01 \pm  0.05 $ \\ \hline\hline
\end{tabular}
\end{footnotesize}
\caption{\small
Measured inclusive jet differential cross section $\mathrm{d}\sigma / \mathrm{d}\ptjet$ as a function of $\ptjet$, for events with at least one jet with 
$\ptjet >30$~GeV and $|\rapjet|<4.4$ in the final state. 
}
\label{tab:comb_pt}
\end{center}
\end{table}



\begin{table}[htbp]
\begin{center}
\begin{footnotesize}
\begin{tabular}{|c| c| c| c|c|} \hline\hline
\multicolumn{5}{|c|}{\small{$\mathrm{d}\sigma / \mathrm{d}\ptjet \ [\rm{pb}/\rm{GeV}]$ \ (leading jet)}} \\ \hline\hline 
 $\ptjet$ & $\zee$ & $\zmm$ & $\zll$ & $\delta^{\rm had} \pm ({\rm{total \ unc.}})$ \\ 
  $[\rm{GeV}]$ & $\sigma \pm {\rm{(stat.)}} \pm {\rm{(syst.)}}$ & $\sigma \pm {\rm{(stat.)}} \pm {\rm{(syst.)}}$ & $\sigma \pm {\rm{(total \ unc.)}} $ & parton $\to$ hadron \\ \hline
30-40 &  $  2.4 \pm 0.1 {}^{+ 0.2 }_{- 0.3 }$ &  $  2.2 \pm 0.1 {}^{+ 0.3 }_{- 0.2 }$ & $ 2.3 {}^{+ 0.3 }_{- 0.2 }$ & $ 1.00 \pm  0.03 $ \\
40-50 &  $  1.5 \pm 0.1 \pm 0.2$ &  $  1.5 \pm 0.1 {}^{+ 0.1 }_{- 0.2 }$ & $ 1.5 \pm 0.2$ & $ 1.00 \pm  0.03 $ \\
50-70 &  $  0.74 \pm 0.04 {}^{+ 0.07 }_{- 0.06 }$ &  $  0.65 \pm 0.03 \pm 0.05$ & $ 0.67 \pm 0.06$ & $ 0.99 \pm  0.02 $ \\
70-90 &  $  0.36 \pm 0.03 {}^{+ 0.04 }_{- 0.03 }$ &  $  0.35 \pm 0.03 \pm 0.03$ & $ 0.35 \pm 0.03$ & $ 0.98 \pm  0.02 $ \\
90-120 &  $  0.15 \pm 0.02 \pm 0.02$ &  $  0.15 \pm 0.01 \pm 0.01$ & $ 0.15 \pm 0.02$ & $ 0.98 \pm  0.01 $ \\
120-150 &  $  0.068 \pm 0.011 {}^{+ 0.008 }_{- 0.007 }$ &  $  0.051 \pm 0.008 \pm 0.004$ & $ 0.056 \pm 0.008$ & $ 1.00 \pm  0.02 $ \\
150-180 &  $  0.034 \pm 0.007 {}^{+ 0.006 }_{- 0.005 }$ &  $  0.031 \pm 0.006 \pm 0.004$ & $ 0.032 \pm 0.006$ & $ 1.01 \pm  0.05 $  \\ \hline\hline
\end{tabular}
\end{footnotesize}
\caption{\small
Measured jet differential cross section $\mathrm{d}\sigma / \mathrm{d}\ptjet$ as a function of the leading-jet $\ptjet$, for events with at least one jet with 
$\ptjet >30$~GeV and $|\rapjet|<4.4$ in the final state. 
}
\label{tab:comb_pt1}
\end{center}
\end{table}



\begin{table}[htbp]
\begin{center}
\begin{footnotesize}
\begin{tabular}{|c| c| c| c|c|} \hline\hline
\multicolumn{5}{|c|}{\small{$\mathrm{d}\sigma / \mathrm{d}\ptjet \ [\rm{pb}/\rm{GeV}]$ \ (second-leading jet)}} \\ \hline\hline 
 $\ptjet$ & $\zee$ & $\zmm$ & $\zll$ & $\delta^{\rm had} \pm ({\rm{total \ unc.}})$ \\ 
  $[\rm{GeV}]$ & $\sigma \pm {\rm{(stat.)}} \pm {\rm{(syst.)}}$ & $\sigma \pm {\rm{(stat.)}} \pm {\rm{(syst.)}}$ & $\sigma \pm {\rm{(total \ unc.)}} $ & parton $\to$ hadron \\ \hline
30-40 &  $  0.66 \pm 0.06 {}^{+ 0.08 }_{- 0.10 }$ &  $  0.55 \pm 0.04 {}^{+ 0.08 }_{- 0.06 }$ & $ 0.58 {}^{+ 0.09 }_{- 0.07 }$ & $ 1.00 \pm  0.04 $ \\
40-50 &  $  0.29 \pm 0.04 {}^{+ 0.05 }_{- 0.04 }$ &  $  0.33 \pm 0.03 {}^{+ 0.04 }_{- 0.05 }$ & $ 0.31 \pm 0.05$ & $ 0.97 \pm  0.02 $ \\
50-70 &  $  0.14 \pm 0.02 \pm 0.02$ &  $  0.13 \pm 0.02 \pm 0.01$ & $ 0.14 \pm 0.02$ & $ 0.97 \pm  0.01 $ \\
70-90 &  $  0.053 \pm 0.012 {}^{+ 0.007 }_{- 0.006 }$ &  $  0.062 \pm 0.011 \pm 0.006$ & $ 0.058 \pm 0.010$ & $ 0.95 \pm  0.02 $ \\
90-120 &  $  0.020 \pm 0.006 \pm 0.002$ &  $  0.024 \pm 0.006 \pm 0.002$ & $ 0.022 \pm 0.005$ & $ 0.95 \pm  0.06 $ \\ \hline\hline
\end{tabular}
\end{footnotesize}
\caption{\small
Measured jet differential cross section $\mathrm{d}\sigma / \mathrm{d}\ptjet$ as a function of the second-leading jet $\ptjet$, for events with at least two jets with 
$\ptjet >30$~GeV and $|\rapjet|<4.4$ in the final state. 
}
\label{tab:comb_pt2}
\end{center}
\end{table}



\begin{table}[htbp]
\begin{center}
\begin{footnotesize}
\begin{tabular}{|c| c| c| c|c|} \hline\hline
\multicolumn{5}{|c|}{\small{$\mathrm{d}\sigma / \mathrm{d}|\rapjet| \ [\rm{pb}]$ \ (inclusive) }} \\ \hline\hline 
 $|\rapjet|$ & $\zee$ & $\zmm$ & $\zll$ & $\delta^{\rm had} \pm ({\rm{total \ unc.}})$ \\ 
          & $\sigma \pm {\rm{(stat.)}} \pm {\rm{(syst.)}}$ & $\sigma \pm {\rm{(stat.)}} \pm {\rm{(syst.)}}$ & $\sigma \pm {\rm{(total \ unc.)}} $ & parton $\to$ hadron \\ \hline
0.0-0.5 &  $  42 \pm 2 \pm 4$ &  $  40 \pm 2 \pm 3$ & $ 40 \pm 3$ & $ 1.00 \pm  0.03 $ \\
0.5-1.0 &  $  39 \pm 2 {}^{+ 3 }_{- 4 }$ &  $  37 \pm 2 \pm 3$ & $ 38 \pm 3$ & $ 1.00 \pm  0.03 $ \\
1.0-1.5 &  $  31 \pm 2 \pm 3$ &  $  31 \pm 1 {}^{+ 3 }_{- 2 }$ & $ 31 \pm 3$ & $ 1.00 \pm  0.03 $ \\
1.5-2.0 &  $  25 \pm 2 \pm 3$ &  $  24 \pm 1 \pm 2$ & $ 24 {}^{+ 3 }_{- 2 }$ & $ 0.99 \pm  0.03 $ \\
2.0-2.5 &  $  16 \pm 1 {}^{+ 1 }_{- 2 }$ &  $  17 \pm 1 \pm 2$ & $ 17 \pm 2$ & $ 0.99 \pm  0.02 $ \\
2.5-3.0 &  $  12 \pm 1 \pm 2$ &  $  8.8 \pm 0.8 \pm 1.4$ & $ 10 \pm 2$ & $ 0.97 \pm  0.02 $ \\
3.0-3.5 &  $  5.7 \pm 0.8 {}^{+ 1.3 }_{- 1.2 }$ &  $  5.2 \pm 0.6 {}^{+ 1.1 }_{- 1.2 }$ & $ 5.4 \pm 1.3$ & $ 0.95 \pm  0.03 $ \\
3.5-4.0 &  $  1.9 \pm 0.5 {}^{+ 0.7 }_{- 0.6 }$ &  $  1.8 \pm 0.4 {}^{+ 0.6 }_{- 0.7 }$ & $ 1.8 \pm 0.7$ & $ 0.91 \pm  0.03 $ \\ \hline\hline
\end{tabular}
\end{footnotesize}
\caption{\small
Measured inclusive jet differential cross section $\mathrm{d}\sigma / \mathrm{d}|\rapjet|$ as a function of $|\rapjet|$, for events with at least one jet with 
$\ptjet >30$~GeV and $|\rapjet|<4.4$ in the final state. 
}
\label{tab:comb_rap}
\end{center}
\end{table}



\begin{table}[htbp]
\begin{center}
\begin{footnotesize}
\begin{tabular}{|c| c| c| c|c|} \hline\hline
\multicolumn{5}{|c|}{\small{$\mathrm{d}\sigma / \mathrm{d}|\rapjet| \ [\rm{pb}]$ \ (leading jet)}} \\ \hline\hline 
 $|\rapjet|$ & $\zee$ & $\zmm$ & $\zll$ & $\delta^{\rm had} \pm ({\rm{total \ unc.}})$\\ 
          & $\sigma \pm {\rm{(stat.)}} \pm {\rm{(syst.)}}$ & $\sigma \pm {\rm{(stat.)}} \pm {\rm{(syst.)}}$ & $\sigma \pm {\rm{(total \ unc.)}} $ & parton $\to$ hadron \\ \hline
0.0-0.5 &  $  34 \pm 2 \pm 3$ &  $  33 \pm 2 \pm 2$ & $ 33 {}^{+ 3 }_{- 2 }$ & $ 1.00 \pm  0.03 $ \\
0.5-1.0 &  $  31 \pm 2 \pm 3$ &  $  29 \pm 1 \pm 2$ & $ 30 \pm 2$ & $ 1.00 \pm  0.03 $ \\
1.0-1.5 &  $  26 \pm 2 \pm 2$ &  $  25 \pm 1 \pm 2$ & $ 25 \pm 2$ & $ 1.00 \pm  0.03 $ \\
1.5-2.0 &  $  19 \pm 1 \pm 2$ &  $  18 \pm 1 {}^{+ 2 }_{- 1 }$ & $ 19 \pm 2$ & $ 1.00 \pm  0.03 $ \\
2.0-2.5 &  $  13 \pm 1 \pm 2$ &  $  13 \pm 1 {}^{+ 2 }_{- 1 }$ & $ 13 \pm 2$ & $ 0.99 \pm  0.02 $ \\
2.5-3.0 &  $  10 \pm 1 \pm 2$ &  $  7 \pm 1 \pm 1$ & $ 8 \pm 1$ & $ 0.97 \pm  0.01 $ \\
3.0-3.5 &  $  4.1 \pm 0.7 {}^{+ 0.9 }_{- 0.8 }$ &  $  4.0 \pm 0.6 {}^{+ 0.8 }_{- 0.9 }$ & $ 4.1 \pm 1.0$ & $ 0.94 \pm  0.01 $ \\
3.5-4.0 &  $  1.2 \pm 0.4 {}^{+ 0.5 }_{- 0.4 }$ &  $  0.9 \pm 0.3 \pm 0.3$ & $ 1.0 \pm 0.4$ & $ 0.92 \pm  0.02 $\\ \hline\hline
\end{tabular}
\end{footnotesize}
\caption{\small
Measured jet differential cross section $\mathrm{d}\sigma / \mathrm{d}|\rapjet|$ as a function of the leading-jet $|\rapjet|$, for events with at least one jet with 
$\ptjet >30$~GeV and $|\rapjet|<4.4$ in the final state. 
}
\label{tab:comb_rap1}
\end{center}
\end{table}



\begin{table}[htbp]
\begin{center}
\begin{footnotesize}
\begin{tabular}{|c|c| c| c| c|} \hline\hline
\multicolumn{5}{|c|}{\small{$\mathrm{d}\sigma / \mathrm{d}|\rapjet| \ [\rm{pb}]$ \ (second-leading jet)}} \\ \hline\hline 
 $|\rapjet|$ & $\zee$ & $\zmm$ & $\zll$ & $\delta^{\rm had} \pm ({\rm{total \ unc.}})$ \\ 
          & $\sigma \pm {\rm{(stat.)}} \pm {\rm{(syst.)}}$ & $\sigma \pm {\rm{(stat.)}} \pm {\rm{(syst.)}}$ & $\sigma \pm {\rm{(total \ unc.)}} $  & parton $\to$ hadron \\ \hline
0.0-0.5 &  $  7.0 \pm 0.8 \pm 0.8$ &  $  6.2 \pm 0.7 \pm 0.6$ & $ 6.5 {}^{+ 0.9 }_{- 0.8 }$ & $ 1.00 \pm  0.03 $ \\
0.5-1.0 &  $  6.7 \pm 0.8 \pm 0.8$ &  $  6.0 \pm 0.7 \pm 0.6$ & $ 6.3 {}^{+ 0.9 }_{- 0.8 }$ & $ 0.99 \pm  0.03 $ \\
1.0-1.5 &  $  4.8 \pm 0.7 \pm 0.6$ &  $  5.0 \pm 0.6 {}^{+ 0.6 }_{- 0.5 }$ & $ 5.0 \pm 0.7$ & $ 1.00 \pm  0.03 $ \\
1.5-2.0 &  $  4.6 \pm 0.7 \pm 0.6$ &  $  3.8 \pm 0.5 \pm 0.4$ & $ 4.1 \pm 0.6$ & $ 0.98 \pm  0.02 $ \\
2.0-2.5 &  $  2.2 \pm 0.5 {}^{+ 0.3 }_{- 0.4 }$ &  $  3.3 \pm 0.5 {}^{+ 0.5 }_{- 0.4 }$ & $ 2.8 \pm 0.5$ & $ 0.98 \pm  0.03 $ \\
2.5-3.0 &  $  1.3 \pm 0.4 \pm 0.2$ &  $  1.9 \pm 0.4 {}^{+ 0.4 }_{- 0.3 }$ & $ 1.6 \pm 0.4$ & $ 0.97 \pm  0.05 $ \\ 
3.0-3.5 &  $  1.2 \pm 0.4 \pm 0.3$ &  $  0.8 \pm 0.2 \pm 0.2$ & $ 0.9 \pm 0.3$ & $ 0.97 \pm  0.05 $ \\ \hline\hline
\end{tabular}
\end{footnotesize}
\caption{\small
Measured jet differential cross section $\mathrm{d}\sigma / \mathrm{d}|\rapjet|$ as a function of the second-leading jet $|\rapjet|$, for events with at least two jets with 
$\ptjet >30$~GeV and $|\rapjet|<4.4$ in the final state. 
}
\label{tab:comb_rap2}
\end{center}
\end{table}



\begin{table}[htbp]
\begin{center}
\begin{footnotesize}
\begin{tabular}{|c|c| c| c| c|} \hline\hline
\multicolumn{5}{|c|}{\small{$\mathrm{d}\sigma / \mathrm{d} \mjj \ [\rm{pb}/\rm{GeV}]$}} \\ \hline\hline 
 $\mjj$ & $\zee$ & $\zmm$ & $\zll$ & $\delta^{\rm had} \pm ({\rm{total \ unc.}})$\\ 
 $[\rm{GeV}]$         & $\sigma \pm {\rm{(stat.)}} \pm {\rm{(syst.)}}$ & $\sigma \pm {\rm{(stat.)}} \pm {\rm{(syst.)}}$ & $\sigma \pm {\rm{(total \ unc.)}} $ & parton $\to$ hadron \\ \hline
60-90 &  $  0.06 \pm 0.01 \pm 0.01$ &  $  0.06 \pm 0.01 \pm 0.01$ & $ 0.06 \pm 0.01$ & $ 1.03 \pm  0.04 $ \\
90-120 &  $  0.11 \pm 0.01 \pm 0.01$ &  $  0.10 \pm 0.01 \pm 0.01$ & $ 0.10 {}^{+ 0.02 }_{- 0.01 }$ & $ 1.01 \pm  0.04 $ \\
120-150 &  $  0.06 \pm 0.01 \pm 0.01$ &  $  0.07 \pm 0.01 \pm 0.01$ & $ 0.07 \pm 0.01$ & $ 1.01 \pm  0.03 $ \\
150-180 &  $  0.057 \pm 0.010 \pm 0.008$ &  $  0.043 \pm 0.007 {}^{+ 0.005 }_{- 0.004 }$ & $ 0.047 \pm 0.008$ & $ 1.00 \pm  0.04 $ \\
180-210 &  $  0.042 \pm 0.009 {}^{+ 0.005 }_{- 0.006 }$ &  $  0.036 \pm 0.007 \pm 0.004$ & $ 0.038 \pm 0.007$ & $ 1.00 \pm  0.02 $ \\
210-240 &  $  0.025 \pm 0.007 {}^{+ 0.004 }_{- 0.003 }$ &  $  0.021 \pm 0.005 {}^{+ 0.002 }_{- 0.003 }$ & $ 0.023 \pm 0.005$ & $ 0.98 \pm  0.04 $ \\
240-270 &  $  0.018 \pm 0.006 {}^{+ 0.002 }_{- 0.003 }$ &  $  0.017 \pm 0.005 \pm 0.002$ & $ 0.017 \pm 0.004$ & $ 0.94 \pm  0.06 $ \\
270-300 &  $  0.015 \pm 0.005 \pm 0.003$ &  $  0.017 \pm 0.005 \pm 0.002$ & $ 0.016 \pm 0.004$ & $ 0.95 \pm  0.05 $ \\ \hline\hline
\end{tabular}
\end{footnotesize}
\caption{\small
Measured differential cross section $\mathrm{d}\sigma / \mathrm{d}\mjj$ as a function of the dijet invariant mass, for events with at least two jets with 
$\ptjet >30$~GeV and $|\rapjet|<4.4$ in the final state. 
}
\label{tab:comb_mjj}
\end{center}
\end{table}



\begin{table}[htbp]
\begin{center}
\begin{footnotesize}
\begin{tabular}{|c|c| c| c| c|} \hline\hline
\multicolumn{5}{|c|}{\small{$\mathrm{d}\sigma / \mathrm{d}|\rapjj| \ [\rm{pb}]$}} \\ \hline\hline 
 $|\rapjj|$ & $\zee$ & $\zmm$ & $\zll$ & $\delta^{\rm had} \pm ({\rm{total \ unc.}})$ \\ 
          & $\sigma \pm {\rm{(stat.)}} \pm {\rm{(syst.)}}$ & $\sigma \pm {\rm{(stat.)}} \pm {\rm{(syst.)}}$ & $\sigma \pm {\rm{(total \ unc.)}} $ & parton $\to$ hadron \\ \hline
0.0-0.5 &  $  5.3 \pm 0.7 \pm 0.6$ &  $  5.6 \pm 0.6 \pm 0.6$ & $ 5.5 {}^{+ 0.8 }_{- 0.7 }$ & $ 0.98 \pm  0.04 $ \\
0.5-1.0 &  $  6.1 \pm 0.8 \pm 0.7$ &  $  6.6 \pm 0.7 \pm 0.7$ & $ 6.4 {}^{+ 0.9 }_{- 0.8 }$ & $ 1.02 \pm  0.04 $ \\
1.0-1.5 &  $  5.1 \pm 0.7 \pm 0.6$ &  $  5.0 \pm 0.6 {}^{+ 0.6 }_{- 0.5 }$ & $ 5.1 \pm 0.7$ & $ 1.01 \pm  0.05 $ \\
1.5-2.0 &  $  4.5 \pm 0.7 \pm 0.6$ &  $  3.6 \pm 0.5 \pm 0.4$ & $ 3.9 \pm 0.6$ & $ 1.00 \pm  0.03 $ \\
2.0-2.5 &  $  2.7 \pm 0.5 \pm 0.4$ &  $  3.0 \pm 0.5 {}^{+ 0.4 }_{- 0.3 }$ & $ 2.9 \pm 0.5$ & $ 0.99 \pm  0.04 $ \\
2.5-3.0 &  $  1.8 \pm 0.4 \pm 0.3$ &  $  1.6 \pm 0.3 \pm 0.2$ & $ 1.7 \pm 0.4$ & $ 0.96 \pm  0.02 $ \\
3.0-3.5 &  $  1.6 \pm 0.4 \pm 0.3$ &  $  1.0 \pm 0.3 \pm 0.2$ & $ 1.2 \pm 0.3$ & $ 0.95 \pm  0.03 $ \\ \hline\hline
\end{tabular}
\end{footnotesize}
\caption{\small
Measured differential cross section $\mathrm{d}\sigma / \mathrm{d}|\rapjj|$ as a function of the dijet rapidity separation, for events with at least two jets with 
$\ptjet >30$~GeV and $|\rapjet|<4.4$ in the final state. 
}
\label{tab:comb_yjj}
\end{center}
\end{table}



\begin{table}[htbp]
\begin{center}
\begin{footnotesize}
\begin{tabular}{|c| c| c| c|c|} \hline\hline
\multicolumn{5}{|c|}{\small{$\mathrm{d}\sigma / \mathrm{d}|\phijj| \ [\rm{pb}]$}} \\ \hline\hline 
 $|\phijj|$ [rad.] & $\zee$ & $\zmm$ & $\zll$ & $\delta^{had} \pm ({\rm{total \ unc.}})$\\ 
          & $\sigma \pm {\rm{(stat.)}} \pm {\rm{(syst.)}}$ & $\sigma \pm {\rm{(stat.)}} \pm {\rm{(syst.)}}$ & $\sigma \pm {\rm{(total \ unc.)}} $  & parton $\to$ hadron \\ \hline
$0 - \pi/8$ &  $  1.8 \pm 0.5 \pm 0.3$ &  $  1.7 \pm 0.4 \pm 0.3$ & $ 1.7 \pm 0.4$ & $ 0.94 \pm  0.04 $ \\
$\pi/8 - \pi/4$ &  $  2.7 \pm 0.6 {}^{+ 0.5 }_{- 0.4 }$ &  $  2.9 \pm 0.5 {}^{+ 0.4 }_{- 0.3 }$ & $ 2.8 {}^{+ 0.6 }_{- 0.5 }$ & $ 0.98 \pm  0.05 $ \\
$\pi/4 - 3\pi/8$ &  $  2.0 \pm 0.5 \pm 0.3$ &  $  2.5 \pm 0.5 \pm 0.3$ & $ 2.3 \pm 0.5$ & $ 1.01 \pm  0.07 $ \\
$3\pi/8 - \pi/2$ &  $  2.5 \pm 0.6 \pm 0.4$ &  $  3.2 \pm 0.5 \pm 0.4$ & $ 2.9 {}^{+ 0.6 }_{- 0.5 }$ & $ 0.97 \pm  0.03 $ \\
$\pi/2 -  5\pi/8$ &  $  4.0 \pm 0.7 {}^{+ 0.5 }_{- 0.6 }$ &  $  3.8 \pm 0.6 \pm 0.5$ & $ 3.9 \pm 0.7$ & $ 0.97 \pm  0.02 $ \\
$5\pi/8 - 3\pi/4$ &  $  4.4 \pm 0.8 \pm 0.6$ &  $  4.6 \pm 0.7 \pm 0.6$ & $ 4.5 {}^{+ 0.8 }_{- 0.7 }$ & $ 0.98 \pm  0.04 $ \\
$3\pi/4 - 7\pi/8$&  $  7.9 \pm 1.0 \pm 0.9$ &  $  6.8 \pm 0.8 \pm 0.7$ & $ 7.0 \pm 1.0$ & $ 0.98 \pm  0.03 $ \\
$7\pi/8 - \pi$&  $  11.4 \pm 1.2 \pm 1.4$ &  $  10.0 \pm 1.0 {}^{+ 1.1 }_{- 1.0 }$ & $ 10.4 {}^{+ 1.4 }_{- 1.3 }$ & $ 1.00 \pm  0.08 $ \\ \hline\hline
\end{tabular}
\end{footnotesize}
\caption{\small
Measured differential cross section $\mathrm{d}\sigma / \mathrm{d}|\phijj|$ as a function of the dijet azimuthal separation, for events with at least two jets with 
$\ptjet >30$~GeV and $|\rapjet|<4.4$ in the final state. 
}
\label{tab:comb_fjj}
\end{center}
\end{table}



\begin{table}[htbp]
\begin{center}
\begin{footnotesize}
\begin{tabular}{|c|c| c| c| c|} \hline\hline
\multicolumn{5}{|c|}{\small{$\mathrm{d}\sigma / \mathrm{d}|\rjj| \ [\rm{pb}]$}} \\ \hline\hline 
 $\rjj$ & $\zee$ & $\zmm$ & $\zll$ & $\delta^{\rm had} \pm ({\rm{total \ unc.}})$\\ 
          & $\sigma \pm {\rm{(stat.)}} \pm {\rm{(syst.)}}$ & $\sigma \pm {\rm{(stat.)}} \pm {\rm{(syst.)}}$ & $\sigma \pm {\rm{(total \ unc.)}} $ & parton $\to$ hadron \\ \hline
0.4-0.8 &  $  1.8 \pm 0.5 \pm 0.3$ &  $  1.6 \pm 0.4 \pm 0.3$ & $ 1.7 \pm 0.4$ & $ 0.91 \pm  0.02 $ \\
0.8-1.2 &  $  1.5 \pm 0.4 \pm 0.2$ &  $  1.9 \pm 0.4 \pm 0.2$ & $ 1.7 \pm 0.4$ & $ 1.04 \pm  0.09 $ \\
1.2-1.6 &  $  1.8 \pm 0.5 \pm 0.3$ &  $  2.2 \pm 0.4 \pm 0.3$ & $ 2.1 \pm 0.4$ & $ 0.99 \pm  0.03 $ \\
1.6-2.0 &  $  2.2 \pm 0.5 \pm 0.3$ &  $  2.7 \pm 0.5 \pm 0.3$ & $ 2.5 \pm 0.5$ & $ 1.02 \pm  0.07 $ \\
2.0-2.4 &  $  3.4 \pm 0.7 {}^{+ 0.5 }_{- 0.4 }$ &  $  3.5 \pm 0.6 \pm 0.4$ & $ 3.5 \pm 0.6$ & $ 1.02 \pm  0.07 $ \\
2.4-2.8 &  $  5.7 \pm 0.9 \pm 0.7$ &  $  5.4 \pm 0.7 \pm 0.6$ & $ 5.6 \pm 0.8$ & $ 0.99 \pm  0.02 $ \\
2.8-3.2 &  $  7.8 \pm 1.0 \pm 0.9$ &  $  8.5 \pm 0.9 {}^{+ 0.9 }_{- 0.8 }$ & $ 8.2 {}^{+ 1.1 }_{- 1.0 }$ & $ 1.01 \pm  0.02 $ \\
3.2-3.6 &  $  5.5 \pm 0.8 \pm 0.7$ &  $  4.7 \pm 0.7 \pm 0.5$ & $ 5.0 {}^{+ 0.8 }_{- 0.7 }$ & $ 0.99 \pm  0.03 $ \\
3.6-4.0 &  $  2.5 \pm 0.6 \pm 0.4$ &  $  1.2 \pm 0.3 \pm 0.2$ & $ 1.5 \pm 0.3$ & $ 0.96 \pm  0.03 $ \\
4.0-4.4 &  $  1.5 \pm 0.4 \pm 0.3$ &  $  1.5 \pm 0.4 \pm 0.2$ & $ 1.5 \pm 0.4$ & $ 0.97 \pm  0.05 $ \\ \hline\hline
\end{tabular}
\end{footnotesize}
\caption{\small
Measured differential cross section $\mathrm{d}\sigma / \mathrm{d}|\rjj|$ as a function of the dijet angular  separation ($y - \phi$ space), for events with at least two jets with 
$\ptjet >30$~GeV and $|\rapjet|<4.4$ in the final state. 
}
\label{tab:comb_rjj}
\end{center}
\end{table}

%% file: extrapol.tex
%
%

\begin{table}[htbp]
\begin{center}
\begin{footnotesize}
\begin{tabular}{|c|| c| c|| c| c|} \hline\hline
\multicolumn{5}{|c|}{\small{$\sigma_{\njet}$}} \\ \hline\hline
 $\njet$ & $\delta^{\rm QED}$ (e-channel) &  $\mathcal{A}$ (e-channel) &  $\delta^{\rm QED}$ ($\mu$-channel) & $\mathcal{A}$ ($\mu$-channel) \\ \hline   
$\geq$ 1 jet & $ 1.024 \pm 0.001 $ & $ 1.143 \pm 0.003 $ & $ 1.024 \pm 0.001 $ &  $ 1.046  \pm  0.003 $ \\
$\geq$ 2 jets & $ 1.021 \pm 0.001 $ & $ 1.144 \pm 0.002 $ & $ 1.022 \pm 0.001 $ &  $ 1.045  \pm  0.002 $ \\
$\geq$ 3 jets & $ 1.021 \pm 0.002 $ & $ 1.144 \pm 0.003 $ & $ 1.021 \pm 0.002 $ &   $ 1.045  \pm  0.002 $\\
$\geq$ 4 jets & $ 1.016 \pm 0.003 $ & $ 1.151 \pm 0.007 $ & $ 1.021 \pm 0.002 $ &  $ 1.048  \pm  0.005 $\\ \hline\hline
\multicolumn{5}{|c|}{\small{$\sigma_{\njet}/\sigma_{\njet -1} $}} \\ \hline\hline
 $\njet$ & $\delta^{\rm QED}$ (e-channel) &  $\mathcal{A}$ (e-channel) &  $\delta^{\rm QED}$ ($\mu$-channel) & $\mathcal{A}$ ($\mu$-channel) \\ \hline   
$\geq$ 1 jet  & $1.005 \pm  0.003  $& $1.001 \pm 0.003 $ & $1.005  \pm 0.003 $ & $0.994 \pm  0.001$\\
$\geq$ 2 jets & $0.997 \pm  0.001  $& $1.001 \pm 0.003 $ & $0.998  \pm 0.002 $ & $0.999 \pm  0.001$\\
$\geq$ 3 jets & $1.000 \pm  0.001  $& $1.001 \pm 0.003 $ & $0.999  \pm 0.001 $ & $1.001 \pm  0.002$\\
$\geq$ 4 jets & $0.996 \pm  0.004  $& $1.006 \pm 0.006 $ & $1.001  \pm 0.002 $ & $1.003 \pm  0.003$\\ \hline\hline 
\multicolumn{5}{|c|}{\small{$\mathrm{d}\sigma / \mathrm{d}\ptjet $ \ (inclusive) }} \\ \hline\hline 
 $\ptjet$ [GeV] & $\delta^{\rm QED}$ (e-channel) &  $\mathcal{A}$ (e-channel) &  $\delta^{\rm QED}$ ($\mu$-channel) & $\mathcal{A}$ ($\mu$-channel) \\ \hline   
30-40 & $ 1.029 \pm 0.001 $ & $ 1.142 \pm 0.001 $ & $ 1.029 \pm 0.001 $ & $ 1.048  \pm  0.003 $  \\
40-50 & $ 1.023 \pm 0.003 $ & $ 1.144 \pm 0.002 $ & $ 1.022 \pm 0.005 $ &  $ 1.048  \pm  0.004 $ \\
50-70 & $ 1.019 \pm 0.001 $ & $ 1.143 \pm 0.006 $ & $ 1.021 \pm 0.001 $ &  $ 1.046  \pm  0.003 $ \\
70-90 & $ 1.019 \pm 0.003 $ & $ 1.145 \pm 0.005 $ & $ 1.019 \pm 0.003 $ &  $ 1.043  \pm  0.003 $\\
90-120 & $ 1.019 \pm 0.003 $ & $ 1.143 \pm 0.002 $ & $ 1.020 \pm 0.003 $ &  $ 1.040  \pm  0.002 $\\
120-150 & $ 1.017 \pm 0.004 $ & $ 1.144 \pm 0.008 $ & $ 1.020 \pm 0.004 $ & $ 1.040  \pm  0.005 $ \\ 
150-180 & $ 1.016 \pm 0.002 $ & $ 1.142 \pm 0.011 $ & $ 1.017 \pm 0.004 $ &  $ 1.036  \pm  0.006 $ \\ \hline\hline
\multicolumn{5}{|c|}{\small{$\mathrm{d}\sigma / \mathrm{d}|\rapjet| $ \ (inclusive) }} \\ \hline\hline 
 $|\rapjet|$ & $\delta^{\rm QED}$ (e-channel) &  $\mathcal{A}$ (e-channel) &  $\delta^{\rm QED}$ ($\mu$-channel) & $\mathcal{A}$ ($\mu$-channel) \\ \hline   
0.0-0.5 & $ 1.024 \pm 0.003 $ & $ 1.135 \pm 0.002 $ & $ 1.024 \pm 0.002 $ &  $ 1.035  \pm  0.001 $\\
0.5-1.0 & $ 1.024 \pm 0.001 $ & $ 1.139 \pm 0.003 $ & $ 1.024 \pm 0.001 $ &  $ 1.039  \pm  0.002 $\\
1.0-1.5 & $ 1.024 \pm 0.001 $ & $ 1.140 \pm 0.005 $ & $ 1.024 \pm 0.002 $ &  $ 1.047  \pm  0.005 $\\
1.5-2.0 & $ 1.024 \pm 0.002 $ & $ 1.148 \pm 0.005 $ & $ 1.024 \pm 0.001 $ &  $ 1.056  \pm  0.006 $\\
2.0-2.5 & $ 1.022 \pm 0.001 $ & $ 1.160 \pm 0.011 $ & $ 1.022 \pm 0.001 $ &  $ 1.059  \pm  0.005 $\\
2.5-3.0 & $ 1.020 \pm 0.001 $ & $ 1.158 \pm 0.003 $ & $ 1.020 \pm 0.001 $ &  $ 1.067  \pm  0.010 $\\
3.0-3.5 & $ 1.019 \pm 0.001 $ & $ 1.152 \pm 0.004 $ & $ 1.020 \pm 0.006 $ &  $ 1.068  \pm  0.011 $\\
3.5-4.0 & $ 1.025 \pm 0.007 $ & $ 1.163 \pm 0.017 $ & $ 1.016 \pm 0.002 $ &  $ 1.065  \pm  0.008 $\\ \hline\hline
\end{tabular}
\end{footnotesize}
\caption{\small
Multiplicative correction factors, applied to the data in the electron and muon channels, that  
extrapolate the measured cross sections to the common acceptance region 
$\ptjet > 20$~GeV and $|\eta| < 2.5$ for the lepton kinematics, defined
at the decay vertex of the Z boson before QED radiation.
}
\label{tab:accept1}
\end{center}
\end{table}


\begin{table}[htbp]
\begin{center}
\begin{footnotesize}
\begin{tabular}{|c|| c| c|| c| c|} \hline\hline
\multicolumn{5}{|c|}{\small{$\mathrm{d}\sigma / \mathrm{d}\ptjet$ \ (leading jet)}} \\ \hline\hline
 $\ptjet$ [GeV] & $\delta^{\rm QED}$ (e-channel) &  $\mathcal{A}$ (e-channel) &  $\delta^{\rm QED}$ ($\mu$-channel) & $\mathcal{A}$ ($\mu$-channel) \\ \hline   
30-40 & $ 1.031 \pm 0.001 $ & $ 1.142 \pm 0.001 $ & $ 1.031 \pm 0.002 $ & $ 1.049  \pm  0.003 $ \\
40-50 & $ 1.023 \pm 0.005 $ & $ 1.143 \pm 0.004 $ & $ 1.022 \pm 0.006 $ & $ 1.048  \pm  0.004 $ \\
50-70 & $ 1.020 \pm 0.001 $ & $ 1.143 \pm 0.005 $ & $ 1.021 \pm 0.001 $ & $ 1.046  \pm  0.004 $ \\
70-90 & $ 1.020 \pm 0.004 $ & $ 1.146 \pm 0.006 $ & $ 1.019 \pm 0.003 $ & $ 1.043  \pm  0.003 $ \\
90-120 & $ 1.019 \pm 0.003 $ & $ 1.142 \pm 0.003 $ & $ 1.020 \pm 0.003 $ & $ 1.040  \pm  0.002 $  \\
120-150 & $ 1.017 \pm 0.004 $ & $ 1.144 \pm 0.010 $ & $ 1.020 \pm 0.004 $ & $ 1.038  \pm  0.003 $ \\
150-180 & $ 1.016 \pm 0.002 $ & $ 1.141 \pm 0.014 $ & $ 1.016 \pm 0.004 $ & $ 1.036  \pm  0.008 $\\ \hline
\multicolumn{5}{|c|}{\small{$\mathrm{d}\sigma / \mathrm{d}|\rapjet|$ \ (leading jet)}} \\ \hline\hline  
 $|\rapjet|$ & $\delta^{\rm QED}$ (e-channel) &  $\mathcal{A}$ (e-channel) &  $\delta^{\rm QED}$ ($\mu$-channel) & $\mathcal{A}$ ($\mu$-channel) \\ \hline   
0.0-0.5 & $ 1.024 \pm 0.003 $ & $ 1.133 \pm 0.004 $ & $ 1.025 \pm 0.002 $ & $ 1.034  \pm  0.001 $\\
0.5-1.0 & $ 1.025 \pm 0.001 $ & $ 1.137 \pm 0.003 $ & $ 1.024 \pm 0.001 $ & $ 1.037  \pm  0.002 $\\ 
1.0-1.5 & $ 1.025 \pm 0.001 $ & $ 1.141 \pm 0.005 $ & $ 1.025 \pm 0.003 $ & $ 1.047  \pm  0.005 $\\
1.5-2.0 & $ 1.025 \pm 0.002 $ & $ 1.150 \pm 0.005 $ & $ 1.024 \pm 0.001 $ & $ 1.057  \pm  0.006 $\\ 
2.0-2.5 & $ 1.023 \pm 0.001 $ & $ 1.161 \pm 0.010 $ & $ 1.022 \pm 0.001 $ & $ 1.063  \pm  0.006 $\\
2.5-3.0 & $ 1.020 \pm 0.001 $ & $ 1.164 \pm 0.006 $ & $ 1.020 \pm 0.002 $ & $ 1.073  \pm  0.010 $\\
3.0-3.5 & $ 1.019 \pm 0.002 $ & $ 1.159 \pm 0.006 $ & $ 1.021 \pm 0.009 $ & $ 1.076  \pm  0.015 $\\
3.5-4.0 & $ 1.025 \pm 0.008 $ & $ 1.170 \pm 0.016 $ & $ 1.017 \pm 0.002 $ & $ 1.074  \pm  0.012 $\\ \hline
\multicolumn{5}{|c|}{\small{$\mathrm{d}\sigma / \mathrm{d}\ptjet$ \ (second-leading jet)}} \\ \hline\hline 
 $\ptjet$ [GeV] & $\delta^{\rm QED}$ (e-channel) &  $\mathcal{A}$ (e-channel) &  $\delta^{\rm QED}$ ($\mu$-channel) & $\mathcal{A}$ ($\mu$-channel) \\ \hline   
30-40 & $ 1.023 \pm 0.001 $ & $ 1.141 \pm 0.002 $ & $ 1.024 \pm 0.001 $ & $ 1.045  \pm  0.002 $ \\
40-50 & $ 1.022 \pm 0.003 $ & $ 1.147 \pm 0.003 $ & $ 1.021 \pm 0.003 $ & $ 1.046  \pm  0.002 $ \\
50-70 & $ 1.018 \pm 0.001 $ & $ 1.146 \pm 0.008 $ & $ 1.020 \pm 0.001 $ & $ 1.043  \pm  0.001 $ \\
70-90 & $ 1.015 \pm 0.002 $ & $ 1.142 \pm 0.009 $ & $ 1.019 \pm 0.005 $ & $ 1.045  \pm  0.005 $ \\
90-120 & $ 1.024 \pm 0.009 $ & $ 1.148 \pm 0.008 $ & $ 1.021 \pm 0.008 $ & $ 1.040  \pm  0.003 $ \\ \hline
\multicolumn{5}{|c|}{\small{$\mathrm{d}\sigma / \mathrm{d}|\rapjet|$ \ (second-leading jet)}} \\ \hline\hline  
 $|\rapjet|$ & $\delta^{\rm QED}$ (e-channel) &  $\mathcal{A}$ (e-channel) &  $\delta^{\rm QED}$ ($\mu$-channel) & $\mathcal{A}$ ($\mu$-channel) \\ \hline   
0.0-0.5 & $ 1.021 \pm 0.002 $ & $ 1.142 \pm 0.003 $ & $ 1.020 \pm 0.002 $ & $ 1.038  \pm  0.002 $ \\
0.5-1.0 & $ 1.021 \pm 0.002 $ & $ 1.148 \pm 0.004 $ & $ 1.023 \pm 0.002 $ & $ 1.042  \pm  0.002 $ \\
1.0-1.5 & $ 1.021 \pm 0.002 $ & $ 1.137 \pm 0.005 $ & $ 1.024 \pm 0.002 $ & $ 1.047  \pm  0.003 $ \\
1.5-2.0 & $ 1.021 \pm 0.002 $ & $ 1.142 \pm 0.004 $ & $ 1.023 \pm 0.002 $ & $ 1.053  \pm  0.003 $ \\
2.0-2.5 & $ 1.021 \pm 0.007 $ & $ 1.155 \pm 0.008 $ & $ 1.021 \pm 0.002 $ & $ 1.048  \pm  0.004 $ \\
2.5-3.0 & $ 1.022 \pm 0.003 $ & $ 1.140 \pm 0.008 $ & $ 1.019 \pm 0.004 $ & $ 1.048  \pm  0.011 $ \\
3.0-3.5 & $ 1.018 \pm 0.004 $ & $ 1.143 \pm 0.007 $ & $ 1.017 \pm 0.005 $ & $ 1.051  \pm  0.011 $ \\ \hline\hline
\end{tabular}
\end{footnotesize}
\caption{\small
Multiplicative correction factors, applied to the data in the electron and muon channels, that  
extrapolate the measured cross sections to the common acceptance region 
$\ptjet > 20$~GeV and $|\eta| < 2.5$ for the lepton kinematics, defined
at the decay vertex of the Z boson before QED radiation.
}
\label{tab:accept2}
\end{center}
\end{table}


\begin{table}[htbp]
\begin{center}
\begin{footnotesize}
\begin{tabular}{|c|| c| c|| c| c|} \hline\hline
\multicolumn{5}{|c|}{\small{$\mathrm{d}\sigma / \mathrm{d} \mjj $}} \\ \hline\hline
 $\mjj$ [GeV] & $\delta^{\rm QED}$ (e-channel) &  $\mathcal{A}$ (e-channel) &  $\delta^{\rm QED}$ ($\mu$-channel) & $\mathcal{A}$ ($\mu$-channel) \\ \hline  
60-90 & $ 1.025 \pm 0.004 $ & $ 1.148 \pm 0.006 $ & $ 1.025 \pm 0.005 $ &  $ 1.044  \pm  0.004 $  \\
90-120 & $ 1.023 \pm 0.002 $ & $ 1.141 \pm 0.005 $ & $ 1.025 \pm 0.002 $ & $ 1.046  \pm  0.004 $  \\
120-150 & $ 1.022 \pm 0.002 $ & $ 1.138 \pm 0.004 $ & $ 1.022 \pm 0.002 $ & $ 1.047  \pm  0.006 $ \\
150-180 & $ 1.016 \pm 0.002 $ & $ 1.146 \pm 0.006 $ & $ 1.021 \pm 0.002 $ & $ 1.043  \pm  0.008 $ \\
180-210 & $ 1.017 \pm 0.003 $ & $ 1.149 \pm 0.007 $ & $ 1.019 \pm 0.004 $ & $ 1.042  \pm  0.002 $ \\
210-240 & $ 1.016 \pm 0.002 $ & $ 1.141 \pm 0.010 $ & $ 1.020 \pm 0.004 $ & $ 1.049  \pm  0.006 $ \\
240-270 & $ 1.022 \pm 0.006 $ & $ 1.140 \pm 0.013 $ & $ 1.022 \pm 0.007 $ & $ 1.045  \pm  0.009 $ \\
270-300 & $ 1.026 \pm 0.015 $ & $ 1.154 \pm 0.016 $ & $ 1.018 \pm 0.005 $ & $ 1.041  \pm  0.009 $ \\ \hline
\multicolumn{5}{|c|}{\small{$\mathrm{d}\sigma / \mathrm{d}|\rapjj| $}} \\ \hline\hline
 $|\rapjj|$ & $\delta^{\rm QED}$ (e-channel) &  $\mathcal{A}$ (e-channel) &  $\delta^{\rm QED}$ ($\mu$-channel) & $\mathcal{A}$ ($\mu$-channel) \\ \hline  
0.0-0.5 & $ 1.021 \pm 0.001 $ & $ 1.146 \pm 0.004 $ & $ 1.023 \pm 0.001 $ & $ 1.041  \pm  0.001 $ \\
0.5-1.0 & $ 1.021 \pm 0.004 $ & $ 1.148 \pm 0.009 $ & $ 1.024 \pm 0.003 $ & $ 1.042  \pm  0.004 $ \\
1.0-1.5 & $ 1.022 \pm 0.002 $ & $ 1.141 \pm 0.004 $ & $ 1.021 \pm 0.003 $ & $ 1.046  \pm  0.004 $  \\
1.5-2.0 & $ 1.022 \pm 0.001 $ & $ 1.141 \pm 0.004 $ & $ 1.022 \pm 0.004 $ & $ 1.044  \pm  0.002 $   \\
2.0-2.5 & $ 1.021 \pm 0.004 $ & $ 1.132 \pm 0.010 $ & $ 1.022 \pm 0.002 $ & $ 1.045  \pm  0.004 $   \\
2.5-3.0 & $ 1.017 \pm 0.003 $ & $ 1.147 \pm 0.008 $ & $ 1.017 \pm 0.002 $ & $ 1.050  \pm  0.003 $    \\
3.0-3.5 & $ 1.019 \pm 0.002 $ & $ 1.145 \pm 0.009 $ & $ 1.023 \pm 0.008 $ & $ 1.052  \pm  0.007 $   \\ \hline
\end{tabular}
\end{footnotesize}
\caption{\small
Multiplicative correction factors, applied to the data in the electron and muon channels, that  
extrapolate the measured cross sections to the common acceptance region 
$\ptjet > 20$~GeV and $|\eta| < 2.5$ for the lepton kinematics, defined
at the decay vertex of the Z boson before QED radiation.
}
\label{tab:accept3}
\end{center}
\end{table}


\begin{table}[htbp]
\begin{center}
\begin{footnotesize}
\begin{tabular}{|c|| c| c|| c| c|} \hline\hline
\multicolumn{5}{|c|}{\small{$\mathrm{d}\sigma / \mathrm{d}|\phijj| $}} \\ \hline\hline
 $|\phijj|$ [rad.] & $\delta^{\rm QED}$ (e-channel) &  $\mathcal{A}$ (e-channel) &  $\delta^{\rm QED}$ ($\mu$-channel) & $\mathcal{A}$ ($\mu$-channel) \\ \hline  
$0 - \pi/8$  &     $ 1.020 \pm 0.006 $ & $ 1.138 \pm 0.007 $ & $ 1.018 \pm 0.004 $ & $ 1.034 \pm 0.002 $ \\
$\pi/8 - \pi/4$  & $ 1.020 \pm 0.004 $ & $ 1.146 \pm 0.007 $ & $ 1.022 \pm 0.007 $ & $ 1.038 \pm 0.004 $ \\
$\pi/4 - 3\pi/8$ & $ 1.017 \pm 0.002 $ & $ 1.144 \pm 0.005 $ & $ 1.021 \pm 0.002 $ & $ 1.037 \pm 0.004 $ \\
$3\pi/8 - \pi/2$ & $ 1.021 \pm 0.002 $ & $ 1.137 \pm 0.005 $ & $ 1.021 \pm 0.002 $ & $ 1.040 \pm 0.002 $ \\
$\pi/2 - 5\pi/8$&$ 1.021 \pm 0.004 $ & $ 1.149 \pm 0.014 $ & $ 1.021 \pm 0.001 $ & $ 1.043 \pm 0.003 $ \\
$5\pi/8 - 3\pi/4$ &$ 1.022 \pm 0.002 $ & $ 1.140 \pm 0.003 $ & $ 1.026 \pm 0.002 $ & $ 1.048 \pm 0.009 $ \\
$3\pi/4 - 7\pi/8$& $ 1.022 \pm 0.002 $ & $ 1.148 \pm 0.004 $ & $ 1.024 \pm 0.002 $ & $ 1.047 \pm 0.003 $ \\
$7\pi/8 - \pi$ &   $ 1.022 \pm 0.001 $ & $ 1.143 \pm 0.002 $ & $ 1.021 \pm 0.002 $ & $ 1.050 \pm 0.003 $\\ \hline
\multicolumn{5}{|c|}{\small{$\mathrm{d}\sigma / \mathrm{d}|\rjj| $}} \\ \hline\hline   
 $\rjj$ & $\delta^{\rm QED}$ (e-channel) &  $\mathcal{A}$ (e-channel) &  $\delta^{\rm QED}$ ($\mu$-channel) & $\mathcal{A}$ ($\mu$-channel) \\ \hline  
0.4-0.8 & $ 1.018 \pm 0.006 $ & $ 1.142 \pm 0.006 $ & $ 1.017 \pm 0.005 $ & $ 1.041 \pm 0.011 $ \\
0.8-1.2 & $ 1.016 \pm 0.006 $ & $ 1.145 \pm 0.012 $ & $ 1.021 \pm 0.006 $ & $ 1.029 \pm 0.007 $ \\ 
1.2-1.6 & $ 1.021 \pm 0.003 $ & $ 1.147 \pm 0.016 $ & $ 1.019 \pm 0.003 $ & $ 1.038 \pm 0.004 $ \\
1.6-2.0 & $ 1.022 \pm 0.005 $ & $ 1.142 \pm 0.004 $ & $ 1.025 \pm 0.003 $ & $ 1.037 \pm 0.004 $ \\
2.0-2.4 & $ 1.022 \pm 0.002 $ & $ 1.147 \pm 0.004 $ & $ 1.024 \pm 0.003 $ & $ 1.043 \pm 0.005 $ \\
2.4-2.8 & $ 1.023 \pm 0.001 $ & $ 1.144 \pm 0.003 $ & $ 1.025 \pm 0.001 $ & $ 1.044 \pm 0.002 $ \\
2.8-3.2 & $ 1.022 \pm 0.003 $ & $ 1.139 \pm 0.005 $ & $ 1.023 \pm 0.001 $ & $ 1.046 \pm 0.003 $ \\
3.2-3.6 & $ 1.019 \pm 0.001 $ & $ 1.144 \pm 0.011 $ & $ 1.020 \pm 0.003 $ & $ 1.048 \pm 0.002 $ \\
3.6-4.0 & $ 1.020 \pm 0.004 $ & $ 1.140 \pm 0.013 $ & $ 1.020 \pm 0.003 $ & $ 1.051 \pm 0.006 $ \\
4.0-4.4 & $ 1.020 \pm 0.003 $ & $ 1.147 \pm 0.009 $ & $ 1.021 \pm 0.003 $ & $ 1.056 \pm 0.007 $ \\ \hline
\end{tabular}
\end{footnotesize}
\caption{\small
Multiplicative correction factors, applied to the data in the electron and muon channels, that  
extrapolate the measured cross sections to the common acceptance region 
$\ptjet > 20$~GeV and $|\eta| < 2.5$ for the lepton kinematics, defined
at the decay vertex of the Z boson before QED radiation.
}
\label{tab:accept4}
\end{center}
\end{table}


\clearpage

%% file: chi2.tex
\begin{table}[htbp]
 \begin{footnotesize}
 \begin{center}
 \begin{tabular}{|l||c|c|c|} \hline
 \multicolumn{4}{|c|}{{\small{$\chi^2$ results (w.r.t. NLO pQCD predictions)}}} \\ \hline\hline\hline
 \multicolumn{4}{|c|}{{e-channel}} \\ \hline
     measurement & degrees of freedom   &  $\chi^2$/d.o.f & $\chi^2$/d.o.f  \\
      & (d.o.f)   &  (fiducial) & (extrapolated) \\ \hline 
  $\sigma_{\njet}$                               & 4 & 1.43  & 1.43  \\
  $\sigma_{\njet}/\sigma_{\njet - 1}$            & 4 & 1.54  & 1.59  \\   
  $\mathrm{d}\sigma/\mathrm{d}\ptjet$ (inclusive)                  & 7 & 0.17  & 0.18 \\
  $\mathrm{d}\sigma/\mathrm{d}|\rapjet|$ (inclusive)               & 8 & 0.79  & 0.74 \\
  $\mathrm{d}\sigma/\mathrm{d}\ptjet$ (leading jet)                & 7 & 0.28  & 0.29 \\
  $\mathrm{d}\sigma/\mathrm{d}|\rapjet|$ (leading jet)             & 8 & 1.19  & 1.16 \\
  $\mathrm{d}\sigma/\mathrm{d}\ptjet$ (second-leading jet)         & 5 & 0.05  & 0.06 \\
  $\mathrm{d}\sigma/\mathrm{d}|\rapjet|$ (second-leading jet)      & 7 & 0.79  & 0.85 \\
  $\mathrm{d}\sigma/\mathrm{d}\mjj$                                & 8 & 0.98  & 0.98 \\
  $\mathrm{d}\sigma/\mathrm{d}|\rapjj|$                            & 7 & 0.32  & 0.34 \\  
  $\mathrm{d}\sigma/\mathrm{d}|\phijj|$                            & 8 & 0.43  & 0.44 \\ 
  $\mathrm{d}\sigma/\mathrm{d}\rjj$                                & 10 & 0.14 & 0.16 \\ \hline\hline
 \multicolumn{4}{|c|}{{$\mu$-channel}} \\ \hline
     measurement & degrees of freedom   &  $\chi^2$/d.o.f & $\chi^2$/d.o.f  \\
      & (d.o.f)   &  (fiducial) & (extrapolated) \\ \hline 
  $\sigma_{\njet}$                               & 4 & 0.09  & 0.11  \\
  $\sigma_{\njet}/\sigma_{\njet - 1}$            & 4 & 0.07  & 0.08  \\   
  $\mathrm{d}\sigma/\mathrm{d}\ptjet$ (inclusive)                  & 7 & 1.78 & 1.77 \\
  $\mathrm{d}\sigma/\mathrm{d}|\rapjet|$ (inclusive)               & 8 & 0.43 & 0.41 \\
  $\mathrm{d}\sigma/\mathrm{d}\ptjet$ (leading jet)                & 7 & 1.17 & 1.13 \\
  $\mathrm{d}\sigma/\mathrm{d}|\rapjet|$ (leading jet)             & 8 & 0.46 & 0.46 \\
  $\mathrm{d}\sigma/\mathrm{d}\ptjet$ (second-leading jet)         & 5 & 1.20 & 1.23 \\
  $\mathrm{d}\sigma/\mathrm{d}|\rapjet|$ (second-leading jet)      & 7 & 0.32 & 0.32 \\
  $\mathrm{d}\sigma/\mathrm{d}\mjj$                                & 8 & 0.61 & 0.61  \\
  $\mathrm{d}\sigma/\mathrm{d}|\rapjj|$                            & 7 & 0.96 & 0.96  \\  
  $\mathrm{d}\sigma/\mathrm{d}|\phijj|$                            & 8 & 0.54 & 0.55  \\ 
  $\mathrm{d}\sigma/\mathrm{d}\rjj$                                & 10 & 1.54 & 1.55 \\ \hline\hline
\end{tabular}
\label{tab:cnlo}
\caption{\small
Results of $\chi^2$ tests to the electron and muon data with 
respect to the NLO pQCD predictions. The results are tabulated for the 
original cross section measurements and after extrapolating to the Born level in a common region 
for the lepton kinematics.} 
\end{center}
\end{footnotesize}
\end{table}

\begin{table}[htbp]
 \begin{footnotesize}
 \begin{center}
 \begin{tabular}{|l||c|c|c|} \hline
 \multicolumn{4}{|c|}{{\small{$\chi^2$ results (w.r.t. ALPGEN predictions)}}} \\ \hline\hline\hline
 \multicolumn{4}{|c|}{{e-channel}} \\ \hline
     measurement & degrees of freedom   &  $\chi^2$/d.o.f & $\chi^2$/d.o.f  \\
      & (d.o.f)   &  (fiducial) & (extrapolated) \\ \hline 
  $\sigma_{\njet}$                               & 4 & 0.99   & 0.99 \\
  $\sigma_{\njet}/\sigma_{\njet - 1}$            & 4 & 1.55   &  1.55       \\   
  $\mathrm{d}\sigma/\mathrm{d}\ptjet$ (inclusive)                  & 7 & 0.13   & 0.13  \\
  $\mathrm{d}\sigma/\mathrm{d}|\rapjet|$ (inclusive)               & 8 & 0.97   & 0.97 \\
  $\mathrm{d}\sigma/\mathrm{d}\ptjet$ (leading jet)                & 7 & 0.17   & 0.17  \\
  $\mathrm{d}\sigma/\mathrm{d}|\rapjet|$ (leading jet)             & 8 & 1.33   & 1.33  \\
  $\mathrm{d}\sigma/\mathrm{d}\ptjet$ (second-leading jet)         & 5 & 0.07   & 0.07  \\
  $\mathrm{d}\sigma/\mathrm{d}|\rapjet|$ (second-leading jet)      & 7 & 0.63   & 0.63  \\
  $\mathrm{d}\sigma/\mathrm{d}\mjj$                                & 8 & 0.87   & 0.87  \\
  $\mathrm{d}\sigma/\mathrm{d}|\rapjj|$                            & 7 & 0.42   & 0.42  \\  
  $\mathrm{d}\sigma/\mathrm{d}|\phijj|$                            & 8 & 0.44   & 0.44  \\ 
  $\mathrm{d}\sigma/\mathrm{d}\rjj$                                & 10 & 0.25  & 0.25  \\ \hline\hline
 \multicolumn{4}{|c|}{{$\mu$-channel}} \\ \hline
     measurement & degrees of freedom   &  $\chi^2$/d.o.f & $\chi^2$/d.o.f  \\
      & (d.o.f)   &  (fiducial) & (extrapolated) \\ \hline 
  $\sigma_{\njet}$                               & 4 & 0.08  &  0.08 \\
  $\sigma_{\njet}/\sigma_{\njet - 1}$            & 4 & 0.11   & 0.11   \\   
  $\mathrm{d}\sigma/\mathrm{d}\ptjet$ (inclusive)                  & 7 & 1.87 & 1.87  \\
  $\mathrm{d}\sigma/\mathrm{d}|\rapjet|$ (inclusive)               & 8 & 0.71 & 0.71  \\
  $\mathrm{d}\sigma/\mathrm{d}\ptjet$ (leading jet)                & 7 & 1.29 & 1.29  \\
  $\mathrm{d}\sigma/\mathrm{d}|\rapjet|$ (leading jet)             & 8 & 0.60 & 0.60  \\
  $\mathrm{d}\sigma/\mathrm{d}\ptjet$ (second-leading jet)         & 5 & 0.89 & 0.89  \\
  $\mathrm{d}\sigma/\mathrm{d}|\rapjet|$ (second-leading jet)      & 7 & 0.50 & 0.50  \\
  $\mathrm{d}\sigma/\mathrm{d}\mjj$                                & 8 & 0.58 & 0.58  \\
  $\mathrm{d}\sigma/\mathrm{d}|\rapjj|$                            & 7 & 0.90 & 0.90  \\  
  $\mathrm{d}\sigma/\mathrm{d}|\phijj|$                            & 8 & 0.43 & 0.43  \\ 
  $\mathrm{d}\sigma/\mathrm{d}\rjj$                                & 10 & 1.59 & 1.59  \\ \hline\hline
\end{tabular}
\label{tab:calpgen}
\caption{\small
Results of $\chi^2$ tests to the electron and muon data with 
respect to the ALPGEN predictions. The results are tabulated for the 
original cross section measurements and after extrapolating to the Born level in a common region 
for the lepton kinematics.} 
\end{center}
\end{footnotesize}
\end{table}

\begin{table}[htbp]
 \begin{footnotesize}
 \begin{center}
 \begin{tabular}{|l||c|c|c|} \hline
 \multicolumn{4}{|c|}{{\small{$\chi^2$ results (w.r.t. Sherpa predictions)}}} \\ \hline\hline\hline
 \multicolumn{4}{|c|}{{e-channel}} \\ \hline
     measurement & degrees of freedom   &  $\chi^2$/d.o.f & $\chi^2$/d.o.f  \\
      & (d.o.f)   &  (fiducial) & (extrapolated) \\ \hline 
  $\sigma_{\njet}$                               & 4 &  0.65  & 0.68  \\
  $\sigma_{\njet}/\sigma_{\njet - 1}$            & 4 & 1.51  & 1.50  \\   
  $\mathrm{d}\sigma/\mathrm{d}\ptjet$ (inclusive)                  & 7 & 0.29   & 0.29 \\
  $\mathrm{d}\sigma/\mathrm{d}|\rapjet|$ (inclusive)               & 8 & 1.19   & 1.13 \\
  $\mathrm{d}\sigma/\mathrm{d}\ptjet$ (leading jet)                & 7 & 0.30   & 0.30  \\
  $\mathrm{d}\sigma/\mathrm{d}|\rapjet|$ (leading jet)             & 8 & 1.85   & 1.82  \\
  $\mathrm{d}\sigma/\mathrm{d}\ptjet$ (second-leading jet)         & 5 & 0.27   & 0.29  \\
  $\mathrm{d}\sigma/\mathrm{d}|\rapjet|$ (second-leading jet)      & 7 & 1.24   & 1.22  \\
  $\mathrm{d}\sigma/\mathrm{d}\mjj$                                & 8 & 1.25   & 1.22  \\
  $\mathrm{d}\sigma/\mathrm{d}|\rapjj|$                            & 7 & 0.80   & 0.80  \\  
  $\mathrm{d}\sigma/\mathrm{d}|\phijj|$                            & 8 & 0.59   & 0.60  \\ 
  $\mathrm{d}\sigma/\mathrm{d}\rjj$                                & 10 & 0.36  & 0.36  \\ \hline\hline
 \multicolumn{4}{|c|}{{$\mu$-channel}} \\ \hline
     measurement & degrees of freedom   &  $\chi^2$/d.o.f & $\chi^2$/d.o.f  \\
      & (d.o.f)   &  (fiducial) & (extrapolated) \\ \hline 
  $\sigma_{\njet}$                               & 4 & 0.37  & 0.38  \\
  $\sigma_{\njet}/\sigma_{\njet - 1}$            & 4 & 0.31  & 0.32  \\   
  $\mathrm{d}\sigma/\mathrm{d}\ptjet$ (inclusive)                  & 7 & 2.10 & 2.08  \\
  $\mathrm{d}\sigma/\mathrm{d}|\rapjet|$ (inclusive)               & 8 & 0.73 & 0.74  \\
  $\mathrm{d}\sigma/\mathrm{d}\ptjet$ (leading jet)                & 7 & 1.24 & 1.21  \\
  $\mathrm{d}\sigma/\mathrm{d}|\rapjet|$ (leading jet)             & 8 & 1.02 & 1.06  \\
  $\mathrm{d}\sigma/\mathrm{d}\ptjet$ (second-leading jet)         & 5 & 1.40 & 1.40  \\
  $\mathrm{d}\sigma/\mathrm{d}|\rapjet|$ (second-leading jet)      & 7 & 0.62 & 0.62  \\
  $\mathrm{d}\sigma/\mathrm{d}\mjj$                                & 8 & 1.01 & 1.02  \\
  $\mathrm{d}\sigma/\mathrm{d}|\rapjj|$                            & 7 & 1.86 & 1.85  \\  
  $\mathrm{d}\sigma/\mathrm{d}|\phijj|$                            & 8 & 1.02 & 1.03  \\ 
  $\mathrm{d}\sigma/\mathrm{d}\rjj$                                & 10 & 2.70 & 2.70  \\ \hline\hline
\end{tabular}
\label{tab:csherpa}
\caption{\small
Results of $\chi^2$ tests to the electron and muon data with 
respect to the Sherpa predictions. The results are tabulated for the 
original cross section measurements and after extrapolating to the Born level in a common region 
for the lepton kinematics.} 
\end{center}
\end{footnotesize}
\end{table}

%% file: atlas_authlist.tex
\begin{flushleft}
{\Large The ATLAS Collaboration}

\bigskip

G.~Aad$^{\rm 48}$,
B.~Abbott$^{\rm 111}$,
J.~Abdallah$^{\rm 11}$,
A.A.~Abdelalim$^{\rm 49}$,
A.~Abdesselam$^{\rm 118}$,
O.~Abdinov$^{\rm 10}$,
B.~Abi$^{\rm 112}$,
M.~Abolins$^{\rm 88}$,
H.~Abramowicz$^{\rm 153}$,
H.~Abreu$^{\rm 115}$
% ...
\bigskip

% ...
$^{10}$ Institute of Physics, Azerbaijan Academy of Sciences, Baku, Azerbaijan\\
$^{11}$ Institut de F\'isica d'Altes Energies and Departament de F\'isica de la Universitat Aut\`onoma  de Barcelona and ICREA, Barcelona, Spain\\
% ...
$^{48}$ Fakult\"{a}t f\"{u}r Mathematik und Physik, Albert-Ludwigs-Universit\"{a}t, Freiburg i.Br., Germany\\
$^{49}$ Section de Physique, Universit\'e de Gen\`eve, Geneva, Switzerland\\
% ...
$^{88}$ Department of Physics and Astronomy, Michigan State University, East Lansing MI, United States of America\\
% ...
$^{111}$ Homer L. Dodge Department of Physics and Astronomy, University of Oklahoma, Norman OK, United States of America\\
$^{112}$ Department of Physics, Oklahoma State University, Stillwater OK, United States of America\\
% ...
$^{115}$ LAL, Univ. Paris-Sud and CNRS/IN2P3, Orsay, France\\
% ...
$^{118}$ Department of Physics, Oxford University, Oxford, United Kingdom\\
% ...
$^{153}$ Raymond and Beverly Sackler School of Physics and Astronomy, Tel Aviv University, Tel Aviv, Israel\\
% ...
\end{flushleft}

%% file: zjets.bbl
\begin{thebibliography}{99.}
\bibitem{coord} 
The ATLAS reference system is a Cartesian 
right-handed coordinate system, with the nominal collision point at the origin. The anti-clockwise beam 
direction defines the positive $z$-axis, while the positive $x$-axis is defined as pointing from the collision 
point to the centre of the LHC ring and the positive $y$-axis points upwards.  The azimuthal angle $\phi$ is 
measured around the beam axis, and the polar angle $\theta$ is measured with respect to the $z$-axis.
The pseudorapidity is defined as $\eta = - {\rm ln}({\rm tan}(\theta/2))$. 
The rapidity is defined as $y = 0.5 \times {\rm ln}[(E+p_z)/(E-p_z)]$, where $E$ denotes the energy and $p_z$ is the component of the momentum along the beam direction.
\bibitem{hadron} The hadronic final state in the MC generators is defined using all particles
with lifetime above 10${}^{-11}$~s, after excluding the $Z/\gamma^*$ decay products.
\end{thebibliography}
